\documentclass[a4paper,10pt]{article}
\usepackage{cite,amsmath,amsfonts,amsthm,fullpage}
\usepackage{youngtab}
\newcommand{\Pf}{\mathop\mathrm{Pf}\nolimits}
\newcommand{\sgn}{\mathop\mathrm{sgn}\nolimits}
\newcommand{\Pa}{\mathop\mathrm{P}\nolimits}
\newcommand{\SCP}{\mathop\mathrm{SCP}\nolimits}
\newcommand{\DP}{\mathop\mathrm{DP}\nolimits}
\newcommand{\FP}{\mathop\mathrm{FP}\nolimits}
\newcommand{\FDP}{\mathop\mathrm{FDP}\nolimits}
\newcommand{\SCDP}{\mathop\mathrm{SCDP}\nolimits}
\newcommand{\bt}{\mathbf{t}}
\newcommand{\bbt}{{\bar{\mathbf{t}}}}
\newcommand{\bs}{\mathbf{s}}

\theoremstyle{plain}

\newtheorem{Lemma}{Lemma}
\newtheorem{Proposition}{Proposition}

\theoremstyle{remark}
\newtheorem{Remark}{Remark}

\def\l{\langle}
\def\r{\rangle}
\def\g{\Gamma}
\def\t{\texttt{T}}

\def\Tr{\mathrm {Tr}}

\def\det{\mathrm {det}}

\def\ln{\mathrm {ln}}

\def\diag{\mathrm {diag}}

\def\Hf{\mathop\mathrm{Hf}\nolimits}

\def\bp{\begin{Proposition}\rm}
\def\ep{\end{Proposition}}
\def\bc{\begin{corollary}}
\def\ec{\end{corollary}}
\def\bl{\begin{Lemma}\em}
\def\el{\end{Lemma}}
\def\be{\begin{equation}}
\def\ee{\end{equation}}
\def\br{\begin{Remark}\rm\small}
\def\er{\end{Remark}}
\def\brs{\begin{remarks}.\\ \rm\
\begin{enumerate}}
\def\ers{\end{enumerate}\end{remarks}}
\def\bea{\begin{eqnarray}}
\def\eea{\end{eqnarray}}

\def\Tr{\mathrm {Tr}}

\def\det{\mathrm {det}}

\def\sgn{\mathrm {sgn}}
\def\ln{\mathrm {ln}}

\def\diag{\mathrm {diag}}

\def\&{&{\hskip -20pt}}

\newcount\YDcount\YDcount=0
\def\YDsize{10pt}

\def\YD#1{%
\ifnum#1=0
 \ifnum\YDcount=0 \ifx\varnothing\undefined\emptyset\else\varnothing\fi
 \else\vskip1.4pt\egroup\YDcount=0\fi
\else
 \ifnum\YDcount=0 \YDcount=1\vcenter\bgroup\vskip1pt
 \else\nointerlineskip\fi
 \vbox{\hrule\hbox{\vrule height\YDsize
 \loop\hskip\YDsize\vrule\ifnum\YDcount<#1\advance\YDcount1\repeat}\hrule
 \kern-0.4pt}\expandafter\YD
\fi}

\begin{document}
\author{ A. Yu.
Orlov\thanks{Oceanology Institute, Nahimovskii Prospekt 36,
Moscow, Russia, email: orlovs55@mail.ru
, and National Research University Higher School of Economics,
International Laboratory of Representation
Theory and Mathematical Physics,
20 Myasnitskaya Ulitsa, Moscow 101000, Russia, email: orlovs@ocean.ru
}\and T. Shiota\thanks{Kyoto university
} \and K.
Takasaki\thanks{Kyoto university}}
\title{Pfaffian structures and certain solutions to
  BKP hierarchies II. Multiple integrals.}

\maketitle

\begin{abstract}

We introduce a useful and rather simple classes of 
BKP tau functions which which we shall shall call ``easy tau functions''.
We consider the ``large BKP hiearchy'' related to $O(2\infty +1)$ which was introduced in \cite{KvdLbispec}  
(which is closely related to the DKP $O(2\infty) $hierarchy introduced in \cite{JM}).
Actually ``easy tau functions'' of the small BKP was already considered in \cite{HLO}, here
we are more interested in the large BKP and also the mixed small-large BKP tau functions \cite{KvdLbispec}.
Tau functions under consideration are equal to sums over partitions and to multi-integrals.
In this way they may be appliciable  in models of random partitions and models of random matrices.
Here in the part II we consider multi-intergals and series of $N$-ply integrals in $N$.
Relations to matrix models is explained. This part of our work may be viewed as a developement of 
the paper by J.van de Leur
\cite{L1} related to orthogonal and symplectic ensembles of random matrices.

\end{abstract}

\bigskip


\section{Introduction}

This is the second part of the paper "Pfaffian structures and certain solutions to
  BKP hierarchies"  devoted to the special family of tau functions which may be called ``easy tau functions''; 
for the first part see \cite{OST-I}.
In the first part we consider sums over partitions originating from both small and large BKP hierarchies introduced
respectively in \cite{JM} and \cite{KvdLbispec}.
We shall refer the large BKP hierarchy just as the BKP one, and tau functions of the large BKP as BKP tau function.
Here we consider certain classes of multiple integrals which depend on parameters and which may by treated as BKP tau functions 
where the parameters play the role of higher times.
Let us note that sums and integrals related to the small BKP hierarchy was previousely considered in \cite{HLO}.

In both parts of our work we use the fermionic approach to tau functions suggested in \cite{JM} and used in \cite{KvdLbispec} to study large BKP,
multicomponent BKP and mixed of large and small BKP tau functions.

Sums of the previous part of our paper \cite{OST-I} and the integrals considered below may be related in two different ways.
(A) The first way is the straightforward specification of the integration measure which may be chosen as the sum of Dirac delta functions.
(B) The second is the presentation of an integral in form of the asymptotic series (in this case the integral may be viewed as the Borel 
summation of the series). In this second 
way we equate multiple integrals to multiple series and this may be viewed as a sort
of Fourier transform.
In the fermionic method which we use the second way follows directly from the formula
 \be
\psi(x)=\sum_{n\in\mathbb{Z}}\,\psi_n x^n
  \ee
which allows to re-write integrals over $x$ which enter expectation values in terms of sums.

Some multiple integrals considered below originate from studies of
ensembles of random matrices. The link between matrix integrals
and soliton theory was found in \cite{GMMMO} for $\beta=2$
ensembles, and in \cite{AvM-Pfaff},\cite{AMS} for orthogonal
$\beta=1$ and symplectic $\beta=4$ ensembles of random matrices.
Here we complete the list of multi-integrals and sums which may be
recognized as (large) BKP tau functions. In particular we
consider partition functions for circular $\beta=1,2,4$ ensembles
as BKP tau functions. 
In case of applications to matrix models BKP higher times play the role
of the so-called coupling constants; then it is convenient to write the perturbation series in coupling constants 
in form of a sum over partitions. 
The perturbation series may be asymptotic one in this case the multi-integral if it is well-defined
play a role of Borel summation of the series.

\section{Multiple integrals  \label{Multiple integrals-section}}

We shall consider two types of multi-integrals. The first type is a $N$-integral where each of these integrals
is evaluated along the same contour, say $\gamma$, namely, this is a integral over $\gamma^{\times N}$. Examples of 
such integrals applicable for the presentation of the orthogonal and symplectic ensembles are widely known. The 
contour may be a circle $S^1$ (for circular ensembles), or, it may be a real line $\mathbb{R}$.  
The second type are integrals over $\mathbb{C}^N$
$N=2n$ where each integral over $\mathbb{C}$  is actually evaluated over a domain in the complex plane, say, upper halfplane.  
To describe ensembles like the real Ginibre one we need the both types of integrals.

To describe integrals below we need certain data denoted by ${\bar A}:=(A,a)$ (compare with \cite{OST-I}) where $A$ and $a$ are
respectively functions of two and one variables  provided  $A(z,z')=-A(z',z)$  (instead of functions 
distribution may be also considered). 

The notation ${\bar A}({\bf z})$ is analogous to  (\ref{A-c}), denoting the
Pfaffian (see Appendix \ref{tools-section} for the definition) of an skew symmetric matrix ${\tilde A}$:
  \be\label{Pf-tilde-A}
{\bar A}({\bf z}):=\,\Pf[{\tilde A}]
  \ee
  whose entries are defined, depending on the parity of $N$, in terms
  of a skew symmetric kernel $A(z,w) $ (possibly, a distribution) and a function
  (or a distribution) $a(z)$ as  follows:

 For $N=2n$ even
  \be\label{A-even-N}
{\tilde A}_{ij}=-{\tilde A}_{ji}:=A(z_i,z_j),\quad 1\le i<j \le 2n
  \ee

  For $N=2n-1$ odd
 \be\label{A-odd-N} {\tilde A}_{ij}=-{\tilde A}_{ji}:=
\begin{cases}
A(z_i,z_j) &\mbox{ if }\quad 1\le i<j \le 2n-1 \\
a(z_i) &\mbox{ if }\quad 1\le i < j=2n
 \end{cases}
  \ee
In addition we define ${\bar A}_0 =1$.

\subsection{Integrals along contours\label{integrals along contours}}

\paragraph{Integrals along $\gamma^{\times N}$. } 

First we generalize some results of \cite{L1}, where nice fermionic expressions were found for the 
partition functions of orthogonal and symplectic ensembles in $\gamma=\mathbb{R}$ case.

Let $d\mu$ be a measure supported on a contour $\gamma$ on the
complex plane. We suppose that $\varsigma$ is a parameter along the contour. 
Our main examples of $\gamma$ are as follows :

 (A) An interval on the real axes $-\infty \le z < \infty$. Then $\varsigma(z)=z\in\mathbb{R}$

 (B) A segment of the unit  circle: given  by
  $z=e^{i\varphi},\, 0\le \varphi \le\theta$, \ $0\le \theta \le 2\pi$. In this case $\varsigma(z)=\arg(z)$
\hfill  \break

  \noindent

We shall study $N$-fold integrals over the cone $\Lambda_N$
 \be\label{Lambda_N}
{\bf z}=(z_1,\dots,z_N)\in\Lambda_N \quad \mbox{iff}  \quad z_1,\dots,z_N\in\gamma \quad \mbox{and} 
\quad \varsigma(z_1)>\cdots >\varsigma(z_N)
  \ee
 defined as follows
\be\label{I^{(1)}}
I^{(1)}(\bt^*,N,{\bar A}):=\int_{\Lambda_N}\,\Delta_N(z)\,{\bar A}({\bf z}) \,
\prod_{i=1}^N\, d\mu(z_i,\bt^*)
 \ee
where 
 the notation ${\bar A}({\bf z})$ was defined earlier in \eqref{Pf-tilde-A} and where
 \be
\Delta_N(z)=\prod_{i<j}^N (z_i-z_j)
 \ee
and
 \be
d\mu(z,\bt^*):=\,e^{\sum_{n=1}^\infty t^*_nz^n+t^*_0\log z -\sum_{n=1}^\infty t^*_{-n}z^{-n}}d\mu(z)
 \ee
Here the set $\bt^*=\{t^*_n,\,n\in\mathbb{Z}\}$ are parameters (sometimes called coupling constants).
We assume that the measure $d\mu(z,\bt^*)$ is chosen in a way that the integral \eqref{I^{(1)}} is convergent.

Using various specifications of ${\bar A}$ we in particular obtain \footnote{Via fermionic construction used both 
in \cite{OST-I} and in the present paper sums $S^{(1)}_i$ may be related to $I^{(1)}_i$. In our notations below we 
keep the numeration adopted in \cite{OST-I} which is rather conventional.}
 \be\label{I{(1)}1}
I^{(1)}_1(\bt^*,N):=\int_{\Lambda_N}\,\Delta_N(z)
\,\prod_{i=1}^N\, d\mu_1(z_i,\bt^*)
 \ee
  \be\label{I{(1)}2}
   I^{(1)}_2(\bt^*,N=2n):=\int_{\Lambda_n}\,\Delta_n(z^2)^2
\,\prod_{i=1}^n\, d\mu_2(z_i,\bt^*)
 \ee
  \be\label{I{(1)}4}
   I^{(1)}_4(\bt^*,N=2n):=\int_{\Lambda_n}\,\Delta_n(z)^4
\,\prod_{i=1}^n\, d\mu_4(z_i,\bt^*)
 \ee
where 
 \be
d\mu_1(z,\bt^*)=d\mu(z,\bt^*)=e^{\sum_{n=1}^\infty t^*_nz^n+t^*_0\log z -\sum_{n=1}^\infty t^*_{-n}z^{-n}}d\mu(z)
 \ee
 \be
d\mu_2(z,\bt^*)=
(-1)^{t^*_0}e^{2\sum_{n=1}^\infty t^*_{2n}z^{2n}+2t^*_0\log z -2\sum_{n=1}^\infty t^*_{-2n}z^{-2n}} d\mu(z)
 \ee
 \be
d\mu_4(z,\bt^*)=e^{2\sum_{n=1}^\infty t^*_nz^n+2t^*_0\log z -2\sum_{n=1}^\infty t^*_{-n}z^{-n}}d\mu(z)
 \ee
These three integrals are related to the well-known random ensembles if $\gamma=\mathbb{R}$ or $\gamma=S^1$, see \cite{Mehta}.
In case $\gamma=\mathbb{R}$ integrals $I^{(1)}_1$ ,$I^{(1)}_2$ and $I^{(1)}_4$ describe 
respectively the models of random symmetric, of real symmetric and of symplectic matrices. In case
$\gamma=S^1$ integrals $I^{(1)}_1$, $I^{(1)}_2$ and $I^{(1)}_4$ describe respectively $\beta=1,2,4$ circular ensembles.

Other examples:
  \be\label{I^{(1)}I56789}
   I^{(1)}_{m}(\bt^*,N):=\int_{\Lambda_N}\,\Delta^{(m)}_N(z)\Delta_N(z)
\,\prod_{i=1}^N\, d\mu_m(z_i,\bt^*)\,,\quad m=5,6,7,8,9
 \ee
where by $\Delta^{(m)}$ we denote the following various Vandermond-like products:
 \be
\Delta^{(5)}_N(z)=\Delta^{(5)}_N(z,f,c):=\prod_{i<j}^N \frac{f(z_i)-f(z_j)}{f(z_i)+f(z_j)+cf(z_i)f(z_j)}
 \ee
($c$ is a constant and $f$ is a arbitrary function). The case $c=0$, $f(z)=z$ is related to the so-called Bures ensembles \cite{OsipovSommers} and for $c=0$
let us use the notation
  \be
\Delta^{*}_N(x):=\prod_{i<j}^N \frac{z_i-z_j}{z_i+z_j}
  \ee
used earlier in \cite{HLO}. Then
 \be
\Delta^{(6)}_N(z):=\prod_{i<j}^N \frac{f(z_i)-f(z_j)}{1-f(z_i)f(z_j)}
 \ee
 \be
\Delta^{(7)}_N(z):=\prod_{i<j}^N
\frac{f(z_i)-f(z_j)}{f(z_i)+f(z_j)}\,\,\Hf\left(\frac 1{f(z_i)+f(z_j)} \right)
 \ee
 where $\Hf$ denotes the Hafnian (see Appendix \ref{tools-section},
 where one needs to pay attention to the case of $N$ odd where we add a variable
 $z_{N+1}$ to the set of $z_1,\dots,z_N$ and then put $z_{N+1}=0$).

 \be\label{Delta^{(8)}}
\Delta^{(8)}_N(z) := \Delta_N(f(z))=\prod_{i<j}^N\left(f(z_i)-f(z_j)\right)
 \ee

 \be\label{Delta^{(9)}}
\Delta^{(9)}_N(z):=\Pf \frac{1}{f(z_i)-f(z_j)}=\frac{1}{\Delta_N(f(z))}\,
\mbox{Symm}\,\left(\prod_{i,j\in J_1}(f(z_i)-f(z_j))^2\prod_{i,j\in J_2}(f(z_i)-f(z_j))^2  \right)
 \ee
where the last equality is known due to the works on quantum Hall effect \cite{Todorov} where $z_i$ is related to 
the particle $i$. 
The two subsets, $J_1$ and $J_2$ , each have $N/2$ particles for $N$ even and
$(N-1)/2$ and $(N +1)/2$ for odd $N$ . $Symm$ indicates the
symmetrization  over the distributions of the particles
(variables $z_i$, $i=1,\dots,N$) into these subsets.

At last ${\Delta}_n^{(10)}(z)$ is the Vandermond determinant
$\,$
 \be \label{Delta^{(10)}}
{\Delta}_n^{(10)}(z):={\Delta}_{2n}(z_1,\frac{1}{z_1},\dots,z_n,\frac{1}{z_n} )
={\Delta_n(z)^2}{\prod_{i,j=1,\dots,n}(1-z_iz_j)}\,
\prod_{i=1}^n\,z_i^{1-2n}
 \ee  
Let us note that up to a sign factor ${\Delta}_n^{(10)}(z)$ coincides with its absolute value in
case all $|z_i|=1$.

Now let us present the specifications of data ${\bar A}$ giving rise to the integrals \eqref{I{(1)}1}-\eqref{I^{(1)}I56789}.
For \eqref{A5}-\eqref{A8} we have used the series of papers by Ishikawa and co-authors (see \cite{Ishikawa},\cite{IKO} and references therein) as a 
source of Pfaffian relations. Eq-s. \eqref{Delta^{(9)}},\eqref{A9} borrowed from \cite{Todorov}.
 \be\label{A1}
A_1(z_i,z_j)=\frac 12 \sgn\left(\varsigma(z_i)-\varsigma(z_j)\right), \quad a_1(z)=1
 \ee
 \be\label{A2}
A_2(z_i,z_j) = \delta(z_i+z_j)\sgn\left(\varsigma(z_i)-\varsigma(z_j)\right)
 \ee
 \be\label{A4}
A_4(z_i,z_j)= \frac 12 \left(\partial_{z_i}\delta(z_i-z_j)-(i \leftrightarrow j )\right)
 \ee
 \be\label{A5}
A_5(z_i,z_j)= \frac{f(z_i)-f(z_j)}{f(z_i)+f(z_j)+cf(z_i)f(z_j)}
 \ee
 \be\label{A6}
A_6(z_i,z_j)= \frac{f(z_i)-f(z_j)}{1-f(z_i)f(z_j)}
 \ee
 \be\label{A7}
A_7(z_i,z_j)= \frac{f(z_i)-f(z_j)}{(f(z_i)+f(z_j))^2}
 \ee
 \be\label{A8}
A_8(z_i,z_j)= \frac{\left(f(z_i)^n-f(z_j)^n\right)^2}{(f(z_i)-f(z_j))}
 \ee
\be\label{A9}
A_9(z_i,z_j)= \frac{1}{f(z_i)-f(z_j)}
 \ee
\be\label{A10}
A_{10}(z_i,z_j)= \frac 12\left(\delta\left(z_i-\frac{1}{z_j}\right)-(i \leftrightarrow j )\right)
 \ee

\bp

For each choice of contour $\gamma$, data $d\mu$ and ${\bar A}$ provided that the integral $I^{(1)}(\bt^*,N,{\bar A})$
exists, this integral is a tau function of the large 2-BKP hierarchy with respect to the time variables $\bt^*$ and the discrete
variable $N$.

\ep

\paragraph{ Applications and remarks}. Applications of the integrals $I^{(1)}$ we know are as follows

(0) The series \eqref{I^{(1)}} for a special choice of data ${\bar A}$ may be identified with the partition function of the
Mehta-Pandey interpolating ensembles, see \cite{Mehta}, Chapter 14. This link will be explained below.

(1) In case $\gamma=\mathbb{R}$ the integral $I^{(1)}_1$ (up to a factor equal to the volume of $O(N)$ group) coincides
with the partition function of the orthogonal Wigner-Dyson ensemble (see \cite{Mehta}, Chapter 7) with a generalized 
(not necessarily Gaussian) probability weight parametrized by $\bt^*$. The link of this model with
integrable systems (Pfaff lattice) was discovered in \cite{AvM-Pfaff}. The nice expression for $I^{(1)}_1$ as fermionic vacuum 
expectation value was found in \cite{L1} and in this way it was shown that $I^{(1)}_1$ is a tau function of the large BKP hierarchy.
 In case $\gamma=S^1$ the integral $I^{(1)}_4$ is a partition function for the $\beta=1$ circular ensemble, see \cite{Mehta}, 
Section 10.1. In the present paper we consider an arbitrary $\gamma$ and add the dependence on $t^*_{n},\,n<0$.

(2) The integral $I^{(1)}_2$ (up to a factor) coincides with the partition function of the ensemble of anti-symmetric Hermitian
matrices, see \cite{Mehta}, Chapter 13. The relation of this ensemble to integrable systems follows from the fact that in the variables
$x=z^2$ it coincides with the known ``one-matrix model'' which known to be Toda chain tau function \cite{GMMMO}.

(4)  In case $\gamma=\mathbb{R}$ the integral $I^{(1)}_4$ (up to a factor) coincides
with the partition function of the symplectic Wigner-Dyson ensemble (see \cite{Mehta}, Chapter 8) with a generalized 
(not necessarily Gaussian) probability weight parametrized by $\bt^*$. The nice expression for $I^{(1)}_4$ as fermionic vacuum 
expectation value was found in \cite{L1} and in this way it was shown that $I^{(1)}_4$ is a tau function of the large BKP hierarchy.
 In case $\gamma=S^1$ the integral $I^{(1)}_4$ is a partition function for the $\beta=4$ circular ensemble, see \cite{Mehta}, Section 10.2
In the present paper we consider an arbitrary $\gamma$ and add the dependence on $t^*_{n},\,n<0$.

(5) The integrals $I^{(1)}_5$  where $f(z)=z$ and $c=0$ describe the so-called ${\hat A}_0$ statistical model, see formula (30) in 
\cite{Kostov}\footnote{In \cite{Kostov} the grand partition function for ${\hat A}_0$ model was considered and it was shown that it 
is a KdV tau function },  and also the so-called Bures ensembles which appears in quantum 
chaos problems where random density matrix appear \cite{OsipovSommers}. The integrals $I^{(1)}_5$ where $f(z)=e^{iz}$ and $c=0$
contains
  \[
 \Delta_N(z)\Delta^{(5)}_N(z,f,0)\prod_{i=1}^N dz_i=
\prod_{k<j}^N (z_k-z_j)\,\prod_{k<j}^N \tanh \pi (z_k-z_j) \prod_{i=1}^N dz_i  
  \]
which up to a normalization constant coincides with the Plancheral measure for the group $SL(N,\mathbb{R})$ (and for the symmetric 
space $SL(N,\mathbb{R})/SO(N)$),  see Section 17.2.8 in \cite{KV} where put $\lambda_i=\frac 12 z_i$.


(6)-(7) Unknown

(8) Integral $I^{(1)}_8$ where $f(z)=z$ may be considered as the ground state wave function for 
fractional quantum Hall state with filling factor 5/2 (Moore–Read state), see for instance \cite{Paredes}, re-written
in moment  representation (moments $\bt$ were introduced in \cite{ZabrodinWiegmannMineev} to describe quantum Hall droplets
in the quasiclassical limit).

(9) Unkown

(10) Unitary ensemble (under the change $u_i=z_i+z^{-1}_i$). 

\subsection{Integrals over complex plane \label{integrals over complex plane}}

Here we shall consider more general case of multi-integrals over complex planes. First let us introduce the ordering of 
a given number of points $z_1,\dots,z_N$ on the complex plane. We impose
 \be
\Re z_i \ge \Re z_{i+1}
 \ee
and if $\Re z_i = \Re z_{i+1}$ then $\Im z_i \ge \Im z_{i+1}$. Then we introduce
 \be \label{mathbb{M}} 
{\bf z}=(z_1,\dots,z_N)\in \mathbb{M}_N \qquad {\mbox {iff}}\quad
\begin{cases}
\Re z_i > \Re z_{i+1} & \\
\Im z_i > \Im z_{i+1}  &\mbox{ if } \quad \Re z_i = \Re z_{i+1}.
 \end{cases}
 \ee

By ${\bar z}$ we shall denote complex conjugated to $z$ (it should not be mixed it with the special notation ${\bar A}:=(A,a)$).

Now let us consider 2N-ply integrals similar to \eqref{I^{(1)}} where the integral over $\Lambda_N$ is replaced by the
integral over $\mathbb{M}_N$:
 \be\label{I^{c}}
I^{c}(\bt,N)=
\int_{\mathbb{M}_N} \Delta_N(z) {\bar A}({\bf z})\prod_{i=1}^{N} e^{\varphi(z_i,\bt^*)} d\mu(z_i,{\bar z}_i)
 \ee
where $d\mu(z,{\bar z})=d\mu({\bar z},z)$ is an arbitrary (perhaps, complex) symmetric measure.

\paragraph{Ginibre and interpolating ensembles. The models.} 

Details concerning the so-called Ginibre ensembles may be found in \cite{Mehta}, Ch. 15.

Consider the Ginibre ensembles ($i=1,2,4$ belows is related respectively to real, complex and real quaternionic 
Ginibre enesembles) with a deformed measure
\be\label{Gin-deformed}
J^{Gin}_i(\bt,N) =\int  d\mu_i(Z,\bt)\,,\quad i=1,2,4
\ee
where $Z$
 \be
d\mu_i(Z,\bt)=e^{\sum_{m=1}^\infty t_m\Tr Z^m} d_i\mu(Z)
 \ee
where the measure $d_i\mu(Z)$, $i=1,2,4$ is defined respectively by \eqref{Ginibre-measure-1}, \eqref{Ginibre-measure-2} and
\eqref{Ginibre-measure-4}.

\paragraph{Ginibre and interpolating ensembles and large BKP tau functions}

\bp
Integrals \eqref{Gin-deformed} are large BKP tau functions where $\bt$ are higher times.

\ep

Let us restrict ourselves to the case 
  \be
A(z,{\bar z},w,{\bar w})=\frac 12 \left(\mu(z,{\bar z})\delta^{(2)}(z-{\bar w})-\mu(w,{\bar w})\delta^{(2)}(w-{\bar z})\right)
  \ee
 that is
  \be\label{z{2i-1}={bar z}{2i}}
z_{2i-1}={\bar z}_{2i}, \quad i=1,\dots, N=2n
  \ee
In this way we obtain partition function for the celebrated Ginibre ensembles, see see \cite{KhoruzhSommers} for a review.
 Below we re-enumerate variables $z_i$ as follows:
$z_{2i}\to z_i$, $z_{2i-1}\to z_i$, $i=1,\dots,n$  which a natural in view of \eqref{z{2i-1}={bar z}{2i}}.  
Quaternionic Ginibre ensemble with a deformed measure
 \be\label{I^{Gin}_{qu-r}}
I^{Gin}_{qu-r}(\bt^*,N=2n)=\int_{\mathbb{M}_n} \,\Delta_{2n}(z_1,{\bar z}_1,z_2,{\bar z}_2,\dots,z_n,{\bar z}_n)
\prod_{i=1}^n|z_i-{\bar z}_i|e^{\varphi(z_i,\bt^*)+\varphi({\bar z}_i,\bt^*)-|z_i|^2} d^2 z_i
 \ee
where now in view of the re-numeration above we write that $\mathbb{M}_n$ consists of the sets of ${\bf z}$ where 
$\Re z_i>\Re z_{i+1}$ and $\Im z_i > 0$,  and 
 \be
\mu(z,{\bar z})=|z-{\bar z}|e^{-|z|^2}e^{\varphi(z,\bt^*)+\varphi({\bar z},\bt^*)} 
  \ee
where
 \be\label{varphi(z,bt^*)}
\varphi(z,\bt^*)=\sum_{n=1}^\infty t^*_nz^n+t^*_0\log z -\sum_{n=1}^\infty t^*_{-n}z^{-n}
 \ee
In case $D$ is the upper half-plane and $t^*_n=t\delta_{n,2}$ we obtain an interpolating ensemble where 
$t$ is an interpolating parameter, see \cite{AkemannPhillipsShifrin}.

In the similar way we obtain the complex part of the real Ginibre ensemble. This time we take
 \be
d\mu(z,{\bar z})= 
\mbox{erfc}\left(\frac{|z-{\bar z}|}{\sqrt 2} \right) e^{-\frac 12 z^2-\frac 12 {\bar z}^2}
e^{\varphi(z,\bt^*)+\varphi({\bar z},\bt^*)}d^2 z 
 \ee
The whole partition function which takes into account both complex and real parts of the spectrum of this ensemble is 
easily obtained.
 \be\label{I^{Gin}_{r}}
I^{Gin}_{r}(\bt^*,N)=
\sum_{m=0}^{\left[\frac N2 \right]}\int_{\mathbb{M}_{2m,N-2m}} \,
\Delta_{2n}(z_1,{\bar z}_1,\dots ,z_m,{\bar z}_m,x_{m+1},\dots,x_N)\,
d\Omega^C_{2m}\,d\Omega^R_{N-2m}
 \ee
where the integration domain $\mathbb{M}_{2m,N-2m}$ is as follows  $\Re z_1>\cdots >\Re z_m$, $x_{2m+1}>\cdots>x_N$, 
$z_i\in\mathbb{C}_+$ (upper halfplane), $x_i\in\mathbb{R}$, and where
 \[
  d\Omega^C_{2m}=
\prod_{i=1}^m \mbox{erfc}\left(\frac{|z_i-{\bar z}_i|}{\sqrt 2} \right) 
 e^{\varphi(z_i,\bt^*)+\varphi({\bar z}_i,\bt^*)} d^2 z_i\,,\qquad
 d\Omega^R_{N-2m}=
\prod_{i=2m+1}^N 
 e^{\varphi(x_i,\bt^*)} d x_i
 \]
(to get real Ginibre ensemble itself we put $t^*_k=-\frac 12 \delta_{k,2}$.)

\subsection{Integrals as fermionic vacuum expectation values}

We have
 \be\label{I{(1)}-f}
 I^{(1)}(\bt^*,N,{\bar A})=
\l N+t^*_0|\,\g(\bt^*_+)\, e^{\int_{\Lambda_2}A(z_1,z_2)\psi(z_1)\psi(z_2)dz_1dz_2
+\int_\gamma a(z)\psi(z)\phi_0 dz} \,{\bar\g}(\bt^*_-)\,|t^*_0\r
 \ee
where for $\Lambda_2$ see \eqref{Lambda_N}.

In particular
\be\label{I{(1)}_i-f}
 I^{(1)}_i(\bt^*,N)\,=\,
\l N+t^*_0|\,\g(\bt^*_+)\, e^{\int_{\Lambda_2}A_i(z_1,z_2)\psi(z_1)\psi(z_2)dz_1dz_2
+\int_\gamma a_i(z)\psi(z)\phi_0 dz} \,{\bar\g}(\bt^*_-)\,|t^*_0\r
 \ee
where $(A_i,a_i)$ are given by \eqref{A1}-\eqref{A9}. For instance
 \be\label{I{(1)}1-f}
I^{(1)}_1(\bt^*,N)\,=\,
\l N+t^*_0|\,\g(\bt^*_+)\, e^{\frac 12 \int_\gamma\int_\gamma 
\sgn\left(\varsigma(z_1)-\varsigma(z_2)\right)\psi(z_1)\psi(z_2)dz_1dz_2
+\int_\gamma \psi(z)\phi_0 dz} \,{\bar\g}(\bt^*_-)\,|t^*_0\r
 \ee
where, say, for the $\beta=1$ circular ensemble we take $\gamma=S^1$ and $\varsigma(z)=\arg (z)$. Then
  \be\label{I{(1)}2-f}
   I^{(1)}_2(\bt^*,N=2n)\,=\,
\l N+t^*_0|\,\g(\bt^*_+)\, e^{\frac 12 \oint \sgn\left(\varsigma(z)-\varsigma(-z)\right)\psi(z)\psi(-z)dz
+\oint \psi(z)\phi_0 dz} \,{\bar\g}(\bt^*_-)\,|t^*_0\r
 \ee
 For circular $\beta=4$ ensemble we take  $\gamma=S^1$ and $A(z_1,z_2)=\delta'(\arg(z_1)-\arg(z_2))$
  \be\label{I{(1)}4-f}
   I^{(1)}_4(\bt^*,N=2n)=\,
\l N+t^*_0|\,\g(\bt^*_+)\, e^{\frac 12 \oint \frac{\partial\psi(z)}{dz}\psi(z)dz
+\oint \psi(z)\phi_0 dz} \,{\bar\g}(\bt^*_-)\,|t^*_0\r
 \ee

Now turn to the integrals over domains in {\bf complex plane}. We have
\be\label{I^{c}-f}
I^{c}(\bt^*,N)\,=\,
\l N+t^*_0|\,\g(\bt^*_+)\, 
e^{\int_{\mathbb{M}_2}A(z_1,{\bar z}_1,z_2,{\bar z}_2)\psi(z_1)\psi(z_2)d\mu(z_1,{\bar z}_1)d\mu(z_2,{\bar z}_2)
+\int_{\mathbb{C}} a(z)\psi(z)\phi_0 d\mu(z,{\bar z})} \,{\bar\g}(\bt^*_-)\,|t^*_0\r
 \ee
(see \eqref{mathbb{M}} for $\mathbb{M}_2$.)
 
Quaternionic Ginibre ensemble with the deformed measure
 \be\label{I^{Gin}_{qu-r}-f}
I^{Gin}_{qu-r}(\bt^*,N=2n)\,=\,
\l N+t^*_0|\,\g(\bt^*_+)\, e^{\frac 12 \int_{\mathbb{C}}(z-{\bar z})\psi(z)\psi({\bar z})e^{-|z|^2}d^2z} 
\,{\bar\g}(\bt^*_-)\,|t^*_0\r
 \ee

Real Ginibre ensemble with the deformed measure
 \be\label{I^{Gin}_{r}-f}
I^{Gin}_{r}\,=\,
 \ee
 \[
\l N+t^*_0|\,\g(\bt^*_+)\, 
e^{ 
\int_{\mathbb{C}_+} \mbox{erfc}\left(\frac{|z-{\bar z}|}{\sqrt 2} \right) 
\psi(z)\psi({\bar z})d^2z +
\frac 12 \int_{\mathbb{R}}\int_{\mathbb{R}} \sgn(x_1-x_2) \psi(x_1)\psi(x_2)dx_1dx_2
+\int_{\mathbb{R}} \psi(x)\phi_0 dx
} 
\,{\bar\g}(\bt^*_-)\,|t^*_0\r
 \]
where the real Ginibre ensemble itself is related to the case $t^*_m=-\frac 12\delta_{m,2}$.

\subsection{Integrals in Pfaffian form}

 \be\label{I^{(1)}=Pfaff}
I^{(1)}(\bt^*,N,{\bar A})\,=\,b \Pf \left[{\bar M}^{(1)}(\bt^*)\right]_{i,j=1,\dots,N}
 \ee
where ${\bar M}$ is the moment $N \times N$ matrix which is defined as follows:

For $N=2n$ even
 \be
{\bar M}^{(1)}_{ij}=-{\bar M}^{(1)}_{ji} = M^{(1)}_{ij}(\bt^*)
 \ee
where
 \be
M^{(1)}_{ij}(\bt^*)=\int_{\Lambda_2}\,z_1^iz_2^j A(z_1,z_2) d\mu(z_1,\bt^*)d\mu(z_2,\bt^*)
 \ee

  For $N=2n-1$ odd
 \be\label{M-odd-N} {\bar M}^{(1)}_{ij}=-{\bar M}^{(1)}_{ji}:=
\begin{cases}
M^{(1)}_{i,j}(\bt^*) &\mbox{ if }\quad 1\le i<j \le 2n-1 \\
M^{(1)}_i(\bt^*) &\mbox{ if }\quad 1\le i < j=2n
 \end{cases}
  \ee
where
 \be
M^{(1)}_n(\bt^*)=\int_{\gamma}\,z^n a(z)d\mu(z,\bt^*)
 \ee

In \eqref{I^{(1)}=Pfaff}
 \be
b=b(\bt^*_+,\bt^*_-)=e^{\sum_{m=1}^\infty mt^*_mt^*_{-m}}
 \ee

For the other cases we have similar formulae. 

For general complex case we have
 \be\label{I^{(c)}=Pfaff}
I^{(c)}(\bt^*,N,{\bar A})\,=\,b \Pf \left[{\bar M}^{(c)}(\bt^*)\right]_{i,j=1,\dots,N}
 \ee
where
 \be
M^{(c)}_{nm}(\bt^*_-)=\int_{\mathbb{M}_2}\,z_1^nz_2^m A(z_1,{\bar z}_1,z_2,{\bar z}_2)
d\mu(z_1,{\bar z}_1,\bt^*)d\mu(z_2,{\bar z}_2,\bt^*)\,,
\quad M^{(c)}_n(\bt^*_-)=\int_{\mathbb{C}}\,z^n a(z,{\bar z})d\mu(z,{\bar z},\bt^*)
 \ee

For the Ginibre cases
 \be\label{I^{Gin}=Pfaff}
I^{Gin}_{qu-r}(\bt^*,N,{\bar A})\,=\,b \Pf \left[{\bar M}^{Gin}_{qu-r}(\bt^*)\right]_{i,j=1,\dots,N}\,,\quad
I^{Gin}_{r}(\bt^*,N,{\bar A})\,=\,b \Pf \left[{\bar M}^{Gin}_r(\bt^*)\right]_{i,j=1,\dots,N}
 \ee
where for the Ginibre quaternionic case we have
 \be
\left(M^{Gin}_{qu-r}\right)_{nm}(\bt^*)=\int_{\mathbb{C}_+}\,z^n{\bar z}^m(z-{\bar z}) 
e^{\varphi(z,\bt^*)+\varphi({\bar z},\bt^*)-|z|^2}d^2z\, ,
 \ee
while for Ginibre real case
\be
\left(M^{Gin}_r\right)_{nm}(\bt^*)=
 \ee
 \[
\int_{\mathbb{C}_+} \mbox{erfc}\left(\frac{|z-{\bar z}|}{\sqrt 2} \right) 
z^n {\bar z}^m e^{\varphi(z,\bt^*)+\varphi({\bar z},\bt^*)} d^2z+
\frac 12 \int_{\mathbb{R}}\int_{\mathbb{R}} \sgn(x_1-x_2) x_1^n x_2^m 
e^{\varphi(x_1,\bt^*)+\varphi(x_2,\bt^*)}dx_1dx_2 
 \]
(for $\mathbb{M}_2$ see \eqref{mathbb{M}}), and
 \be
\left(M^{Gin}_r\right)_n(\bt^*)=\int_{\mathbb{R}}\,x^n e^{\varphi(x,\bt^*)} dx
 \ee

\subsection{Perturbation series - series in the Schur functions}

To read this section we shall need some notations used in \cite{OST-I}, see Appendix \ref{Sums}.

Having the fermionic expresions we can re-write our integrals as series of the Schur functions over partitions, 
see \cite{OST-I} as follows
 \be
I(\bt^*_+,\bt^*_-,N,{\bar A}^*)=\sum_{\lambda\in \Pa}\,{\bar A}^*_\lambda(\bt^*_-) s_\lambda(\bt^*_+)
\,=\,S^{(1)}(\bt^*_+,N,U=0,{\bar{A^*}}(\bt^*_-))
 \ee
  where we use notations of \cite{OST-I}, see formula \eqref{S^{(1)}} and formulae \eqref{A-c}-\eqref{A-alpha-odd-n} 
in Appendix \ref{Sums},
where for ${\bar A}=(A,a)$ we take ${\bar {A^*}}=(A^*,a^*)$, defined as follows:

For integrals $I^{(1)}(\bt^*,N,{\bar A}^*)$ from \eqref{I{(1)}_i-f} we have
 \be
A^*_{nm}(\bt^*_-)=\int_{\Lambda_2}\,z_1^nz_2^m A(z_1,z_2)e^{\varphi(z_1,\bt^*_-)+\varphi(z_2,\bt^*_-)}dz_1dz_2\,,
\qquad a^*_n(\bt^*_-)=\int_{\gamma}\,z^n e^{\varphi(z,\bt^*_-)} a(z)dz
 \ee
where for $\Lambda_2$ see \eqref{Lambda_N}.

For integral \eqref{I^{c}-f}
 \be
A^*_{nm}(\bt^*_-)=\int_{\mathbb{M}_2}\,z_1^nz_2^m A(z_1,{\bar z}_1,z_2,{\bar z}_2)
e^{\varphi(z_1,\bt^*_-)+\varphi(z_2,\bt^*_-)}d^2z_1d^2z_2\,,
\qquad a^*_n(\bt^*_-)=\int_{\mathbb{C}}\,z^n e^{\varphi(z_1,\bt^*_-)} a(z,{\bar z})d^2z
 \ee
(for $\mathbb{M}_2$ see \eqref{mathbb{M}}).

For quaternionic Ginibre ensemble \eqref{I^{Gin}_{qu-r}} from \eqref{I^{Gin}_{qu-r}-f} we get
 \be
A^*_{nm}(\bt^*_-)=\int_{\mathbb{C}_+}\,z^n{\bar z}^m(z-{\bar z}) 
e^{\varphi(z,\bt^*_-)+\varphi({\bar z},\bt^*_-)-|z|^2}d^2z\, 
 \ee
(for $\mathbb{M}_2$ see \eqref{mathbb{M}})

For real Ginibre ensemble \eqref{I^{Gin}_{r}} from \eqref{I^{Gin}_{r}-f} we get
 \be
A^*_{nm}(\bt^*_-)=
 \ee
 \[
\int_{\mathbb{C}_+} \mbox{erfc}\left(\frac{|z-{\bar z}|}{\sqrt 2} \right) 
z^n {\bar z}^m e^{\varphi(z,\bt^*_-)+\varphi({\bar z},\bt^*_-)} d^2z+
\frac 12 \int_{\mathbb{R}}\int_{\mathbb{R}} \sgn(x_1-x_2) x_1^n x_2^m 
e^{\varphi(x_1,\bt^*_-)+\varphi(x_2,\bt^*_-)}dx_1dx_2 
 \]
(for $\mathbb{M}_2$ see \eqref{mathbb{M}}), and
 \be
a^*_n(\bt^*_-)=\int_{\gamma}\,x^n e^{\varphi(x,\bt^*_-)} dx
 \ee

Let us consider further examples

\paragraph{Circular $\beta=4$ ensemble.}

Fermionic representation was written down in \eqref{I{(1)}4-f}.
In this case
 \be
 A^*_{nm}({\bt}^*_-)=\frac{n-m}{2}\,\oint \, x^{n+m-1}\,
 \,d\mu_4(\bt^*_-,x)
 \ee
and
 \be
a^*_n(\bt^*_-)=\oint\, x^n d\mu_4(x,\bt^*_-)
 \ee
where 
  \be
d\mu_4(\bt^*_-,x):=\,x^{2t^*_0} e^{-2\sum_{m=1}^\infty\,x^{-m}t_{-m}^*} dx
  \ee
and where for the symplectic ensemble we take $\gamma=R$ while for the
circular ensemble $\gamma=S^1$. In the last case
 \be
A^*_{nm}(0,\bt^*_-)=\frac{n-m}{2}s_{n+m-t^*_0}(-2 \bt^*_-)
 \ee
where $s_n:=s_{(n)}$ is a one-row Schur funtion (or, the same,
complete symmetric function).

\paragraph{Ginibre quaternionic ensemble.}

Take $\bt^*_-=0$. We obtain
 \be
A_{nm}=-A_{mn}=m!\delta_{n+1,m}\,,\quad n<m
 \ee
which yields a formal perturbation series over partitions as follows
 \be
I^{Gin}_{qu-r}=\sum_\lambda \prod_{i=1}^N h_i!s_{\{h\}}(\bt^*_+)
 \ee
If we put $t^*_k\to t^*_k-\frac 12 \delta_{k,2}$, 
then for $n<m$ we obtain
 \be
A_{nm}=-A_{mn}=2i\int_{-\infty}^{+\infty}\int_{-\infty}^{+\infty}(x+iy)^n(x-iy)^m e^{-2y^2}y dxdy=
2\pi i\int_0^\infty d\rho \int_0^{2\pi} \rho^{n+m}e^{i(n-m)\phi-2\rho^2\sin^2\phi}d\rho^2 d\phi
 \ee
which may be expressed in terms of hypergeometric functions.

Ginibre real ensemble in details will be considered elsewhere.

\subsection{Series in zonal functions\label{Series in zonal functions}}

Series in zonal functions are difficult to analyze, nevertheless we write down some of them here. Let us write it in terms
of Jacks polynomials.

First consider circular $\beta=\frac 2\alpha$ ($\beta=1,4$) ensemble with $N$-ply integral. This is the integral 
$I^{(1)}_{\beta}$ of (\ref{I{(1)}1}) and (\ref{I{(1)}4})
where $\Lambda_N = \left(S^1\right)^{\times N}$.  Here we used

 \be
\oint \cdots \oint \, \prod_{i\neq j
}^N\,(1-x_ix_j^{-1})^{\frac1\alpha} \,
J_\lambda^{(\alpha)}(x)\,J_\lambda^{(\alpha)}(x)\ \,
\prod_{i=1}^N d x_i=\prod_{1 \le i<j\le N}\frac{\Gamma(\xi
_i-\xi_j+\frac1\alpha)\Gamma(\xi
_i-\xi_j-\frac1\alpha+1)}{\Gamma(\xi _i-\xi_j)\Gamma(\xi
_i-\xi_j+1)}=:c_\lambda^{(\alpha)}
 \ee
where $\xi_i:=\lambda_i+\frac1\alpha(N-1),\, 1\le i\le N$,
\cite{Mac}. This formula together with
 \be
e^{\frac 2\alpha \sum_{m=1}^\infty mt_mt_{-m}}
=\sum_{\lambda} d_\lambda^{(\alpha)}J^{(\alpha)}_\lambda(\bt_+)J^{(\alpha)}_\lambda(\bt_-)
 \ee

 \be
I^{(1)}_\beta(\bt^*,N)=\sum_{\lambda} c_\lambda^{(\alpha)}d_\lambda^{(\alpha)} J^{(\frac 12)}(\bt^*_+)J^{(\frac12)}(\bt^*_-)
 \ee

\paragraph*{Some formulas on Jack polynomials} see \cite{AMOS}:

\def\N{\mathcal N}
\def\G{\mathcal G}
\def\mybox#1#2#3#4{\raise.4pt\hbox{\vrule height0ptdepth#1}\vtop to#1{\hrule
	\hbox to#2{\vphantom{h}\hfil$#3$\hfil}\vss
	\hbox to#2{$\,#4$\hfil}\vss
	\hbox{\hfil}\hrule}\raise.4pt\rlap{\vrule height0ptdepth#1}\ignorespaces}
Let $\G_s$ be the multiplication operator
which increases all the pseudo-moments by $s$, i.e.,
$\lambda=(\lambda_1,\dots,\lambda_r)\mapsto\lambda+(s^r)=
(\lambda_1+s,\dots,\lambda_r+s)$:
$$
\G_s\colon J_\lambda^{(1/\beta)}(z_1,\dots,z_r)\mapsto
J_{\lambda+(s^r)}^{(1/\beta)}(z_1,\dots,z_r)=
\prod_{i=1}^rz_i^s\cdot J_\lambda^{(1/\beta)}(z_1,\dots,z_r)\,.
$$
The second operator $\N_{NM}$ changes the number of particles
from $M$ to $N$:
\begin{multline*}
\N_{NM}\colon J_\lambda^{(1/\beta)}(t_1,\dots,t_M)\mapsto
J_\lambda^{(1/\beta)}(z_1,\dots,z_N)=\\=
\oint\prod_{j=1}^M\frac{dt_j}{t_j}
\prod_{\scriptstyle1 \le i\le N\atop\scriptstyle 1\le j\le M}
\biggl(1-\frac{z_i}{t_j}\biggr)^{-\beta}
\prod_{\scriptstyle 1\le i,j\le M\atop\scriptstyle i\ne j}
\biggl(1-\frac{t_i}{t_j}\biggr)^\beta J_\lambda^{(1/\beta)}(t_1,\dots,t_M)\,,
\end{multline*}
where the integral is over a small cycle around the origin.

Repeatedly applying these operators to the $s_1$-particle vacuum
we obtain all the Jack polynomials, i.e.,
if
$$
\lambda:=\sum_{a=1}^{n-1}(s_a^{r_a})=\vcenter{\hbox{%
	\mybox{.8in}{.5in}{s_{n-1}}{r_{n-1}}
	\mybox{.7in}{.6in}{s_{n-2}}{r_{n-2}}
	\mybox{.5in}{.6in}{s_{n-3}}{r_{n-3}}
	\lower15pt\hbox{$\ \ldots\ldots$}
	\mybox{.35in}{.4in}{s_1}{r_1}
	}}
$$
the corresponding Jack polynomial has the following integral representation:
\bea
\lefteqn{J^{(1/\beta)}_\lambda(z_1,\dots,z_N) 
=\N_{r_nr_{n-1}}\G_{s_{n-1}}\N_{r_{n-1}r_{n-2}}\cdots
\G_{s_2}\N_{r_2r_1}\G_{s_1}\cdot1={}}\nonumber\\
&=&\oint\prod_{a=1}^{n-1}\Biggl[\kern.4pt\prod_{i=1}^{r_a}\frac{dt_i^{(a)}}{t_i^{(a)}}
\prod_{\scriptstyle 1\le i\le r_{a+1}\!\atop\scriptstyle 1\le j\le r_a}
\Biggl(1-\frac{t_i^{(a+1)}}{t_j^{(a)}}\Biggr)^{\!-\beta}
\prod_{\scriptstyle 1\le i,j\le r_a\!\atop\scriptstyle i\ne j}
\Biggl(1-\frac{t_i^{(a)}}{t_j^{(a)}}\Biggr)^{\kern-.8pt\beta}
\prod_{i=1}^{\,r_a}\bigl(t_i^{(a)}\bigr)^{s_a}\Biggr]\,,\nonumber
\eea
where $z_j:=t_j^{(n)}$, $N=r_n$.
In particular, for a rectangular $\lambda=(s^r)$ we have
$$
J^{(1/\beta)}_{(s^r)}(z_1,\dots,z_N)=\oint\prod_{i=1}^r\frac{dt_i}{t_i}
\prod_{i=1}^N\prod_{j=1}^r\biggl(1-\frac{z_i}{t_j}\biggr)^{-\beta}
\prod_{i\ne j}\biggl(1-\frac{t_i}{t_j}\biggr)^\beta\prod_{i=1}^r t_i^s\,.
$$
If $\beta=2$, the latter becomes a $\tau$-function:

Let $\theta:=\theta(x):=\sum_{k=1}^\infty x^kt_k$,
$\theta_i:=\theta(x_i)$ with $t_k:=\frac1k\sum_{j=1}^Nz_j^k$.
Since $e^{2\theta}=e^{2\sum_{k=1}^\infty\sum_{j=1}^N x^kz_j^k/k}=
\prod_{j=1}^Ne^{2\sum_{k=1}^\infty x^kz_j^k/k}=
\prod_{j=1}^N(1-xz_j)^{-2}$ we have
\bea
\bigl\langle 2n\bigr|\Gamma(t)\exp\oint\psi(x)\psi'(x)x^adx\bigl|0
\bigr\rangle
&=&\bigl\langle 2n\bigr|\exp\oint e^{2\theta}\psi(x)\psi'(x)x^adx\bigl|0
\bigr\rangle\nonumber\\
&=&\oint\bigl\langle2n\bigr|\prod_{i=1}^n\psi(x_i)\psi'(x_i)x_i^ae^{2\theta_i}
\bigl|0\bigr\rangle\prod dx_i\nonumber\\
&=&\oint\Delta(x)^4\prod_{i=1}^n\bigl(x_i^ae^{2\theta_i}\bigr)\prod dx_i\nonumber\\
&=&\oint\Delta(x)^4\prod_{i=1}^nx_i^a\prod_{i,j}(1-x_iz_j)^{-2}\prod dx_i\,.
\nonumber
\eea

\subsection{Asymmetric two-matrix ensembles and Interpolating Mehta-Pandey
Ensembles \label{asymmetric-section}}

This section was written as a proof of the conjecture of
E.Kanzieper and V.Osipov \cite{Kanzieper} that interpolating
ensembles \cite{PandeyMehta}  are related to integrable systems.
Interpolating ensembles have wide applications in description of
quantum chaos. They were also studied in relation to non-colliding
Brownian motion, see \cite{KatoriTanemura}, \cite{KatoriK}.

 \paragraph{Asymmetric two matrix models.}

Let a $N$ by $N$ matrix $X$ is one of
 \bea
\label{R}
X &=& R \quad \mbox{real\, symmetric}
\\
\label{Ra}
X &=& R^a \quad \mbox{real\, anti-symmetric}
\\
\label{Q}
X &=& Q \quad \mbox{real\, self-dual}
\\
\label{Qa}
X &=& Q^a \quad \mbox{real\, anti-self-dual}
 \eea 

We introduce the four types of two-matrix models
  \be\label{Z{HX}}
Z^{HX}_N(l,\bt,\bt')= \int\int dHdX \,e^{\,c\, Tr\,HX \, +\,
 Tr\,V(H,l,\bt)\,
 +\, Tr\,V(X,l,\bt')}
 \ee
where $H$ is a Hermitian $N$ by $N$ matrix while a $N$ by $N$ matrix $X$ is to be specified according to 
\eqref{R}-\eqref{Qa}. In \eqref{Z{HX}}
 \be
V(H,l,\bt)=l\,\ln\,\det H\,+\,\sum_{n=1}^\infty \left(t_nH^n -{\bar
t}_nH^{-n}\right)\,
 \ee
 \be
V(X,l,\bt')=l\,\ln\,\det X \,+\,\sum_{n=1}^\infty \left(t_n'X^n -{\bar
t}_n'X^{-n}\right)\,
 \ee
where $\bt=(t_1,t_2,\dots)$,  ${\bt}'=({ t}_1',{ t}_1',\dots)$ and
an integer $l$ are parameters.
 
\br
In case $t_n=\delta_{n,2}t_2$, $t_n'=\delta_{n,2}t_2'$ these models coincide with the interpolating enesembles
introduced by Mehta and Pandey, see Chapter 14 \cite{Mehta}.
\er

Repeating the calculation by Pandey and Mehta in Ch 14 \cite{Mehta} integrals \eqref{Z{HX}} may be reduced to the 
integral \eqref{I^{(1)}}.

We obtain
 \bp

\be\label{I^{(1)}}
Z^{HX}_N(l,\bt,\bt')=c\int_{\Lambda_N}\,\Delta_N(z)\,{\bar A}({\bf z}) \,
\prod_{i=1}^N\, d\mu(z_i,\bt^*)
 \ee
where

For the real symmetric case
 \be
A_{ij}= \mbox{erf}\left[  
\left(\frac{\alpha^2-1}{8v^2\alpha^2} \right)^{1/2}(x_i-x_j)\right]\,,
\quad a_{i,N-1}=
 \ee
For the antisymmetric case
 \be
A_{ij}=\exp \left(-\frac{\alpha^2-1}{8v^2\alpha^2}(x_i^2+x_j^2)\right) \mbox{erf}\left[  
\left(\frac{\alpha^2-1}{8v^2\alpha^2} \right)^{1/2}(x_i-x_j)\right]\,,
\quad a_{i,N-1}=\exp -\frac{\alpha^2-1}{8v^2\alpha^2} x_i^2
 \ee
For self-dual case
 \be
A_{ij}=(x_i-x_j)
\exp \left(-{(x_i-x_j)^2}\frac{1-\alpha^2}{8v^2\alpha^2}\right)\,,\quad \alpha^2 < 1
 \ee
For anti-self-dual case for $t'_n=\delta_{n,2}$
 \be
A_{ij}=(x_i-x_j)
\exp (\left(-{(x_i-x_j)^2}\frac{\alpha^2-1}{8v^2\alpha^2}\right)\,,\quad \alpha^2 > 1
 \ee
 \ep

\bp
Integrals \eqref{Z{HX}} are large BKP tau functions with respect to parameters $N,l,\bt$.
\ep

In case
 \be c=t_2=-\frac{1}{4v^2}\, ,\quad {
t}_2'=-\frac{1}{4v^2}\left(1+\frac{1}{1-\alpha^2} \right),
 \ee
all other $t_n$ and ${t}_n'$ vanish, we obtain
 \be\label{interp}
\int\int dHdA e^{-\frac{1}{4v^2}Tr\, HA \, -\frac{1}{4v^2}Tr\, H^2
\, -\frac{1}{4v^2}\frac{2-\alpha^2}{1-\alpha^2}Tr\, A^2}
 \ee
 \be\label{interpol-U-O}
 \equiv  \int\int d(\Re H)d(\Im H) e^{-\frac{1}{4v^2}Tr\,(\Re H)^2 \,
-\frac{1}{4v^2\alpha^2}Tr\, (\Im H)^2 }
 \ee
The last integral was an object of intensive study (see for
instance \cite{FNH}) because it describes the ensemble which
interpolates between Gauss unitary and Gauss orthogonal ones
\cite{PandeyMehta}. It is known (see Chapter 14 of \cite{Mehta})
that integral \eqref{interpol-U-O} may be written as
  \be
\int_R\cdots\int_R \, \Delta_N(x)\,\Pf\, [A(x_i,x_j)]
\,\prod_{i=1}^N \, dx_i
  \ee
with some matrix $A$ described below. This is exactly the
expression \eqref{I^{(1)}}, therefore, the interpolating ensemble is an
example of BKP tau function.

For the consideration of Harish-Chandra-Itzykson-Zuber (HCIZ)
integral in case where $H$ is Hermitian and $A$ is symmetric was
done in \cite{Mehta}. It is clear that exactly as in the case
where both $H,A$ are Hermitian we have
 \be
\int_{V\in\mathbb{U}(N)}\,e^{\,c\, Tr\,VXV^{-1}Y} \,
dV=\frac{\det\,
\left[e^{cx_iy_j}\right]}{\Delta_N(x)\Delta_N(y)}
 \ee
 where $H=UXU^{-1}$, $A=OYO^{-1}$ and $V=O^{-1}U$ and
$X=\diag(x_i)$, $U\in\mathbb{U}(N),\,O\in\mathbb{O}(N)$.

Then it follows that
 \[
Z^{HA}_N(\bt,{\bar \bt},\bt')=
 \]
 \be
\int_{\mathbb{R}^N}\int_{\mathbb{R}^N}\,\Delta_N(x)\,\det\,
\left[e^{cx_iy_j}\right]\,\sgn\,\Delta_N(y)\,\prod_{i=1}^N\,
d\nu(x_i) \, d\mu(y_i)
 \ee
 where
  \be\label{mu-nu-interp}
d\mu(y)=e^{\sum_{n=1}^\infty \left(t_n' {y}^n-{\bar t}'_n
y^{-n}\right)}dy\, ,\quad d\nu(x)=x^{l}\,e^{\sum_{n=1}^\infty
\left(t_n x^n-{\bar t}_nx^{-n}\right)}dx
  \ee

 Now we apply the following Lemma by Mehta \cite{Mehta}
  \bl \em
 \be
\int \cdots \int
\,\prod_{i=1}^N\,d\mu(y_i)\,\det\left[\theta_i(y_j)
\right]\,\sgn\,\Delta(y)=N! \Pf\left[
a_{ij}\right]_{i,j=1,\dots,2m}
 \ee
 where $2m=N $ if $N$ is even and $2m=N+1$ if $N$ is odd, and
  \be\label{a-ij-Mehta}
a_{ij}=\int\int_{x\le y}\,
d\mu(x)d\mu(y)\left[\theta_i(x)\theta_j(y)-\theta_j(x)\theta_i(y)\right],\quad
i,j=1,\dots,N
  \ee
  When $N$ is odd we have in addition $a_{N+1,N+1}=0$ and
  \be
a_{i,N+1}=-a_{N+1,i}=\int\, \theta_i(y)d\mu(y),\quad
i=1,\dots,N
  \ee
 \el
In our case $d\mu(y)$ is given by \eqref{mu-nu-interp} and
 \be\label{theta-IZ}
\theta_i(y_j):=e^{cx_iy_j}
 \ee
As a result we obtain
 \[
a_{ij}=a(x_i,x_j,\bt'):=
 \]
  \be\label{a-ij-for-H-rs}
 \int\int_{x\le y}\,e^{\sum_{n=1}^\infty
 \left(t_n'(x^n +y^n)-{\bar t}_n'(x^{-n} +y^{-n})\right)}
 \left(e^{cx_ix+cx_jy}-e^{cx_iy+cx_jx}\right)dx dy
  \ee
   \be\label{a-for-H-rs}
a_{i,N+1}= a(x_i,\bt'):= \int \,e^{\sum_{n=1}^\infty
 \left(t_n' y^n - {\bar t}_n' y^{-n}\right)} e^{cx_i y} dy
  \ee

Here it is supposed that there exists a certain domain $D$ of
parameters $\bt'$ where $a(\bt',x_1,x_2)$ exist for all
$x_{1,2}\in\mathbb{R}$. At last for we obtain

\bp 
  
  {\em The partition function for asymmetric matrix model
  \eqref{Z-H-rs} is the following 2-BKP tau function with respect to the
  variables $\bt=(t_1,t_2,\dots)$, ${\bar\bt}=({\bar t}_1,{\bar t}_2,\dots)$:
  \be\label{prop-Z-H-rs}
Z^{HA}_N(\bt,{\bar\bt},\bt')=\langle N+l|\,\g(\bt)\,g(\bt')
\,{\bar\g}({\bar\bt})\, |l\rangle
  \ee
  where
  \be
g(\bt')=e^{\int\int\,a(x_1,x_2,\bt')\psi(x_1)\psi(x_2)dx_1
dx_2}e^{\sqrt{2}\int a(x,\bt') \psi(x)dx\,\phi_0}
  \ee
where $a(\bt',x_1,x_2)$ is given by \eqref{a-for-H-rs}, $\bt'\in
D$. It may be also written in the following form
  \be\label{W-n-action}
 Z^{HA}_N(\bt,{\bar\bt},\bt')
=\langle N+l|\,\g(\bt) \, e^{\sum_{n=1}^\infty\,
(t_n'-t''_n)W_n} \, g(\bt'') \, e^{\sum_{n=1}^\infty\,
(t_n''-t'_n)W_n}\,{\bar\g}({\bar\bt})\,|l\rangle
  \ee
where $\bt'':=(t''_1,t''_2,\dots)\in D$ and
 \[
W_n=\frac{1}{2\pi
ic^n}\oint\,\psi(z)\frac{d^n\psi^\dag(z)}{dz^n}dz
 \]
}  
  \ep
Representation \eqref{W-n-action} results from the re-writing of
\eqref{a-for-H-rs} as follows
 \[
a(x_i,x_j,\bt')= \int\int_{x\le y}\,e^{\sum_{n=1}^\infty
 t_n'c^{-n}
(\partial_{x_i}^n +\partial_{x_j}^n)}
\left(e^{cx_ix+cx_jy}-e^{cx_iy+cx_jx}\right)dx dy
 \]
 and from formula $e^{\sum_{n=1}^\infty\, t_n'W_n}\psi(x)e^{-\sum_{n=1}^\infty\,
t_n'W_n}=e^{\sum_{n=1}^\infty t_n'c^{-n} (-\partial_{x})^n }\cdot
\psi(x)$.

\br More generally for the following series
  \be\label{KP-hyp-tau-matrix-arg}
  \tau^{KP}(I_N,HA):=\sum_{\lambda\atop\ell(\lambda)\le
  N}e^{U_0-U_\lambda}s_\lambda(I_N)s_\lambda(HA),\quad
  U_\lambda:=\sum_{i=1}^N\,U_{\lambda_i-i+N}
  \ee
(which is KP tau function \eqref{KP-hyp-tau} where higher times
are chosen as $t_m=\frac1m Tr \,(HA)^m$ and ${\bar t}_m=\frac1m
Tr\, I_N^m$, $I_N$ is $N$ by $N$ unity matrix) we have
\cite{O2004},\cite{HO2006}
 \be
\int_{V\in\mathbb{U}(N)}\,\tau^{KP}(I_N,VXV^{-1}Y)dV=c_N
\frac{\det[\theta_i(y_j)]_{i,j=1,\dots,N}}{\Delta_N(x)\Delta_N(y)}
 \ee
 where $c_N$ is some constant and where
 \be\label{theta-KPtau}
\theta_i(y_j)=\sum_{n=0} e^{U_{1-N}-U_{n
+1-N}}x_i^ny_j^n\quad\footnote{As one can see $\theta_1(y_1)$ up
to a constant coincides with $\tau^{KP}(I_N,HA)$ where
$H=x_1,\,A=y_1$}
 \ee
Now, similarly to the case of two-matrix models considered in
\cite{HO2006}, \cite{O2004} (where both $H,A$ were Hermitian) we
can replace the Itzykson-Zuber interaction term $e^{\,c\, Tr\,HA
}$ by $ \tau^{KP}(I_N,HA)$. The partition function of the
resulting asymmetric two-matrix model is a DKP tau function
\eqref{prop-Z-H-rs} where instead of \eqref{theta-IZ} one put
 \eqref{theta-KPtau}. Namely
  \be\label{Z-H-rs-tau} Z_N^{HA}(\bt,\bt',U):= \int\int dHdA
\,\tau^{KP}(I_N,HA)\,e^{Tr\,\sum_{n=1}^\infty t_nH^n\, +\,
Tr\,\sum_{n=1}^\infty { t}_n'A^n}=
 \ee
 \be
=c_N\int_{\mathbb{R}^N}\int_{\mathbb{R}^N}\,\Delta_N(x)\,\det\,
\left[\theta_i(y_j)\right]\,\sgn\,\Delta_N(y)\,\prod_{i=1}^N\,
d\nu(x_i,\bt,{\bar \bt})\, d\mu(y_i,\bt')
 \ee
 Thus $Z_N^{HA}(\bt,\bt',U)$ is the DKP tau function defined by
\eqref{prop-Z-H-rs} (or, the same, by \eqref{W-n-action})
 where $a(\bt',x_i,x_j)=a_{ij}$ is given by \eqref{a-ij-Mehta}
where $\theta_i(y_j)$ now is given by \eqref{theta-KPtau}.

Examples of series \eqref{KP-hyp-tau-matrix-arg} were considered
in \cite{OSch2}. One of the examples is the hypergeometric
function of matrix argument ${}_p{{F}}_s$, case $\mathbb{C}$, see
\cite{GR}, (see also \cite{KV}, Section 17.4). The matrix argument
is $HA$. In this case
  \[
  Z_N^{HA}({\bf{a};\bf{b}};\bt,\bt'):=
  \]
\be\label{Z-H-rs-F} \int\int dHdA
\,{}_p{{F}}_s\left.\left(a_1+N,\dots ,a_p+N\atop b_1+N, \cdots
,b_s+N\right| HA\right)\,e^{Tr\,\sum_{n=1}^\infty t_nH^n\, +\,
Tr\,\sum_{n=1}^\infty { t}_n'A^n}=
 \ee
 \be
c_N\int_{\mathbb{R}^N}\int_{\mathbb{R}^N}\,\Delta_N(x)\,\det\,
\left[{}_p{F}_s\left.\left(a_1+1,\dots ,a_p+1\atop b_1+1, \cdots
,b_s+1\right|
x_iy_j\right)\right]\,\sgn\,\Delta_N(y)\,\prod_{i=1}^N \,
d\nu(x_i)\, d\mu(y_i)
 \ee
In the last formula ${}_p{F}_s$ is the ordinary generalized
hypergeometric tau function of one variable and $c_N$ is some
constant. Here
 \[
\theta_i(y_j)={}_p{F}_s\left.\left(a_1+1,\dots ,a_p+1\atop b_1+1,
\cdots ,b_s+1\right| x_iy_j\right)
 \]
For instance ${}_0{{F}}_0(HA)=e^{Tr\,HA}$ yields Itzykson-Zuber
interaction term as in \eqref{Z-H-rs} , while
${}_1{{F}}_0(a|HA)=\det\,(I_N-HA)^{-a}$ results in Cauchy-type
interaction for asymmetric  two-matrix model (the model with
Cauchy-type interaction was introduced in \cite{HO2006} for the
"symmetric" case where both $H$ and $A$ are Hermitian).

The other example of series \eqref{KP-hyp-tau-matrix-arg} is a
hypergeometric function introduced by Milne \cite{Milne}, see
\cite{OSch1}. In this case $\theta_i(y_j)$ coincides with a basic
hypergeometric function of argument $x_iy_j$.

\er

\paragraph*{ $Z^{HA}_N(\bt,\bt')$ as DKP-nBKP tau function.} Let
$c=\sqrt{-1}$ and let $\bf{t}'$ be a set of non-vanishing times
with odd subscripts $\bt'=({t}_1',0,{t}_3',0,\bar{t}_5,\dots )$,
then, we have the following
  \be\label{H-A-DKP-NBKP}
 \int\int dHdA \,e^{\,c\, Tr\,HA \, +
 \, Tr\,\sum_{n=1}^\infty t_nH^n\, +\, Tr\,\sum_{n=1,3,5,\dots} {\bar
 t}_nA^n}
   \ee
  \be\label{H-A-DKP-NBKP-tau}
 =\langle N|\Gamma(\bt){ \Gamma}_B({\bf
 t'})\mathbb{T}_B(U)e^{\int \psi(x)\phi(x) dx}|0\rangle =
  \ee
   \be\label{integral-D-D^*}
\int_{\mathbb{R}}\cdots \int_{\mathbb{R}} \, \prod_{i>j}
(x_i-x_j)\,\prod_{i=1}^N e^{\sum_{n=1}^\infty t_n
x_i^n}\prod_{i=1}^N \left(e^{\sum_{n=1,3,5,\dots}^\infty  t_n'
\partial_{x_i}^n}\, \prod_{i>j}
\frac{x_i-x_j}{x_i+x_j}\,\right) dx_i
   \ee
where $U=(U_1,U_2,\dots)$, $U_n:=\log\Gamma(n+1)$. Thanks to the
fermionic representation  \eqref{H-A-DKP-NBKP-tau} we see that the
integral \eqref{integral-D-D^*} is a DKP-nBKP tau function where
$\bt$ and $\bt'$ are respectively DKP and nBKP higher times.

   In particular in case $\bt'=0$ we obtain
 \be
 \int\int dHdA \,e^{\,c\, Tr\,HA \, +
 \, Tr\,\sum_{n=1}^\infty t_nH^n}
=\int_{\mathbb{R}} \cdots \int_{\mathbb{R}}\, \prod_{i>j}
\frac{(x_i-x_j)^2}{x_i+x_j}\,\prod_{i=1}^N e^{\sum_{n=1}^\infty
t_n x_i^n} dx_i
   \ee
   which is the continues analog \eqref{Bures-density-continues1} of
   \eqref{Bures-density} and may be compare with the so-called Bures ensemble.

Formula \eqref{integral-D-D^*} follows from the equality
\[
a(x_i,x_j,0)= \int\int_{x\le y}\,
\left(e^{cx_ix+cx_jy}-e^{cx_iy+cx_jx}\right)dx dy
 \]
 \[
= -\frac{1}{x_ix_j}\frac{x_i-x_j}{x_i+x_j}=
-\frac{2}{x_ix_j}\l 0|\,\phi(x_i)\phi(x_j)\,|0\r
 \]
and the fact that for the choice of $U$ as above we have
 \[
 \Gamma_B({\bf t'}) \mathbb{T}_B(U)\cdot \phi(x) \cdot
 \mathbb{T}_B(U)^{-1}
\Gamma_B({\bf t'})^{-1}=e^{\sum_{n=1,3,\dots}\,
t_n'\partial_x^n}\cdot \phi(x)
 \]

The following Lemmas were used in the Chapter 14 of \cite{Mehta} devoted to the interpolating ensembles:

 \bl {\em
 \be
\int \cdots \int
\,\prod_{i=1}^n\,d\mu(y_i)\,\det\left[\theta_i(y_j),\nu_i(y_j)
\right]_{i=1,\dots,2n;\,j=1,\dots,n}=n! \Pf\left[
b_{ij}\right]_{i,j=1,\dots,2n}
 \ee
 where 
  \be\label{b-ij-Mehta}
b_{ij}=\int\int_{x\le y}\,
d\mu(y)\left[\theta_i(y)\nu_j(y)-\nu_i(y)\theta_j(y)\right],\quad
i,j=1,\dots,N
\ee
}
 \el

 \bl \em
 \be
\int \cdots \int
\,\prod_{i=1}^n\,d\mu(y_i)\,\det\left[\theta_i(y_j),\nu_i(y_j),\chi_i(y_n)
\right]_{i=1,\dots,2n-1;\,j=1,\dots,n-1}=(n-1)! \Pf\left[
c_{ij}\right]_{i,j=1,\dots,2n}
 \ee
 where 
  \be\label{c-ij-Mehta}
c_{ij}=\int\int_{x\le y}\,
d\mu(y)\left[\theta_i(y)\nu_j(y)-\nu_i(y)\theta_j(y)\right],\quad
i,j=1,\dots,2n-1
  \ee
  and
  \be\label{c-i,2n-Mehta}
c_{i,2n}=-c_{2n,i}=\int\, \chi_i(y)d\mu(y),\quad
i=1,\dots,2n-1
  \ee
 \el

Consider the following $2N$ fold integral
  \[
\int\int\cdots\int\int\, e^{}\Delta_N^2(x)\frac{\prod_{i=1}^N
f(x_iy_i)}{\Delta_N(x)\Delta_N(y)}e^{}|\Delta_N(y)|^\beta
\prod_{i=1}^N\, d\mu(x_i)d\mu(y_i)
  \]
First of all let us note that for every reasonable choice of a
function $f$

Consider the following $2N$-fold integral
 \be
\int\int\cdots\int\int\, \Delta_N(x)
\det[f(x_i,y_j)]_{i,j=1,\dots,N}\, \sgn\,\Delta_N(y)
\prod_{i=1}^N\, d\mu(x_i)d\nu(y_i)
  \ee
 Thanks to Lemma ... it is equal to $N$-ply integral
  \[
=N!\int\cdots\int\, \Delta_N(x)\Pf \left[ A(x_i,x_j)\right]
\prod_{i=1}^N\, d\mu(x_i)
  \]
  where
  \[
A(x_i,x_j)= \int\int_{y\le y'}\,
d\nu(y)d\nu(y')\left[f(x_i,y)f(x_j,y')-f(x_j,y)f(x_i,y')\right]
  \]

\section{Grand partition function for $2N$-fold integrals \label{other-section}}.

Let $d\mu^\pm$ be  measures supported respectively on contours
$\gamma^\pm$ on the complex plane. Our main examples of
$\gamma^\pm$ are the same contours (A) and (B) as in  subsection
\ref{integrals}.

 Let us adopt the following notation:
 \be\label{1,2,4}
\Delta_N^{(1)}(x):=\prod_{i>j}^N |x_i-x_j|,\quad
\Delta_N^{(2)}(x):=\prod_{i>j}^N (x_i-x_j)^2,\quad
\Delta_N^{(4)}(x):=\prod_{i>j}^N (x_i-x_j)^4
 \ee
also
 \be\label{11}
  \Delta_N^{(11)}(x):=\prod_{i>j}^N \frac{(x_i-x_j)^2}{x_i+x_j}
 \ee
and
 \be\label{12}
\Delta_{N=2n}^{(12)}(x):=\prod_{i>j}^{2n}
\frac{{x_i-x_j}}{(x_i+x_j)^2}\, \Hf\left(\frac 1{x_i+x_j}
\right),\qquad \Delta_{N=2n-1}^{(12)}(x)=\prod_{i>j}^{2n}
\frac{{x_i-x_j}}{(x_i+x_j)^2}\, \Hf\left(\frac 1{x_i+x_j} \right)
 \ee
 where in the right-hand side of the last equality we add a variable
 $x_{2n}$ to the set of $x_1,\dots,x_{N}$, $N=2n-1$, and then put $x_{2n}=0$.

Consider the following series over $N$ in $2N$-fold integrals:
 \be\label{Jbeta-beta+}
J_{\beta_-,\beta_+}:=\sum_{N=0}^\infty\frac{\nu^{2N}}{N!}\int_{\gamma^-}\int_{\gamma^+}\cdots
\int_{\gamma^-}\int_{\gamma^+}\,\Delta^{(\beta_-)}(x)\Delta^{(\beta_+)}(y)
\,\prod_{i=1}^N\,f(x_i,y_i) d\mu^-(x_i)d\mu^+(y_i)
 \ee
 where $\beta_-$ denotes any of the index among the set $1,2,4,11,12$ which
 are used in \eqref{1,2,4}-\eqref{12}. The same
 convention is chosen for $\beta_+$ which is independent of $\beta_-$ (thus formula
\eqref{Jbeta-beta+} contains 15 cases.)

Series in integrals \eqref{Jbeta-beta+} may be obtained as
particular cases of the series
 \be\label{J-a-a+}
J(a_-^c,a_+^c):=\sum_{N=0}^\infty\frac{\nu^{2N}}{N!}\int_{\gamma^-}\int_{\gamma^+}\cdots
\int_{\gamma^-}\int_{\gamma^+}\,\Delta(x)\Delta(y)\,a_-^c({\bf x})
\,\,a_+^c({\bf y}) \, \,\prod_{i=1}^N\,f(x_i,y_i)
d\mu^-(x_i)d\mu^+(y_i)
 \ee

 \be\label{J1}
J_1:=\sum_{N=0}^\infty\frac{\nu^{2N}}{N!}\int_{\gamma^-}\int_{\gamma^+}\cdots
\int_{\gamma^-}\int_{\gamma^+}\,|\Delta(x)\Delta(y)|
\,\prod_{i=1}^N\,f(x_i,y_i) d\mu^-(x_i)d\mu^+(y_i)
 \ee
 \be\label{J11}
J_{11}:=\sum_{N=0}^\infty\frac{\nu^{2N}}{N!}\int_{\gamma^-}\int_{\gamma^+}\cdots
\int_{\gamma^-}\int_{\gamma^+}\,\Delta^*(x)\Delta(x)\Delta^*(y)\Delta(y)
\,\prod_{i=1}^N\,f(x_i,y_i) d\mu^-(x_i)d\mu^+(y_i)
 \ee
 \be\label{J2}
J_2:=\sum_{N=0}^\infty\frac{\nu^{2N}}{N!}\int_{\gamma^-}\int_{\gamma^+}\cdots
\int_{\gamma^-}\int_{\gamma^+}\,\Delta(x)^2\Delta(y)^2
\,\prod_{i=1}^N\,f(x_i,y_i) d\mu^-(x_i)d\mu^+(y_i)
 \ee
  \be\label{J3}
J_3:=\sum_{N=0}^\infty\frac{\nu^{2N}}{N!}\int_{\gamma^-}\int_{\gamma^+}\cdots
\int_{\gamma^-}\int_{\gamma^+}\,\Delta(x)\Delta(y)\,a_-^c({\bf x})
\,\,a_+^c({\bf y}) \, \,\prod_{i=1}^N\,f(x_i,y_i)
d\mu^-(x_i)d\mu^+(y_i)
 \ee
  \be\label{J4}
J_4:=\sum_{N=0}^\infty\frac{\nu^{2N}}{N!}\int_{\gamma^-}\int_{\gamma^+}\cdots
\int_{\gamma^-}\int_{\gamma^+}\,\Delta(x)^4\Delta(y)^4
\,\prod_{i=1}^N\,f(x_i,y_i) d\mu^-(x_i)d\mu^+(y_i)
 \ee
where, as before,
\[
\Delta(x)=\Delta_N(x)=\prod_{i>j}^N (x_i-x_j),\qquad
\Delta^*(x)=\Delta^*_N(x)=\prod_{i>j}^N \frac{x_i-x_j}{x_i+x_j}
\]
The first ($N=0$) term in the series  is assumed to be 1. The
notation $a^c_\pm({\bf x})$ is analogous to (\ref{A-c}), denoting
the Pfaffian of an antisymmetric matrix ${\tilde a}_\pm$:
  \be\label{a-c'}
a^c_\pm({\bf x}):=\,\Pf[{\tilde a}_\pm]
  \ee
  whose entries are defined, depending on the parity of $N$, in terms
  of a skew symmetric kernel $a(x,w) $ (possibly, a distribution) and a function
  (or distribution) $a(x)$ as  follows:

 For $N=2n$ even
  \be\label{a-alpha-even-n'''}
{\tilde a}_{ij}=-{\tilde a}_{ji}:=a(x_i,x_j),\quad 1\le i<j \le 2n
  \ee

  For $N=2n-1$ odd
\be\label{A-alpha-odd-n'''} {\tilde a}_{ij}=-{\tilde a}_{ji}:=
\begin{cases}
a(x_i,x_j) &\mbox{ if }\quad 1\le i<j \le 2n-1 \\
a(x_i) &\mbox{ if }\quad 1\le i < j=2n
 \end{cases}
  \ee
In addition we define $a_0^c =1$.

To relate these integrals to the 2-BKP hierarchy we introduce
deformations $I_i(N)\to I_i(N;\bt,{\bar\bt})$ through the
following deformation of the measure \be d\nu({ z})\to d\nu({
z}|\bt,{\bar\bt})= b(\bt,\{ z\})b(-{\bar\bt},\{ z^{-1}\})d\nu({z})
\ee where
\begin{equation}
\label{ebb2'} b(\bs,\bt)=\exp \sum_{n\ {\rm odd}} \frac{n}{2}
s_nt_n
\end{equation}
and
\begin{equation}
\label{bracketz'} \{
z\}=(2z,\frac{2z^3}{3},\frac{2z^5}{5},\cdots)\, .
\end{equation}

Below, we show that the generating series obtained by
Poissonization (the grand partition function) \be Z_i(\mu \,
;\bt,{\bar\bt})\, =\,b(\bt,{\bar\bt})\sum_{N=0}^\infty
\,I_i(N;\bt,{\bar\bt}) \,\frac{\mu^N}{N!}\,,\quad i=1,2,3,4, \ee
are particular 2-BKP tau functions (\ref{2-nBKP}).

We also consider the following $2N$-fold integrals:
 \be\label{I5'}
I_5(N;\bt^{(1)},\bt^{(2)},{\bar\bt}^{(1)},{\bar\bt}^{(2)}):=\int
\Delta^*_{N}(z)\Delta^*_{N}(y) \prod_{i=1}^{N}d\nu({ z}_i,
y_i|\bt^{(1)},\bt^{(2)},{\bar\bt}^{(1)},{\bar\bt}^{(2)}),
 \ee
where
 \be
  d\nu({
z},y|\bt^{(1)},\bt^{(2)},{\bar\bt}^{(1)},{\bar\bt}^{(2)})=
  \ee
 \[
b(\bt^{(1)},\{ z\})b(-{\bar\bt}^{(1)},\{ z^{-1}\}) b(\bt^{(2)},\{
y\})b(-{\bar\bt}^{(2)},\{ y^{-1}\})d\nu(z,y)
 \]
(here $d\nu(z,y)$ is an arbitrary bi-measure), and show that the
generating series
 \be\label{Z5'}
Z_5(\mu\, ;\bt^{(1)},\bt^{(2)},{\bar\bt}^{(1)},{\bar\bt}^{(2)})\,
=\,b(\bt^{(1)},{\bar\bt}^{(1)})b(\bt^{(2)},{\bar\bt}^{(2)})\sum_{N=0}^\infty
\,I_5(N;\bt^{(1)},\bt^{(2)},{\bar\bt}^{(1)},{\bar\bt}^{(2)})
\,\frac{\mu^N}{N!}
 \ee
is a particular case of the two-component 2-BKP tau function
(\ref{2c-2-nBKP}).

 \br
 \label{Z2-Z5}
Note that \be Z_2(\mu \, ;\bt,{\bar\bt}) =Z_5(\mu \,
;\bt^{(1)},\bt^{(2)},{\bar\bt}^{(1)},{\bar\bt}^{(2)}) \ee
 if
 \be
 d\nu(z,y) = \delta(z-y)d\nu(z)d\nu(y), \quad
\bt=\bt^{(1)}+\bt^{(2)}, \quad
{\bar\bt}={\bar\bt}^{(1)}+{\bar\bt}^{(2)}. \ee
 \er
The integrals $Z_1(\mu \, ;\bt,{\bar\bt})$, $Z_2(\mu \,
;\bt,{\bar\bt})$, $Z_4(\mu \, ;\bt,{\bar\bt})$ and $Z_5(\mu \,
;\bt^{(1)},\bt^{(2)},{\bar\bt}^{(1)},{\bar\bt}^{(2)})$
 may be obtained as  continuous limits of
$S_1(\bt_\infty,\bt^*)$, $S_2(\bt_\infty,\bt_\infty,\bt^*)$,
$S_4(\bt_\infty,\bt^*)$ and $S_5(\bt_\infty,\bt_\infty,\bt^*)$,
respectively of \cite{OST-I} (see also Appendix \ref{Sums}).

\paragraph*{(A)} Consider DKP tau function
   \[
\tau^{A}(\bt,{\bar \bt}):=
  \]
 \be\label{double-beta=1A}
\langle 0|\,\g(\bt)\, g^{--}\,g\, g^{++}\, {\bar \g}({\bar
\bt})\,|0\rangle
   \ee
where
 \[
g^{--}=e^{\frac a2 \oint\oint \psi(x)
\psi(y)\sgn\left(\arg(x)-\arg(y)\right)d\mu^-(x)d\mu^-(y) }
 \]
  \[
 g^{++}=e^{\frac a2 \oint\oint
\psi^\dag(x)
\psi^\dag(y)\sgn\left(\arg(x)-\arg(y)\right)d\mu^+(x)d\mu^+(y) }
  \]

 In case $g=1$ tau function \eqref{double-beta=1A}  resembles a grand partition function
 for two-(unitary)-matrix models with
   Cauchy-kernel type interaction, see (A-41) in \cite{HO2006}:
    \be\label{2-matrixCauchy-un}
=c^2\left(1+\sum_{N=1}^\infty \left(-\frac{a^2}{\pi i}
\right)^N\left(\frac{1}{N!}\right)^2 \oint \dots \oint
\frac{|\Delta_N(z)\Delta_N(z')|}{\prod_{i,k=1}^N(z_i-z_k')}
\prod_{i=1}^N \frac{d\nu^-(z_i)d\nu^+(z_i')}{(z_iz_i')^{\tfrac12
(N-1)}} \right)
    \ee
where
 \be\label{measure-def}
d\nu^\mp=d\nu^\mp(z,\bt,{\bar \bt})=z^{\pm l}
e^{\pm\sum_{n=1}^\infty(z^nt_n-z^{-n}{\bar t}_n)}d\mu^\mp(z)
 \ee
 and where $c=\exp\sum_{n=1}^\infty mt_m{\bar t}_m$.

\paragraph*{(B)} Tau function
   \[
\tau^{B}(\bt,{\bar \bt}):=
  \]
 \be\label{double-beta=1B}
\langle 0|\,\g(\bt)\,e^{\frac a2 \int_{R}\int_{R} \psi(x)
\psi(y)\sgn (x-y)d\mu(x)d\mu(y) }g e^{\frac a2 \int_R\int_R
\psi^\dag(x) \psi^\dag(y)\sgn (x)-y)d\mu(x)d\mu(y) }{\bar
\g}({\bar \bt})\,|0\rangle
   \ee
  in case $g=1$ resembles a grand partition function for two-(Hermitian)-matrix models
  with  Cauchy-kernel type interaction, see (A-41) in \cite{HO2006}.
    \be\label{2-matrixCauchy-Her}
=c^2+c^2\sum_{N=1}^\infty \left(-\frac{a^2}{\pi i}
\right)^N\left(\frac{1}{N!}\right)^2 \int_R \dots \int_R
\frac{|\Delta_N(z)\Delta_N(z')|}
{\prod_{i,k=1}^N(z_i-z_k')}\prod_{i=1}^N d\nu(z_i)d\nu(z_i')
    \ee
where $c$ and the dependence of $d\nu^\mp$ in variables
$l,\bt,{\bar \bt}$ is the same formula \eqref{measure-def} as in
the case (A).

 Remark 1. As we mentioned both expression \eqref{2-matrixCauchy-un} and
 \eqref{2-matrixCauchy-Her} are similar to grand partition functions
 for two matrix models - unitary matrices for the case (A) and
 Hermitian ones for the case (B). The difference is that in
 expressions \eqref{2-matrixCauchy-un} and
 \eqref{2-matrixCauchy-Her} we have absolute values of Vandermond
 determinants instead of their own values. I do not know are there applications for
 integrals \eqref{2-matrixCauchy-un} and
 \eqref{2-matrixCauchy-Her}.

  Remark 2. In certain cases both integrals \eqref{2-matrixCauchy-un} and
  \eqref{2-matrixCauchy-Her} may be identified with the following DKP tau
    function from \cite{OST-I}
  \[
\tau(\bt,U( \bt^*))=1+\sum_{\alpha,\beta}e^{U_{\{-\beta-1\}}(
\bt^*)-U_{\{\alpha\}}(\bt^*)}s_{(\alpha|\beta)}(\bt)
  \]
where $(\bt,U(\bt^*))$ are specified either by \eqref{choice2}
(this leads to $\tau(\bt,U({ {\bt}^*}))=\tau^A( \bt^*)$) or by
\eqref{choice1} (this leads to $\tau(\bt,U({ {\bt}^*}))=\tau^B({
\bt}^*)$). These identifications are achieved respectively due to
\eqref{stinftyq} and to \eqref{stinfty}.

  Remark 3. In case in formulae \eqref{2-matrixCauchy-un} and
  \eqref{2-matrixCauchy-Her} we put $g=\exp \sum
  c_{ij}\psi_i\psi_j^\dag$ the interaction term (which is the Cauchy kernel\footnote
  {The Cauchy type interaction for two matrix models was considered in
  \cite{HO2006}, see also \cite{BertolaCauchy}} for $g=1$) and the dependence
  of the measures
  $d\nu^\mp$ on the variables will be different and expressed in
  terms of 2D Toda lattice Baker functions defined by the tau
  function $\tau^{TL}:=\langle
  l|\Gamma(\bt)g{\bar \Gamma}({\bar \bt})|l\rangle$.

\paragraph*{(C)} One can observe that
 \be
g_0^{--}=e^{\sum_{n>m}\, \psi_n\psi_m}=e^{-\frac 12 \oint \,
\psi(z)\psi(z^{-1})\,\frac{1+z}{1-z}\,\frac{dz}{z}}
 \ee
 where
 \be\label{ratio=series}
 \frac{1+z}{1-z}:=
1+2z+2z^2+\cdots
 \ee
Then
 \be
 g_0^{--}|0\rangle =e^{\sum_{n>m\ge 0}\, \psi_n\psi_m}|0\rangle\,=
 \,e^{\sum_{n>m}\, \psi_n\psi_m}|0\rangle=e^{-\frac 12
\oint \,
\psi(z)\psi(z^{-1})\,\frac{1+z}{1-z}\,\frac{dz}{z}}|0\rangle
 \ee
Also we have for \eqref{}
 \[
g_0^{-,0}=e^{\sqrt{2}\,\sum_m\,\psi_m \phi_0}=e^{\frac{1}{\pi i
\sqrt{2}}\oint\,\psi(z) \phi_0 \frac{dz}{z-1}}
 \]
 Then we obtain the following $N$-fold integral representation for
 \eqref{restricted-tau-N}:
  \be\label{int-repr-restricted-tau-N}
  \sum_{\ell(\lambda)\le N}\,s_\lambda({\bf t})
 =
\langle N|\,\Gamma(\bt)\,  g_0^{--}g_0^{-,0}\,|0\rangle=c\oint
\cdots \oint\,{\tilde\Delta}_N(z)
\,\prod_{i=1}^N\,e^{t_n(z_i^n+z_i^{-n})}\,\frac{z_i+1}{z_i-1}\,\frac{dz_i}{z_i}
 \ee where ${\tilde\Delta}_N(z)$ is the Vandermond determinant
${\Delta}_{2N}(z_1,\frac{1}{z_1},\dots,z_N,\frac{1}{z_N} )\,$
factor ${\tilde\Delta}_N(z)$ coincides with its absolute value in
case all $|z_i|=1$.

\section{Symmetries and $\beta=2$ circular ensemble.  One-matrix
model. \label{symmetries-section}}

It is well known that one matrix model both for Hermitian and
unitary matrices may be expressed as KP tau functions
\cite{Morozov}. Here we obtain partition functions of these models
as DKP tau functions

 \paragraph*{$\beta=2$ circular ensemble} known also as the model
of unitary matrices. Let us notice that there is a set of
$o(2\infty)$ elements $\{ I_n^\pm, \, n\, \mathrm{ odd \,
integers} \} $
 \be\label{I^pm}
I^-_n:=\sum_{i\in \mathbb{Z}}\, (-)^i\psi_i\psi_{-n-i},\quad
I^+_n:=\sum_{i\in \mathbb{Z}}\, (-)^{i}\psi^\dag_{n-i}\psi^\dag_i,
 \ee
commuting with currents $J_n$ as follows
 \be
[J_{2m-1},I^\pm_n]=0,\qquad [J_{2m},I^\pm_n]=\pm 2I^\pm_{n + 2m}
 \ee
 It results from $[J_i,\psi(x)]=x^i \psi(x),\, [J_i,\psi^\dag (x)]=-x^i \psi^\dag(x)$
  and from the following representation:
   \be
I_n^-=\frac{1}{2\pi i}\oint \, x^n \psi(x)\psi(-x) dx ,\qquad
I_n^+=\frac{1}{2\pi i}\oint \, x^{n+1} \psi^\dag(x)\psi^\dag(-x)
dx
   \ee
   Now
   \be
[I_n^+,I_m^-]= ???
   \ee
 Operators $e^{\sum_{n\, \mathrm{odd}} s_n^\pm I^\pm_n}$ may be
 considered as symmetry operators which commute with odd DKP
 flows. Parameters $s_n^\pm$ play the role of group times. Let us consider a simple
 DKP tau function
  \be
\tau_{2N}(\bt,{\bar \bt},{\bf s}^-):=\langle 2N|\g(\bt)
\,e^{\sum_{n\, \mathrm{odd}} s_n^-I^-_n}{\bar \g}({\bar
\bt})|0\rangle
  \ee
  We introduce a (deformed) measure as
  \be
d\mu(x,{\bf s}^-):=\left(\sum_{n\in\mathbb{Z}}\, s_{2i-1}^-
x^{2i}\right)\frac{dx}{x}
  \ee
  where ${\bf s}^-$ are deformation parameters,
  then
  \[
\sum_{n\, \mathrm{odd}} \,s_n^-I^-_n =\frac{1}{2\pi i}\oint\,
\psi(-x)\psi(x)\,d\mu(x,{\bf s}^-)
  \]
   Introducing $z=x^2$ and notations
   \[
   c_N=(-\pi i)^{-N}\frac{1}{N!}\,\exp\sum_{n=1}^\infty
   nt_n{\bar t_n} \, ,\quad \Delta_N(z):=\,\det
\left(z_i^{N-k}\right)|_{i,k=1,\dots,N}
 \]
  \[
 d\nu(z):=\,\frac 12 \,e^{2\sum_{n=1}^\infty \left(z^nt_n-z^{-n}{\bar t}_n\right)}
\,d\mu\left(z^{\frac 12}\right)
   \]
   we find that our tau function is the following $N$-ply integral
  \[
\tau_{2N}(\bt,{\bar \bt},{\bf s}^-)=c_N \oint\dots\oint\,
\Delta_N^2(z)\,\prod_{i=1}^N  \,d\nu(z_i) \, ,
 \]
  which may be interpreted as the partition function for $\beta=2$ circular
  ensemble.

  Let us mention that
   \[
\tau(\bt,{\bar \bt},{\bf s}^-):=\,\langle
0|\,\g(\bt)\,e^{a\sum_{n\, \mathrm{odd}} s_n^-I^-_n}e^{a\sum_{n\,
\mathrm{odd}} s_n^+I^+_n}\,{\bar \g}({\bar \bt})\,|0\rangle
   \]
    \[
=c_0^2\left(1+\sum_{N=1}^\infty \left(\frac{a}{\pi i}
\right)^{2N}\left(\frac{1}{N!}\right)^2 \oint \dots \oint\,
\frac{\Delta_N^2(z)\Delta_N^2(z')}{\prod_{i,k=1}^N(z_i-z_k')^2}\,\prod_{i=1}^N\,
d\nu(z_i)\,d\nu(z_i')\right)
    \]
For a special choice of measure $d\mu$ it may be equated to a
special KP tau function \eqref{KP-hyp-tau} evaluated at special
value of KP times $\bt={\bar \bt}$ given by
\eqref{choicetinftyq'}, see \eqref{stinftyq}
 \[
\tau^{KP}=\sum_{}\sum{} e^{U_{\{-\beta -1\}}-U_{\{\alpha \}}}
\left(s_{(\alpha|\beta)}(\bt)\right)^2
 \]
where variables $U_n$ are related to the even (coupled) DKP higher
times $t_{2m}, t_{2m}$ as follows
 \[
U_n =U_n(\bt^*,{\bar \bt}^*,c) = \sum_{m=1}^\infty \, ((n+c)^m
t_{2m}^* - (n+c)^{-m} {\bar t}_{2m}^*)
 \]
 where $c$ is an arbitrary non-vanishing non-integer parameter introduced to get
 rid of divergent terms in the expression for $U_0$.

\paragraph*{ One-Matrix Model.} Apart from $\beta=2$ circular
ensemble one obtains partition functions of the so-called matrix
models as DKP tau functions.

Consider DKP tau function
 \be
\langle 2N|\g(\bt)\,e^{\int_\gamma \psi(-x)\psi(x)d\mu(x)} {\bar
\g}({\bar \bt})|0\rangle
 \ee
 Repeating the previous calculation we obtain the $N$-ply integral
  \[
c_N \int_\gamma \dots\int_\gamma \Delta_N^2(z)z^N
e^{\sum_{n=1}^{\infty} (z^nt_{2n}-z^{-n}{\bar t}_{2n})}d\nu(z)\,
,\quad d\nu(z):=\frac 12 d\mu(z^{\frac 12})
  \]

\section{Remarks}

\subsection{Replacing $|0\rangle$ by $|\Omega\rangle$:
From KP and TL to BKP \label{replacing-section}}

There is a simple way how  a
 TL tau function $\tau^{TL}$  (where for simplicity we fix the discrete variable $t_0=0$) may be transformed to a BKP tau
function. This is as follows. We present $\tau^{TL}$ as the
vacuum expectation value. Now, to get the BKP tau function we
replace the vacuum $|0\r$ by the vector $|\Omega\r$ 
\be\label{Omega}
|0\rangle\,\to\,\,\sum_{\lambda\in\Pa}\,|\lambda\rangle =:|\Omega\rangle
\ee
 \be\label{replacement}
\tau^{TL}(\bt,{\bar \bt})=\langle 0|\,\g(\bt)\,g\,{\bar
\g}(\bt)\,|0\rangle\quad \to \quad \tau^{BKP}(\bt|{\bbt})= \langle
0|\,\g(\bt)\,g\,\g({\bar\bt})\,|\Omega\rangle
  \ee
  where $\bt$ is the BKP higher time, while $\bbt$ is a hidden parameter.
  This replacement may be also presented as the action of a certain vertex-like operator on the TL tau function:
  \be
  \tau^{BKP}(\bt|{\bbt}) = e^{\hat\Omega}\cdot \tau^{TL}(\bt,{\bar \bt}+{\bf s})
  \ee
  where 
  \be
  {\hat\Omega = \frac 12 \sum_{m>0} \frac 1m \partial_{s_m}^2 + \sum_{m>0,{\rm odd}} \frac {1}{m}\partial_{s_m} }
  \ee
  is a sort of Laplacian operator.

In this subsection we write down few examples.

\paragraph{Example 1. A multiple integral.} Here, starting from the partition function
of the two matrix model known to be a tau function of the 2KP (TL)
hierarchy we shall obtain new "integrable" multiple integral
("integrable" means related to an integrable hierarchy, in this
particular case, related to the BKP one).

\bl\label{replacement'}
 \em
We have
 \bea\label{instalation}
\langle n-m|\,\g(\bt)\,\psi(x_1)\cdots \psi(x_n)\,\psi^\dag(y_1)
\cdots \psi^\dag(y_m)\,{\bar\g}({\bar\bt})\,|\Omega\rangle= \\
\label{instalation-result}
\tau_o(\bt)
C({\bf x}^n,{\bf y}^m)\, \langle n-m|\,\g(\bt)\,\psi(x_1)\cdots
\psi(x_n)\,\psi^\dag(y_1)\cdots \psi^\dag(y_m)\,{\bar
\g}({\bar
\bt})\,|0\rangle
 \eea
where
 \be
C({\bf x}^n,{\bf
y}^m):=\prod_{i=1}^n\,(1+x_i^{-1})^{-1}\prod_{i=1}^m\,(1-y_i^{-1})^{-1}
\prod_{i<j\le n}\,(1-x_i^{-1}x_j^{-1})^{-1}\prod_{i<j\le
m}\,(1-y_i^{-1}y_j^{-1})^{-1}
 \ee
 \be\label{simplest}
 \tau_o(\bt):=e^{\sum_{n>0} \left(\frac 12 nt_n^2 +t_{2n-1}\right)}
 \ee

 \el
 For the proof first we send ${\bar\g}$ to the left vacuum, this gives the same factors for both 
 \eqref{instalation} and \eqref{instalation-result}. Then we write down Schur-Littlewood formula 
 (see \eqref{SchurSum1} and \eqref{sum-Schur-Pa} in Appendix, 
 or \eqref{simplest'} below), where we consider $\tau_o(\bt-\bt({\bf x}^n)+\bt({\bf y}^m))$  where
 $\bt({\bf x}^n)=\left(t_1({\bf x}), t_2({\bf x}),\dots \right)$ with
  $t_m({\bf x}):=\frac 1m\sum_{i=1}^n x_i^{-m}$. (Such $\tau_o$ is basically the left hand side of \eqref{instalation}, namely
  it is \eqref{instalation} where $\bbt=0$ divided by the vacuum expectation value in \eqref{instalation-result}.)
After application we get the factor $C({\bf x}^n,{\bf y}^n)$ times $\tau_o(\bt)$.
  
  Let us apply Lemma \ref{replacement} to the two-component TL tau
function of \cite{HO2006} which yields the partition function for
two matrix models
 \[
\tau^{2cTL}(\bt^{(1)},\bt^{(2)},{\bar\bt}^{(1)},{\bar\bt}^{(2)})=\l
N,-N|\g^{(1)}(\bt^{(1)})\g^{(2)}(\bt^{(2)})e^{\int
\psi^{(1)}(x){\psi^\dag}^{(2)}(y)
d\mu(x,y)}{\bar\g}^{(1)}({\bar\bt}^{(1)}){\bar\g}^{(2)}({\bar\bt}^{(2)})
|0,0\r =
 \]
  \[
c\int\cdots\int {\Delta}_N(x){\Delta}_N(y) \prod_{i=1}^N
e^{\sum_{m=1}^\infty\,(x_i^m t_m^{(1)}-x_i^{-m}{\bar
 t}_m^{(1)}+y_i^mt_m^{(2)}-y_i^{-m}{\bar
 t}_m^{(2)})} d\mu(x_i,y_i)
 \]
(here $c=c(\bt^{(1)},\bt^{(2)},{\bar\bt}^{(1)},{\bar\bt}^{(2)})$
is an unimportant factor see \cite{HO2006}) replacing the vacuum
$|0,0\r$ by $|\Omega,\Omega\r$. As a result we obtain:

\bp \em
The following $2N$-fold integral
 \be
c\int\cdots\int {\tilde\Delta}_N(x){\tilde\Delta}_N(y)
\prod_{i=1}^N e^{\sum_{m=1}^\infty\,(x_i^m t_m^{(1)}-x_i^{-m}{\bar
 t}_m^{(1)}+y_i^mt_m^{(2)}-y_i^{-m}{\bar
 t}_m^{(2)})} d\mu(x_i,y_i)
 \ee
 where
 \be
{\tilde\Delta}_N(x):=
\prod_{i=1}^N\,\frac{1-x_i^{-1}}{1-y_i^{-1}}\,\prod_{i<j}\,\frac{x_i-x_j}{1-x_ix_j}
 \ee
 is a two-component BKP tau function (namely it is BKP tau functions in
each set $\,\bt^{(i)},\,i=1,2$) equal to
 \be
\l N,-N|\g^{(1)}(\bt^{(1)})\g^{(2)}(\bt^{(2)})e^{\int
\psi^{(1)}(x){\psi^\dag}^{(2)}(y)
d\mu(x,y)}{\bar\g}^{(1)}({\bar\bt}^{(1)}){\bar\g}^{(2)}({\bar\bt}^{(2)})
|\Omega,\Omega\r
 \ee
 \ep

 Let us mark that the variables ${\bar\bt}^{(i)},\,i=1,2$, loose the
 meaning of the BKP higher times.

\paragraph{Example 2. } Another example is obtained if in the
right-hand side of \eqref{replacement} we additionally put
$\bbt=0$. We obtain

\bp \em If
 \[
\sum_{\lambda,\mu\in \Pa}\,s_\lambda(\bt)g_{\lambda\mu}s_\mu(\bbt
)
 \]
 is a TL tau function (the Takasaki series \cite{Takasaki}), then,
 \[
\sum_{\lambda,\mu\in \Pa}\,s_\lambda(\bt)g_{\lambda\mu}
 \]
 is the BKP tau function.

\ep

Let us notice that the BKP tau function
$\tau^{BKP}(\bt)$ may be obtained as a certain linear combination
of any TL tau function $\tau^{TL}(\bt,\bbt)$ by integration with
respect to variables $\bbt=({\bar t}_1,{\bar t}_2,\dots)$
 \be
\tau^{BKP}(\bt)=\int \tau^{TL}_N(\bt,\bbt)e^{-\sum_{m=1}^\infty
m{\bar t}_m {\bar t}_m^*}\tau_o(\bbt^*)\prod_{m=1}^\infty
\frac{d{\bar t}_m d{\bar t}_m^* }{ 4\pi^2 m i}
 \ee
 where each ${\bar t}_m^*$, $m=1,2,\dots$, is complex conjugated to ${\bar
 t}_m$, and where 
 \be\label{simplest'}
 \tau_o(\bt)=\sum_{\lambda\in\Pa} s_\lambda(\bt) = e^{\sum_{m>0} \left(\frac 12 mt_m^2 + t_{2m-1} \right)}
 \ee

Let us notice that there is a scalar product where the Schur
functions are ortho-normal: $\l s_\mu, s_\nu \r =\delta_{\mu,\nu}$
\cite{Mac}. Then $\tau^{BKP}(\bt)=\l
\tau^{TL}(\bt,\bbt),\sum_\mu s_\mu(\bbt) \r$. More generally one
BKP tau function may be obtained from the other by
$\tau^{BKP}_2(\bt)=\l \tau^{TL}(\bt,\bbt),\tau^{BKP}_1(\bbt)
\r$.

\paragraph{Example 3.} Let us introduce the following KP tau
function
 \be
\tau^{KP}_N(\bt,g,\lambda):=\l
N|\,\g(\bt)\,g\,|\lambda,N\r=\sum_{\mu\in\Pa}\,s_\mu(\bt)g_{\mu\lambda}(N),
\qquad g_{\mu\lambda}(N):=\l \mu,N|\,g \,|\lambda,N\r
 \ee
\br

If we choose $g=\exp \sum_{n,m>N} a_{nm}\psi_n\psi^\dag_{-m-1}$ we
obtain a tau function related to the Schubert cell of the Sato
Grassmannian \cite{SS} marked by a partition $\lambda$.

\er

 One may call it 
the generalized Schur function since the
specialization $g=1$ yields the ordinary Schur function. Then in
many formulae which express DKP and BKP tau functions as series in
the Schur functions one can replace the Schur functions
$s_\lambda$ by the generalized Schur functions
$\tau^{KP}_N(\bt,g,\lambda)$.

Instead of the Proposition \ref{BKP-Schut-prop} we obtain

\bp \label{BKP-Schut-prop-gen} \em

\be\label{BKP-tauUA-gen} \tau_{N}^{BKP}(l,\bt,g,U,{\bar A}):\,=
\,\langle N+l|\,\g(\bt)\,g\,\mathbb{T}(U)\,g^{--}(A)\,
g^{-,0}(a)\,| l\rangle
 \ee
 \be\label{BKP-tau-Schur-gen}
=\,c_{N+l} \,\sum_{\lambda\atop \ell(\lambda)\le N} \,
e^{-U_\lambda(N+l)}\, {\bar A}_{\{h+l\}}\,
\tau^{KP}_{N+l}(\bt,g,\lambda)
 \ee
where $h_i=\lambda_i-i+N$, $U_\lambda(N+l)$ and ${\bar A}_{\{h+l\}}$
are the same as in Proposition \ref{BKP-Schut-prop}. In
\eqref{BKP-tauUA-gen} $ g^{--}(A)$ and $g^{-,0}(a)$ are given
respectively by \eqref{g^{--}BKP} and \eqref{g^{-,0}a} and
$\mathbb{T}(U)$ is as in \eqref{mathbb T}. The constant $c_{N+l}$
is defined by \eqref{c_n}.

\ep

Tau function \eqref{BKP-tauUA-gen} vanishes if $N<0$.

 \br

The series \eqref{BKP-tau-Schur-gen} may be also equated to a DKP
tau function.

 \er

Examples  \eqref{restricted-tau+NBKP
U}-\eqref{restricted-tau-N-gen}  are replaced by
 \be\label{restricted-tau+NBKP U-gen}
  \tau_N^{BKP-nBKP}(l,\bt,\bt',g,U): =
  \sum_{\lambda\atop \ell(\lambda)\le N} \, e^{-U_\lambda(N+l)}\,\,
Q{_{l+\lambda^-}\left({\tfrac 12
\bt}'\right)}\,\tau^{KP}_{N+l}(\bt,g,\lambda)
 \ee
 \be\label{restricted-tau+NBKP U=0-gen}
  \tau_N^{BKP-nBKP}(l,\bt,\bt',g,U=0): =
  \sum_{\lambda\atop \ell(\lambda)\le N} \,
Q{_{l+\lambda^-}\left({\tfrac 12
\bt}'\right)}\,\tau^{KP}_{N+l}(\bt,g,\lambda)
 \ee
 \be\label{restricted-tau+N U-gen}
  \tau_N(l,\bt,\bt',g,U): =
  \sum_{\lambda\atop \ell(\lambda)\le N} \, e^{-U_\lambda(N+l)}\,
  \tau^{KP}_{N+l}(\bt,g,\lambda)
 \ee
 \be\label{restricted-tau-N-gen}
  \tau_N(g,\bt): =
  \sum_{\lambda\atop \ell(\lambda)\le N} \,\tau^{KP}_{N+l}(\bt,g,\lambda)
 \ee
Notations are the same as in \eqref{restricted-tau+NBKP
U}-\eqref{restricted-tau-N-gen}.

\paragraph{}

\section*{Acknowledgements}

We are grateful to John Harnad, Johan van de Leur, Vladimir Osipov
and  and most of all to Eugene Kanzieper for discussions of the
topic. One of the authors (A.O) thanks E. Kanzieper and Holon Technology Institute for
hospitality (July 2008) where a part of this work (a main part of
the section "Asymmetric two-matrix ensemble") was done. The work was 
supported by RFBR grants 11-01-00440-а and by Japanese-RFBR grant 10-01-92104 JF.
This work is also partly supported by Grant-in-Aid for Scientific Research
No.~22540186 from the Japan Society for the Promotion
of Science and by the Bilateral Joint Project ``Integrable Systems,
Random Matrices, Algebraic Geometry and Geometric Invariants''
(2010--2011) of the Japan Society for the Promotion of Science and the
Russian Foundation for Basic Research. 
This work has been funded by the  Russian Academic Excellence Project '5-100'.

\appendix

\section{Appendices}\subsection{Pfaffnians and Hafnians \label{tools-section}}

\paragraph{(A) Pfaffians.} We need the notion of Pfaffian. If $A$ an
anti-symmetric matrix of an odd order its determinant vanishes.
For even order, say $k$, the following multilinear form in
$A_{ij},i<j\le k$
 \be\label{Pf'}
\Pf [A] :=\sum_\sigma
{\sgn(\sigma)}\,A_{\sigma(1),\sigma(2)}A_{\sigma(3),\sigma(4)}\cdots
A_{\sigma(k-1),\sigma(k)}
 \ee
where sum runs over all permutation restricted by
 \be
\sigma:\,\sigma(2i-1)<\sigma(2i),\quad\sigma(1)<\sigma(3)<\cdots<\sigma(k-1),
 \ee
 coincides with the square root of $\det A$ and is called the
 Pfaffian of $A$, see, for instance \cite{Mehta}. As one can see the Pfaffian  contains
 $1\cdot  3\cdot 5\cdot \cdots \cdot(k-1)=:(k-1)!!$ terms.

The following equality is known as Schur identity
\begin{equation}
\label{PfSchurA}
 \Pf\left(\left(
\frac{x_i-{x_j}}{x_i+{x_j}} \right)_{1\le i,j\le
2n}\right)=\Delta_{2n}^*(x)
\end{equation}
where
 \be
\Delta_{k}^*(x):= \prod_{1\le i<j\le k}\frac{x_i-{x_j}}{x_i+{x_j}}
 \ee
Let us mark that a special case of this relation is obtained if
$x_{2n}$ vanishes. In this case we write
 \be\label{PfSchur-odd}
\Pf(A)=\Delta^*_{2n-1}(x)
 \ee
 where $A$ is an antisymmetric $2n\times 2n$ matrix defined by
 \be\label{PfSchurA'}
A_{ij}=\begin{cases} \frac{x_i-{x_j}}{x_i+{x_j}} &{\mbox if}\quad
 {1<i<j < 2n} \\
 1 &{\mbox if}\quad i<j=2n,
 \end{cases}
 \ee

\paragraph{Hafinans} The {\em Hafnian} of  a symmetric matrix $A$ of even order $N=2n$
is defined as
 \be\label{Hf}
\Hf (A) :=\sum_\sigma
\,\,A_{\sigma(1),\sigma(2)}A_{\sigma(3),\sigma(4)}\cdots
A_{\sigma(2n-1),\sigma(2n)}
 \ee
where sum runs over all permutation restricted by
 \be
\sigma:\,\sigma(2i-1)<\sigma(2i),\quad\sigma(1)<\sigma(3)<\cdots<\sigma(2n-1),
 \ee
  As one can see the this sum  contains $1\cdot  3\cdot 5\cdot \cdots
  \cdot(2N-1)=:(2N-1)!!$ terms.

\br \em \label{Hfdiag} Let us note that entries on the diagonal of the
matrix $A$ does not contribute the sum \eqref{Hf}.

\er

The following equality was found in \cite{IKO}
\begin{equation}
\label{pfaffhaf}
 \Pf\left(\left(
\frac{x_i-{x_j}}{\left(x_i+{x_j}\right)^2} \right)_{1\le i,j\le
2n}\right)= \prod_{1\le i<j\le 2k}\frac{x_i-{x_j}}{x_i+{x_j}}
\mbox{Hf}\left(\left(\frac{1}{x_i+x_j}\right)_{1\le i,j\le
2n}\right)
\end{equation}
Another proof of this relation was presented in \cite{LOS}. Let us
mark that a special case of this relation is obtained if $x_{2n}$
vanishes. In this case we write
 \be\label{pfaffhaf-odd}
\Pf(B)=\Delta^*_{2n-1}(x)\Hf(C)=:\Delta^{**}_N(x)
 \ee
 where $B$ and $C$ are respectively antisymmetric and symmetric $2n\times 2n$
 matrices whose relevant entries (see Remark \ref{Hfdiag}) are given by
 \be\label{pfaffhafBC}
B_{ij}=\begin{cases} \frac{x_i-{x_j}}{(x_i+{x_j)^2}} &{\mbox
if}\quad
 {1\le i<j < 2n} \\
 \frac 1{x_i} &{\mbox if}\quad i<j=2n,
 \end{cases}
 \qquad
C_{ij}=\begin{cases} \frac 1{x_i+{x_j}} &{\mbox if}\quad
 {1\le i < j < 2n} \\
 \frac 1{x_i} &{\mbox if}\quad i<j=2n,
 \end{cases}
 \ee

\section{Sums of Schur functions \cite{OST-I} \label{Sums}}

In this Appendix we recall some relations from the first part of this work.

\paragraph{Subsets of partitions.}
In the following, we consider sums over partitions and strict
partitions , which will
 be denoted by Greek letters $\alpha$, $\beta$. Recall  \cite{Mac} that a
strict partition $\alpha$ is a set of integers (parts)
$(\alpha_1,\dots,\alpha_k)$ with $\alpha_1>\dots
>\alpha_k\ge 0$. The length of a partition $\alpha$, denoted
$\ell(\alpha)$, is the number of non-vanishing parts, thus it is
either $k$ or $k-1$.

Let $\Pa$ be the set of all partitions. 
We shall need two special subsets of $\Pa$.

The first one consists of all partitions $\lambda=(\lambda_1,\dots,\lambda_{2n})$,
$0\le n\in\mathbb Z$, $\lambda_{2n}\ge0$, which satisfy
$$
\lambda_i+\lambda_{2n+1-i}\ \ \hbox{is independent of\,\ $i$}\,,\quad i=1,\dots,2n\,,
$$
or equivalently
 \be\label{SCP}
h_i+h_{2n+1-i}=2c\ \ \hbox{is independent of\,\ $i$\,\ (hence }=h_1+h_{2n}\ge2n-1),\quad i=1,\dots,2n\,,
 \ee
where $h_i=\lambda_i-i+ 2n\,$, and $2c$ is a natural number conditioned by $2c\ge 2n$.
This subset consists of all partitions $\lambda$ of length $l(\lambda)\le2n$
whose Young diagram satisfies the property that its complement in the
rectangular Young diagram $\overline Y$ corresponding to
$(\lambda_1+\lambda_{2n})^{2n}$ coincides with itself rotated 180 degrees
around the center of $\overline Y$.
This set of partitions will be denoted by $\SCP(c)$ or simply SCP, for
``self-complementary partitions''.
 If we introduce 
\be\label{y-h}
 y_i:=h_i - {c}\,,\quad {c} = \frac{h_1+h_{2n-1}}2 \,,
 \ee
then relation \eqref{SCP} may be rewritten as
\be\label{y+y=0}
y_i + y_{2n+1-i}=0\,.
 \ee

The second subset we need consists of the partitions $\lambda$ which satisfy,
equivalently,
 \be\label{FP}
\lambda_{2i}=\lambda_{2i-1}\,, \quad i=1,2,\dots,
 \ee
or $\lambda=\mu \cup \mu:=(\mu_1,\mu_1,\mu_2,\mu_2,\dots,\mu_{k},\mu_{k})$
($\exists\,\mu=(\mu_1,\mu_2,\dots,\mu_k)\in\Pa$), or that the conjugate
partitions of $\lambda$ are even, i.e., the ones whose parts are even numbers.
This set of partitions will be denoted by FP, for ``fat partitions''.

Following \cite{Mac} we will denote by DP
the set of all strict partitions (partitions with distinct parts),
namely, partitions
$(\alpha_1,\alpha_2,\dots,\alpha_k)$, $1\le k\in\mathbb Z$  with the strict
inequalities
$\alpha_1>\alpha_2>\cdots>\alpha_k>0$.

Strict partitions $\alpha$ with the property
 \be
\alpha_{2i}=\alpha_{2i-1}+1 \quad\hbox{for}\quad 2i-1\le l(\alpha)\,,
 \ee
where we set $\alpha_{2i}=0$ if $l(\alpha)=2i-1$,
will be called fat strict partitions. The set of all fat strict partitions will be denoted by 
FDP.\footnote{This subset was used in \cite{HLO} where it was denoted by $\DP'$.}

The set of all self-complementary strict partitions will be denoted by SCDP.

Let $R_{NM}$ denote the set of all partitions whose Young diagram
may be placed into the rectangle $N \times M$, namely, $R_{NM}$ is
the set of all partitions $\lambda$ restricted by the conditions
$\lambda_1\le M$ and $ \ell(\lambda) \le N$.

\paragraph{Sums over partitions.} Consider the following sums (for $\bt := (t_1, t_3, \dots )$, $\bt^*:=(t_1^*, t_3^*,\dots )$,
  $\bar{\bt} := (\bar{t}_1, \bar{t}_3, \dots )$, $\bt := (t_1, t_3, \dots )$,
  $\bt^*:=(t_1^*, t_3^*, \dots )$,   $\bar{\bt} := (\bar{t}_1, \bar{t}_3, \dots )$), $N$).

 \be
 \label{S^{(1)}}
S^{(1)}(\bt,N;U,{\bar A})\,:=\,\sum_{\lambda\in\Pa \atop \ell(\lambda)\le N}\,{\bar A}_{h(\lambda)}e^{-U_{\{h\}}}
\,s_\lambda(\bt)
 \ee 
where $h(\lambda)=\lambda_i-i+N$. The factors ${\bar A}_h$  on the  right-hand side of \eqref{S^{(1)}} are
   determined in terms a pair $(A, a)=:{\bar A}$ where $A$ is an infinite skew symmetric matrix and $a$
an infinite vector.  For a strict partition $h=
(h_1,\dots,h_N )$, the numbers
${\bar A}_h$ are defined as the Pfaffian of an antisymmetric $2n
\times 2n$ matrix ${\tilde A}$  as follows:
  \be
  \label{A-c}
{\bar A}_{h}:=\,\Pf[{\tilde A}]
  \ee
where for $N=2n$ even
  \be
  \label{A-alpha-even-n}
{\tilde A}_{ij}=-{\tilde A}_{ji}:=A_{h_i,h_j},\quad 1\le
i<j \le 2n
  \ee
and for $N=2n-1$ odd
 \be \label{A-alpha-odd-n} {\tilde
A}_{ij}=-{\tilde A}_{ji}:=
\begin{cases}
A_{h_i,h_j} &\mbox{ if }\quad 1\le i<j \le 2n-1 \\
a_{h_i} &\mbox{ if }\quad 1\le i < j=2n  .
 \end{cases}
  \ee
In addition we set ${\bar A}_0 =1$.

Then 
 \be
U_{\{h\}}\,:=\,\sum_{i=1}^N\,U_{h_i}
 \ee
where $U_n$, $n=0,1,2,\dots$ is a set of given complex numbers. This set is denoted by $U$.

As we see the factor $e^{-U_{\{h\}}}$ can be included into the factor ${\bar A}_{h}$ by redefinition of the data ${\bar A}$
as follows:
 \[
  A_{nm}\to A_{nm}e^{-U_n-U_m}\,,\quad a_n\to a_ne^{-U_n}
 \]
However we prefer to keep $U$ as a set of parameters.

 {\bf Example 0}
We choose the following matrix $A$ is given by
 \be \label{ExA0}
A_{ik}=\,(A_0)_{ik}\,:=\begin{cases} \sgn(i-k) & \hbox{if}\ 1\le i,k \le L\\[3pt] 0 
& \hbox{otherwise}
                              \end{cases}\,,
 \qquad a_k\,=\begin{cases} 1 &\hbox{if}\ k\le L \\[3pt] 0 & \hbox{otherwise}\end{cases}\,.
 \ee 
\br \em The matrix $A_1$ is infinite. However if in series \eqref{S^{(1)}} we put
  $U_n =+\infty$  for  $ n>L$, it will be the same as if we deals with the finite $L$ by $L$ matrix $A$, given by \eqref{ExA0}.
\er

 {\bf Example 1}

 \be \label{ExA1}
A_{ik}=\,(A_1)_{ik}\,:=\, 1 \,,\quad i<k\,, \qquad a_k= 1
 \ee
Then
 \be
({\bar A}_1)_{\{h\}}=1
 \ee

 {\bf Example 2}

The matrix $A$  is a finite $2n$ by $2n$ matrix, and $a=0$, thus the sum \eqref{S^{(1)}} ranges only  
partitions with even number of non-vashing parts. We put
 \be\label{ExA2}
A_{ik}=(A_2)_{ik}\,:=\,-\delta_{i,2c-i}\,,\quad i<k
 \ee
Then 
\be
\label{transweight}
({\bar A}_2)_{\{h\}}=
\begin{cases}1   &\mbox{ iff }\ \lambda\in\SCP(c) \\
0 &\mbox{ otherwise }\ 
\end{cases}
\ee
where $h=(h_1,\dots,h_N)$ is related to $\lambda=(\lambda_1,\dots,\lambda_{N})$ as $ h_i=\lambda_i-i+N\,,i=1,\dots,N=2n$.

 {\bf Example 3} Given set of additional variables $\bt'=(t_1',t_3',t_5',\dots)$
 where we take
 \be\label{ExA3}
 A_{nm}=\,(A_3)_{nm}\,:=\,\frac 12\,
e^{-U_m-U_n}\,Q_{(n,m)}(\tfrac 12{\bt'}), \qquad
a_n=(a_3)_n\,:=\,e^{-U_n}Q_{(n)}(\tfrac 12{\bt'})
 \ee 
 Here, the {\em  projective Schur functions} $Q_\alpha$  are weighted polynomials in the variables $t'_m$,
 $\deg t_m' =m$, labeled by strict partitions
(See \cite{Mac} for their detailed definition.)

\br \em\label{Q(t-infty)}
Let us introduce notation $\bt'_\infty=(1,0,0,\dots)$. It is known that
 $Q_{h}(\tfrac 12{\bt'_\infty})=\Delta^{*}(h)\prod_{i=1}^N\frac{1}{h_i!}$ where 
 \be\label{Delta^{*}}
\Delta^{*}(h)\,:=\,\prod_{i<j}\frac{h_i-h_j}{h_i+h_j}
 \ee
Thus for this choice of $\bt'$ we obtain
 \be
({\bar A}_3)_{\{h\}}=\Delta^{*}(h)\prod_{i=1}^N\frac{1}{h_i!}
 \ee
One may compare it with Example 5 where $f(n)=n$.

\er

  {\bf Example 4}
 \be\label{ExA4}
 A_{nm}=\,(A_4)_{nm}\,:=\,\delta_{n+1,m}-\delta_{m+1,n}.
 \ee

Then  
\be
\label{transweight4}
({\bar A}_4)_{\{h\}}=
\begin{cases}1   &\mbox{ iff }\ \lambda=(\lambda_1,\dots,\lambda_{2n})\in\FP \\
0 &\mbox{ otherwise }\ 
\end{cases}
\ee
where $h=(h_1,\dots,h_N)$ is related to $\lambda=(\lambda_1,\dots,\lambda_{N})$ as $ h_i=\lambda_i-i+N\,,i=1,\dots,N=2n$.

\,

\,

 \br \em\label{Examples-5-6-7} For some applications we may need further examples.
In Examples 5-7  ${\bar A}$ depends on a given function on the lattice denoted by $f$. In particular one can choose $f(n)=n$. 
Below are examples of matrices $A$ whose  Pfaffians are well-known (see \cite{Ishikawa} and references there).

 {\bf Example 5} 
\be\label{ExA5}
 A_{nm}=\,(A_5)_{nm}\,:=\,\frac{f(n)-f(m)}{f(n)+f(m)}
 \ee
Then for $ h_i=\lambda_i-i+N\,,i=1,\dots,N$, we have
\be
({\bar A}_5)_{\{h\}}=\Delta^{(5)}_N\left(f(h)\right)
 \ee
where
 \be
\Delta^{(5)}_N\left(f(h)\right):=\prod_{i<j\le N}\frac{f(h_i)-f(h_j)}{f(h_i)+f(h_j)}
 \ee

    {\bf Example 6}
 \be\label{ExA6}
 A_{nm}=\,(A_6)_{nm}\,:=\,\frac{f(n)-f(m)}{1-f(n)f(m)}
 \ee
Then for $ h_i=\lambda_i-i+N\,,i=1,\dots,N$, we have
 \be
({\bar A}_6)_{\{h\}}=\Delta^{(6)}_N(f(h))
 \ee 
where
 \be\label{Delta^{(6)}}
\Delta^{(6)}_N\left(f(h)\right)\,:=\,\prod_{i<j\le N}\frac{f(h_i)-f(h_j)}{1-f(h_i)f(h_j)}
 \ee

  {\bf Example 7}
 \be\label{ExA7}
 A_{nm}=\,(A_7)_{nm}\,:=\,\frac{f(n)-f(m)}{(f(n)+f(m))^2}.
 \ee
Then for $ h_i=\lambda_i-i+N\,,i=1,\dots,N$, we have
 \be
({\bar A_7})_{\{h\}}=\Delta^{(7)}_N\left(f(h)\right)
 \ee 
where
 \be\label{Delta^{(7)}}
\Delta^{(7)}_N\left(f(h)\right)\,:=\,\Biggl(\prod_{i<j\le N}\frac{f(h_i)-f(h_j)}{(f(h_i)+f(h_j))^2}\Biggr) \,\Hf\left( \frac{1}{f(h_i)+f(h_j)}\right)
 \ee

\er

\,

Having these examples we introduce the notation
  \be
 \label{S^{(1)}i}
S^{(1)}_i(\bt,N;U)\,:=S^{(1)}(\bt,N;U,{\bar A}_i)\,=\,\sum_{\lambda\in\Pa \atop \ell(\lambda)\le N}\,({\bar A_i})_{h(\lambda)}\,e^{-U_\lambda}
\,s_\lambda(\bt)\,,\quad i=1,\dots,6
 \ee 

In particular we obtain 
 \bea
 \label{S^{(1)}0}
S^{(1)}_{0}(\bt,N;M,U)&:=&\sum_{\lambda\in R_{N,M}}\,e^{-U_\lambda}\,s_\lambda(\bt)
\\
 \label{S^{(1)}1}
S^{(1)}_{1}(\bt,N;U)&:=&\sum_{\lambda\in\Pa \atop \ell(\lambda)\le N}\,e^{-U_\lambda}s_\lambda(\bt)
\\
 \label{S^{(1)}2}
S^{(1)}_{2}(\bt,N;U,c)&:=&\sum_{\lambda\in\SCP(c) \atop \ell(\lambda)\le N}\,e^{-U_\lambda}s_\lambda(\bt)
\\
 \label{S^{(1)}3}
S^{(1)}_{3}(\bt,N,\bt';U)&:=&\sum_{\lambda\in\Pa\atop \ell(\lambda)\le N}\,e^{-U_\lambda} 
Q_{\alpha(\lambda)}(\tfrac 12\bt') \,s_\lambda(\bt)
\\
 \label{S^{(1)}4}
S^{(1)}_{4}(\bt,N=2n,U)&:=&\sum_{\lambda\in\FP\atop \ell(\lambda)\le N}\,e^{-U_\lambda}s_\lambda(\bt)
\\
 \label{S^{(1)}i=567}
S^{(1)}_{i}(\bt,N;U,f)&:=&\sum_{\lambda\in\Pa\atop \ell(\lambda)\le N}\,\Delta^{(i)}_N\left(f(h)\right)\,e^{-U_\lambda}s_\lambda(\bt)\,,\quad i=5,6,7
\eea
The coefficients $U_{\{\alpha\}}$ are defined as \be
 U_{\{\alpha\}} := \sum_{i=1}^k U_{\alpha_i},
 \ee
The notation $U_\lambda$ serves for 
 \be
U_\lambda\,:=\,U_{\{h\}}\,,\quad h_i=\lambda_i-i+\ell(\lambda)
 \ee
\begin{Proposition} \em \label{Prop1}
Sums \eqref{S^{(1)}},\eqref{S^{(1)}0}-\eqref{S^{(1)}i=567} are tau functions of the ``large'' BKP hierarchy introduced in 
\cite{KvdLbispec} with respect to the time variables $\bt$. Sums \eqref{S^{(1)}3} are tau functions of the ``small''
BKP hierarchy introduced in \cite{JM} with respect to the time variables $\bt'$.

\end{Proposition}

Actually the sum \eqref{S^{(1)}4} where we put $U=0$ and $\bt=(1,0,0,\dots)$ was used in \cite{BorStrahov}.

\paragraph{Sums over pairs of strict partitions.}

In the Frobenius notations \cite{Mac} we write $\lambda=(\alpha|\beta)=(\alpha_1,\dots,\alpha_k|\beta_1,\dots,\beta_k)$.
where $\alpha=(\alpha_1,\dots,\alpha_k)$, $\alpha_1>\cdots>\alpha_k\ge 0$ and $\beta=( \beta_1,\dots,\beta_k)$ may be 
viewed as strict partitions. It is clear that $\ell(\alpha)=\ell(\beta),\ell(\beta)\pm 1$, and we imply this restriction
in sums over pairs of strict partitions below.

Now we consider
 \be
 \label{S^{(2)}}
S^{(2)}(\bt; U,{\bar A},{\bar B})\,:=\,1+\sum_{\alpha ,\beta\in\DP}\,
e^{U_{\{-\beta-1\}}-U_{\{\alpha\}}}\,{\bar A}_{\alpha}
\,s_{(\alpha|\beta)}(\bt)\,{\bar B}_{\beta}
  \ee
where given infinite skew matrices $A$ and $B$ and given vectors $a$ and $b$, the  factors ${\bar A}_{\alpha}$ and 
${\bar B}_{\alpha}$ are defined in the same way as before. 
 \be
U_{\{\alpha\}}=\sum_{i=1}^k\,U_{\alpha_i}\,,\quad 
U_{\{-\beta-1\}}=\sum_{i=1}^k\,U_{-\beta_i-1}
 \ee

We introduce the following notation
\be
 \label{S^{(2)}ij}
S^{(2)}_{ij}(\bt;U)\,:=\,S^{(2)}(\bt; U,{\bar A}_i,{\bar A}_j)
  \ee
where $i,j=1,\dots,7$ and matrices $A_i$ are taken from the Examples 1-7 above.
In particular we obtain series
 \bea
 \label{S^{(2)}11}
S^{(2)}_{11}(\bt;U)&:=&\sum_{\lambda\in\Pa}\,e^{-U_\lambda}s_\lambda(\bt)
\\
 \label{S^{(2)}22}
S^{(2)}_{22}(\bt;U)&:=&
1+\sum_{\alpha,\beta\in\SCDP}\,
e^{U_{\{\beta \}}-U_{\{\alpha \}}} s_{(\alpha|\beta)}(\bt)
\\
\label{S^{(2)}24}
S^{(2)}_{24}(\bt;U)&:=&1+\sum_{\alpha\in\SCDP
,\beta\in\FDP}\,e^{U_{\{\beta \}}-U_{\{\alpha \}}}
\,s_{(\alpha|\beta)}(\bt)
\\
 \label{S^{(2)}31}
S^{(2)}_{31}(\bt,\bt';U)&:=&1+\sum_{\alpha
,\beta\in\DP}\,e^{U_{\{\beta \}}-U_{\{\alpha \}}}
\,Q_{{\bar\alpha}(\alpha)}(\tfrac 12\bt')
\,s_{(\alpha|\beta)}(\bt)
\\
 \label{S^{(2)}41}
S^{(2)}_{41}(\bt;U)&:=&1+\sum_{\alpha\in\FDP
,\beta\in\DP}\,e^{U_{\{\beta \}}-U_{\{\alpha \}}}
\,s_{(\alpha|\beta)}(\bt)
\\
 \label{S^{(2)}44}
S^{(2)}_{44}(\bt;U)&:=&
1+\sum_{\alpha,\beta\in\FDP}\,
e^{U_{\{\beta \}}-U_{\{\alpha \}}} s_{(\alpha|\beta)}(\bt)
\\
 \label{S^{(2)}33}
S^{(2)}_{33}(\bt,\bt',\bt'';U)&:=&1+\sum_{\alpha
,\beta\in\DP}\,e^{U_{\{\beta \}}-U_{\{\alpha \}}}
\,Q_{{\bar\alpha}(\alpha)}(\tfrac 12\bt')
\,s_{(\alpha|\beta)}(\bt)\,Q_{{\bar\beta}(\beta)}(\tfrac 12\bt'')
\\
 \label{S^{(2)}34}
S^{(2)}_{34}(\bt,\bt';U)&:=&1+\sum_{\alpha\in\DP
,\beta\in\FDP}\,e^{U_{\{\beta \}}-U_{\{\alpha \}}}
Q_{{\bar\alpha}(\alpha)}(\tfrac 12\bt')
\,s_{(\alpha|\beta)}(\bt)
\\
 \label{S^{(2)}ij567}
S^{(2)}_{ij}(\bt;U,f)&:=&1+\sum_{\alpha\in\DP
,\beta\in\DP}\,e^{U_{\{\beta \}}-U_{\{\alpha \}}}
\Delta^{(i)}(f(\alpha)) \,s_{(\alpha|\beta)}(\bt)\, \Delta^{(j)}(f(\beta))\,,\quad i,j=5,6,7
 \eea

Each
 $Q_\alpha(\tfrac 12\bt')$ is known to be a ``small'' BKP 
 \cite{DJKM-B1},\cite{JM} tau function. (This was a nice observation of
 \cite{You},\cite{Nimmo}).  The fact that only
 odd subscripts appear in the BKP higher times $t_{2m-1}$ is
 related to the reduction from the KP hierarchy.

\begin{Proposition} \em \label{Prop3}
Sums \eqref{S^{(2)}},\eqref{S^{(2)}ij} are tau functions of the ``large'' BKP hierarchy introduced in 
\cite{KvdLbispec} with respect to the time variables $\bt$. Sums \eqref{S^{(2)}31} are tau functions of the
BKP hierarchy introduced in \cite{JM} with respect to the time variables $\bt'$. Sums \eqref{S^{(2)}33} are tau functions of the
two-component BKP hierarchy introduced in \cite{JM} with respect to the time variables $\bt'$ and $\bt''$.

\end{Proposition}

\br \em

Let us remind that for the small BKP hierarchy obtained from KP we have the following \cite{HLO} 
 \be
S=\sum_{\alpha\in\DP}\,{\bar A}_\alpha\,Q_\alpha(\bt')
 \ee
By specification of the data ${\bar A}$ we obtain
 \be
 \label{smallBKPoutproduct1}
\sum_{\alpha\in\DP}\,e^{-U_{\{\alpha\}}}Q_\alpha(\tfrac 12 \bt') \,,\quad \sum_{\alpha\in\DP}\,
e^{-U_{\{\alpha\}}}Q_\alpha(\tfrac 12\bt')Q_\alpha(\tfrac 12{\bt''})\,,\quad \sum_{\alpha\in\DP'}\,
e^{-U_{\{\alpha\}}}Q_\alpha(\tfrac 12 \bt')
 \ee

The sums \eqref{smallBKPoutproduct1} are particular examples (see \cite{HLO}) of  BKP tau  functions,
 as introduced in \cite{DJKM-B1}, defining solutions to what was called the small BKP
 hierarchy in \cite{KvdLbispec}.

 The coupled small BKP yields series
 \be
 \label{S5}
S_5({\bt'},{\bt''},D):=\sum_{\alpha,\beta\in\DP \atop
\ell(\alpha)=\ell(\beta)}\,Q_\alpha(\tfrac
12\bt')D_{\alpha,\beta}Q_\beta(\tfrac 12{\bt''})
 \ee

The coefficients $D_{\alpha,\beta}$ in (\ref{S5}) are defined as
determinants:
 \be\label{D-alpha-beta}
D_{\alpha,\beta}=\det\,\left( D_{\alpha_i,\beta_j}\right)
 \ee
 where $D$ is a given infinite matrix. Taking $D_{nm}=e^{U_m-U_n}s_{(n|m)}(\bt)$ we reproduce \eqref{S^{(2)}33}.

\er

\subsection{ Pfaffian representations \label{Pfaffian representations-1}}

For 
 \be
\bt=\bt({\bf x}^{(M)})=:[x_1]+\cdots +[x_M]
 \ee
we have for any $N\ge M=1$
 \be
S^{(1)}(\bt(x_i);N,U,\bar{A})\,=\,\sum_{n=0}^\infty\,a_ne^{-U_n}x_i^n
 \ee
and for any $N\ge M=2$ we have
 \be
S^{(1)}(\bt(x_i,x_j);N,U,\bar{A})\,=\,\frac{1}{x_i-x_j}
\sum_{m>n\ge 0}^\infty\,A_{nm}e^{-U_n-U_m}\left(x_i^mx_j^n-x_i^n x_j^m \right)
 \ee

\bp\label{Pfaff-Miwa}

For $M=N$ we have
\be
S^{(1)}(\bt({\bf x}^{(M)});N,U,\bar{A})\,=\,
\frac{1}{\Delta_N(x)}\Pf [{\tilde S}]
\ee
where for $N=2n$ even
  \be
  \label{S-alpha-even-n}
{\tilde S}_{ij}=-{\tilde S}_{ji}:=(x_i-x_j)S^{(1)}(\bt(x_i,x_j),N,U,\bar{A}),\quad 1\le
i<j \le 2n
  \ee
and for $N=2n-1$ odd
 \be \label{S-alpha-odd-n} {\tilde
S}_{ij}=-{\tilde S}_{ji}:=
\begin{cases}
(x_i-x_j)S^{(1)}(\bt(x_i,x_j),N,U,\bar{A}) &\mbox{ if }\quad 1\le i<j \le 2n-1 \\
S^{(1)}(\bt(x_i),N,U,\bar{A}) &\mbox{ if }\quad 1\le i < j=2n  
 \end{cases}
  \ee
and where
 \be\label{true-sign-Delta_N(x)}
\Delta_N(x)\,:=\,\prod_{0\le i < j \le N}\,(x_i-x_j)
 \ee
 
\ep

We shall omit more spacious formulae for the case $M \neq N$.

 \br\label{AppendixS^{(1)}(bt(x_i,x_j))}
 Let us write down the entries of ${\tilde S}$ to express 
$S^{(1)}_i,\,i=0,\dots,6$

 \be
S^{(1)}_1(\bt(x_i,x_j),N,U)=(x_i-x_j)^{-1}\sum_{m>n\ge 0} e^{-U_n-U_m}(x_i^mx_j^n-x_j^mx_i^n)
 \ee
 \be
S^{(1)}_1(\bt(x_i),N,U)=\sum_{n=0}^\infty e^{-U_n}x_i^n
 \ee
\,
 \be
S^{(1)}_2(\bt(x_i,x_j),2n,U)=(x_i-x_j)^{-1}\sum_{n = 0}^\infty e^{-U_n-U_{c-n}}(x_i^{c-n}x_j^n-x_j^{c-n}x_i)
 \ee
\,
 \be
S^{(1)}_3(\bt(x_i,x_j),N,U)=(x_i-x_j)^{-1}\sum_{m>n\ge 0}^\infty e^{-U_n-U_{m}}Q_{(n,m)}(\bt')(x_i^{n}x_j^m-x_j^{n}x_i^m)
 \ee
 \be
S^{(1)}_3(\bt(x_i),N,U)=\sum_{n\ge 0}^\infty e^{-U_n}Q_{(n)}(\bt') x_i^{n}
 \ee
\,
 \be
S^{(1)}_4(\bt(x_i,x_j),N`,U)=(x_i-x_j)^{-1}\sum_{n\ge 0}^\infty e^{-U_n-U_{n+1}}(x_i^{n}x_j^{n+1}-x_j^{n}x_i^{n+1})
 \ee
 \be
S^{(1)}_4(\bt(x_i),N,U)=\sum_{n\ge 0}^\infty e^{-U_n} x_i^{n}
 \ee

 \er

In particular substituting \eqref{ExA1},\eqref{ExA4}  we obtain
 \be
S^{(1)}_1(\bt,N,U=0)=\frac{1}{\Delta_N(x)}\Pf\,\frac{x_j-x_i}{(1-x_i)(1-x_j)(1-x_ix_j)}
 \ee
 \be
S^{(1)}_4(\bt,N,U=0)=\frac{1}{\Delta_N(x)}\Pf\,\frac{x_j-x_i}{1-x_ix_j}
 \ee
Then it follows that
 \be\label{sum-Schur-Pa}
\sum_{\lambda\in\Pa}\,s_\lambda(\bt({\bf x}^N))\,=\,\prod_{i=1}^N(1-x_i)^{-1}\,\prod_{i<j\le N}(1-x_ix_j)^{-1}
 \ee
which called Schur-Littlewood identity,
 and
 \be\label{sum-Schur-FP}
\sum_{\lambda\in\Pa}\,s_{\lambda\cup\lambda}(\bt({\bf x}^N))\,=\,\prod_{i<j\le N}(1-x_ix_j)^{-1}
 \ee
Formulae \eqref{sum-Schur-Pa} and \eqref{sum-Schur-FP} are known, see Ex-s 4-5 in I-5 of \cite{Mac}, 
\eqref{sum-Schur-Pa} is called Schur-Littlewood identity.

It is convenient to re-write these formulae in a way independent of the choice of $N$:

 \bp

 \be\label{SchurSum1}
\sum_{\lambda\in\Pa}\,s_\lambda(\bt)\,=\,
e^{\frac 12\sum_{m=1}^\infty\,mt_m^2\,+\,\sum_{m=1}^\infty \, t_{2m-1}}
 \ee
and
 \be\label{SchurSum4}
\sum_{\lambda\in\Pa}\,s_{\lambda\cup\lambda}(\bt)\,=\,
e^{\frac 12\sum_{m=1}^\infty\,mt_m^2\,-\,\sum_{m=1}^\infty \, t_{2m}}
 \ee
Relations \eqref{SchurSum1} and \eqref{SchurSum4} will be used later in Section{} to solve
certain combinatorial problem.
From
 \be
s_{\lambda^{tr}}(\bt)=(-1)^{|\lambda|}s_\lambda(-\bt)
 \ee
we obtain
\be\label{SchurSum1-}
\sum_{\lambda\in\Pa}\,(-1)^{|\lambda|} s_\lambda(\bt)\,=\,
e^{\frac 12\sum_{m=1}^\infty\,mt_m^2\,-\,\sum_{m=1}^\infty \, t_{2m-1}}
 \ee
and
 \be\label{SchurSum-even}
\sum_{\lambda\in\Pa_{even}}\, s_{\lambda}(\bt)\,=\,
e^{\frac 12\sum_{m=1}^\infty\,mt_m^2\,+\,\sum_{m=1}^\infty \, t_{2m}}
 \ee

\ep
 
  By the simple re-scaling $t_m\to z^mt_m$ in equations \eqref{SchurSum1}-\eqref{SchurSum-even}
  and equating factors before same powers of $z$ we obtain

\bp

\bea
\label{areafixedsum1}
\sum_{\lambda\in\Pa \atop
|\lambda|=\t}\,s_\lambda(\bt) &=& s_{(\t)}({\tilde\bt})
\,\qquad\qquad\qquad \left[\,{\tilde t}_{2m-1}= t_{2m}\,,\,\, {\tilde t}_{2m}=\frac m2 t_m^2\,\right]
\\
\label{areafixedsum1-}
\sum_{\lambda\in\Pa \atop
|\lambda|=\t}\,(-1)^{|\lambda|} s_{\lambda}(\bt) &= &s_{(\t)}({\tilde\bt}) 
\,\qquad\qquad\qquad \left[\,{\tilde t}_{2m-1}= -t_{2m}\,,\,\, {\tilde t}_{2m}=\frac m2 t_m^2\,\right]
  \\
 \label{areafixedsum4}
\sum_{\lambda\in\Pa \atop
|\lambda|=\t}\,s_{\lambda\cup\lambda}(\bt) & = & s_{(\t)}({\tilde\bt})\,\qquad \qquad\qquad
 \left[\,{\tilde t}_m=\frac m2 t_m^2 - t_{2m}\,\right]
 \\
\label{areafixedsum-even}
\sum_{\lambda\in\Pa_{even} \atop
|\lambda|=\t}\,s_{\lambda}(\bt) & = & s_{(\t)}({\tilde\bt}) 
\,\qquad\qquad \qquad \left[\,{\tilde t}_m=\frac m2 t_m^2 + t_{2m}\,\right]
  \eea
where auxilary sets of times ${\tilde\bt}=({\tilde t}_1,{\tilde t}_2,\dots)$ are specified in the brackets to the right of
equalities. 

\ep

For instance we get \eqref{areafixedsum1} from
 \eqref{SchurSum1} using the equality
  \[
  \sum_{\lambda\in\Pa}\,z^{|\lambda|}s_\lambda(\bt)=e^{\sum_{m=1}^\infty
  \,\frac {z^{2m}}{2} mt_m^2+\sum_{m=1}^\infty\,
  z^{2m-1}t_{2m-1}}=\sum_{\t=0}^\infty\,z^{\t}\, s_{(\t)}({\tilde\bt})
  \]
 where ${\tilde\bt}=
 \left(t_1,\frac{1\cdot t_1^2}{2},t_3,\frac{2\cdot t_2^2}{2},t_5,\frac{3\cdot t_3^2}{2},
 \dots \right)$.

Formula \eqref{areafixedsum1} in case $\bt=(1,0,0,\dots)$ has an interpretation in terms of total numbers of standard tableaux 
of weight $(1^\t)$ and numbers of involutive permutations of $S_\t$, see Ex 12 I.5 of \cite{Mac}, 
\cite{BaikRains-involution}, \cite{BF}.

We get from Proposition \ref{Pfaff-Miwa}

\bp

 \be
S^{(1)}_1(\bt({\bf x}^{(2n)}),2n,U)\,=
\,\frac{1}{\Delta_{2n}(x)}\Pf \left[ \sum_{m>n\ge 0 }e^{-U_n-U_m}(x_j^mx_i^n-x_i^mx_j^n) \right]_{i,j=1,\dots,2n}
 \ee
Choosing $U_m=0,\,m\le L+2n-1$ and $U_m=+\infty,\, m > L+2n-1$ we obtain
 \be
\sum_{\lambda\in P\atop \lambda_1\le L}\,s_{\lambda}(\bt({\bf x}^{(2n)}))\,=\,
\frac{1}{\Delta_{2n}(x)}\Pf \left[\frac{x_j-x_i}{(1-x_i)(1-x_j)(1-x_ix_j)}\left(1-(x_ix_j)^{L+N}+
\frac{x_j^{L+N}-x_i^{L+N}}{x_j-x_i} \right)\right]_{i,j=1,\dots,2n}
 \ee

\ep

 \bp

 \be
S^{(1)}_4(\bt({\bf x}^{(2n)}),2n,U)\,=\,\frac{1}{\Delta_{2n}(x)}\Pf \left[(x_j-x_i)f(x_ix_j,U)\right]_{i,j=1,\dots,2n}
 \ee
where
 \[
  f(z,U)=\sum_{m=0}^\infty e^{-U_m-U_{m+1}}z^m\,=\,S^{(1)}_4(\bt(x_i,x_j),U)
 \]
Choosing $U_m=0,\,m\le L+2n-1$ and $U_m=+\infty,\, m > L+2n-1$ we obtain
 \be
\sum_{\lambda\in P\atop \lambda_1\le L}\,s_{\lambda\cup\lambda}(\bt({\bf x}^{(2n)}))\,=\,
\frac{1}{\Delta_{2n}(x)}\Pf \left[(x_j-x_i)\frac{1-(x_ix_j)^{L+2n-1}}{1-x_ix_j}\right]_{i,j=1,\dots,2n}
 \ee

\ep

Next, as a corollary of Proposition \ref{Pfaff-Miwa} we obtain

 \bp

 \be\label{S^{(1)}2-U=0-Pf}
\sum_{\lambda\in\SCP(c)} s_\lambda(\bt({\bf x}^{(2n)}))\,=\,
\frac{1}{\Delta_{2n}(x)}
\Pf \left(\frac{\left( x_j^{[c+\frac 12]}-x_i^{[c+\frac 12]} \right)^2}{x_j - x_i}\right)_{i,j=1,\dots,2n} 
 \ee
 \be
\sum_{\lambda\in\SCP(c)}(-1)^{\sum_{i=1}^n(\lambda_i-i+2n)} s_\lambda(\bt({\bf x}^{(2n)}))\,=\,
\frac{1}{\Delta_{2n}(x)}
\Pf \left(\left( x_j^{[c+\frac 12]}-x_i^{[c+\frac 12]}\right)
\frac{x_j^{[c+\frac 12]}-(- x_i)^{[c+\frac 12]} }{x_j + x_i}\right)_{i,j=1,\dots,2n} 
 \ee
where $[a]$ is equal to the integer part of $a$. Notice that in case $c=n$ we have only one term related to $\lambda=0$
and thus the both sides of identity \eqref{S^{(1)}2-U=0-Pf} are equal to 1 (compare to Lemma 5.7 in \cite{Ishikawa}).

 \ep

\paragraph{Integral representations for the sums}.

Let us recall the fermionic expression for the sums of Schur functions \eqref{{S^{(1)}}}
\[
S^{(1)}(\bt,N;U,{\bar A})  = \langle N|\Gamma(\bt) \mathbb{T}(U) g^{--}(A) |0\rangle,\quad
\mathbb{T}(U)=\exp \sum_{i\ge 0} \left( U_{-i-1} \psi^\dag_{-i-1}\psi_{-i-1} - U_i\psi_i\psi_i^\dag \right)
\]
where for the Examples 0-6 above we have
\bea
 \label{S^{(1)}0-f-int}
g^{--}((A_0)^c)&:=&
e^{\sum_{ M \ge m > n}\,\psi_m\psi_n +\sum_{m\le M}\,\psi_m\phi_0} 
\\
 \label{S^{(1)}1-f-int}
g^{--}({\bar A}_1)&:=&
e^{\sum_{m>n}\,\psi_m\psi_n +\sum_{m\in\mathbb{Z}}\,\psi_m\phi_0} =
\,e^{\oint\,\psi(x^{-1})\left(\psi(x)+\phi_0 \right)\frac{1}{1-x} dx}
\\
 \label{S^{(1)}2-f-int}
g^{--}({\bar A}_2)&:=&
e^{\sum_{m<n}\,(-1)^m\psi_{2c-m+1}\psi_{m} } =
e^{\oint\,x^{-2c-2}\psi(x)\psi(-x) dx}
\\
 \label{S^{(1)}3-f-int}
g^{--}({\bar A}_3(\bt'))&:=&
e^{\sum_{m>n}\,Q_{(\alpha(n),\alpha(m))}(\tfrac 12\bt')\psi_m\psi_n +
\sum_{m\in\mathbb{Z}}\,Q_{(\alpha(m))}(\tfrac 12\bt')\psi_m\phi_0} =
\\
 \label{S^{(1)}3-t'-infty-f-int}
g^{--}({\bar A}_3(\bt'_\infty))&:=&
e^{\sum_{m>n}\,\frac{m-n}{m+n}\psi_m\psi_n +
\sum_{m\in\mathbb{Z}}\,\psi_m\phi_0} =
\\
 \label{S^{(1)}4-f-int}
g^{--}({\bar A}_4)&:=&
e^{\sum_{m \in\mathbb{Z}}\psi_m\psi_{m-1} } =
e^{\oint\,\psi(x^{-1})\psi(x) dx}
\\
 \label{S^{(1)}5-f-int}
g^{--}({\bar A}_5)&:=&
e^{\sum_{m>n}\,\frac{m-n}{1-mn}\psi_m\psi_n +
\sum_{m\in\mathbb{Z}}\,\psi_m\phi_0} =
 \\
 \label{S^{(1)}6-f-int}
g^{--}({\bar A}_6)&:=&
e^{\sum_{m>n}\,\frac{m-n}{(m+n)^2}\psi_m\psi_n +
\sum_{m\in\mathbb{Z}}\,\psi_m\phi_0} =
\eea

The corollary of the right hand side expressions is the fact that sums 
\eqref{S^{(1)}i} may be re-written as certain multiply integrals ($\frac 12 N$-ply integrals
for $S^{(1)}_2$, $S^{(1)}_4$, and $N$-ply integrals in other cases).

For a sum \eqref{S^{(1)}} it is true in case we can present $A_{nm}$ as moments, or, the same the following inverse 
moment problem:
given $A_{nm}=-A_{mn},\,m,n\ge 0$ to find such an integration domain $D$ and an antisymmetric measure 
$dA(x,y)=-dA(y,x)$ such that
 \be\label{IMP}
A_{nm}=\int_D\, x^ny^m d A(x,y),\quad n,m\ge 0
 \ee
Also
 \be
a_n=\int_\gamma\,x^n da(x)
 \ee

If we have \eqref{IMP} then in case $N=2n$ we can write $N$-ply integral [[to be fixed]]
 \be
S^{(1)}(\bt,2n,U,{A}^c)=e^{-\sum_{i=0}^{N-1} U_{i}}\int_{D^n}\,
\left(\prod_{i=1}^{2n}e^{\xi_r(\bt,x_i)}\cdot\Delta_{2n}(x)\right) \Pf \left[ dA(x_i,x_j) \right] 
 \ee
where $\xi_r(\bt,x)$ is the following $\Psi DO$ operator
 \be
\xi_r(\bt,x)=\sum_{m=1}^\infty\,t_m \left(x r(D) \right)^m\,,\quad D=x\partial_x
 \ee
and $r$ is related to $U$ as follows
 \be
r(n)=e^{U_{n}-U_{n+1}}
 \ee
The case $U=0$ causes $
  \xi_r(\bt,x)=\sum_{m=1}^\infty\,t_m x ^m $ and we obtain more familier expression
 \be
S^{(1)}(\bt,2n,U=0,{A}^c)=\int_{D^n}\,
\prod_{i=1}^{2n}e^{\sum_{m=1}^\infty\,t_m x_i^m }\,\Delta_{2n}(x) \Pf \left[ dA(x_i,x_j) \right] 
 \ee

In case $N=2n+1$ we have more involved expressions.

 In case the solution of the inverse problem is not unique we have a set of different
integral representations for the sum \eqref{S^{(1)}}.

\section{Matrix integrals. Integration measures \label{Integration measures} }

\paragraph{Three Ginibre ensembles}

The complex Ginibre ensemble of complex $N\times N$ matrices $Z$ is defined by the following probability measure
 \be\label{Ginibre-measure-2}
d\mu_2(Z)= e^{-\Tr ZZ^\dag}\prod_{i,j=1,\dots,N}\,\frac{|dZ_{ij}|^2}{2\pi}
 \ee
If we write $Z=U(\Lambda+\Delta)U^\dag$ where $U\in\mathbb{U}(N)$, $\Lambda=\diag (z_1,\dots,z_N)$ and $\Delta$
is stricly upper-triangular, then
 \be
d\mu_2(Z)=C_N e^{-\Tr \left(\Lambda\Lambda^\dag - \Delta\Delta^\dag\right)} \prod_{i<j}^N\,|z_i-z_j|^2 
|d\Delta|^2 |d\Lambda|^2 |d_*U|^2
 \ee

The real quaternionic 
 \be\label{Ginibre-measure-4}
d\mu_4(X)= e^{-\Tr XX^\dag}\prod_{i,j=1,\dots,N}\,\frac{|dX_{ij}|^2}{2\pi}
 \ee

The real Ginibre ensemble of $N\times N$ real matrices $X$ is defined be the following measure
 \be\label{Ginibre-measure-1}
d\mu_1(X):= \prod_{i,j=1,\dots,N} \,e^{ -\frac 12 \Tr X_{ij}^2  }\,\frac{dX_{ij}}{\sqrt{2\pi}}
 \ee

\paragraph{Integrals over orthogonal and symplectic groups.}  Now

For $\mathbb{O}(2n)$ the Haar measure is
 \be
d\mu(O)=\frac{2^{(n-1)^2}}{\pi^n n!} 
\int_{\theta_1 \le \cdots \le \theta_n \le 2\pi}\,\prod_{i<j}^n\,(\cos \theta_i -\cos \theta_j )^2 \,\prod_{i=1}^n d\theta_i
 \ee
where $e^{\theta_1},e^{-\theta_1},\dots,e^{\theta_n},e^{-\theta_n}$ are eigenvalues of $O$.

For $\mathbb{O}(2n+1)$ the Haar measure is
 \be
d\mu(O)=\frac{2^{n^2}}{\pi^n n!} 
\int_{\theta_1 \le \cdots \le \theta_n \le 2\pi}\,\prod_{i<j}^n\,(\cos \theta_i -\cos \theta_j )^2 \,
\prod_{i=1}^n \sin^2\frac{\theta_i}{2}d\theta_i
 \ee
where $e^{\theta_1},e^{-\theta_1},\dots,e^{\theta_n},e^{-\theta_n},1$ are eigenvalues of $O$.

For $\mathbb{S}p(2n)$ the Haar measure is
 \be
d\mu(O)=\frac{2^{n^2}}{\pi^n n!} 
\int_{\theta_1 \le \cdots \le \theta_n \le 2\pi}\,\prod_{i<j}^n\,(\cos \theta_i -\cos \theta_j )^2 \,
\sin^2 {\theta_i}\prod_{i=1}^n d\theta_i
 \ee
where $e^{\theta_1},e^{-\theta_1},\dots,e^{\theta_n},e^{-\theta_n}$ are eigenvalues of $S$.

\section{More details\label{More-details}}

\subsection{\label{integrals}}

Integrals $I^{(1)}_1$,$I^{(1)}_2$ and $I^{(1)}_4$ may be considered as $\beta=1,2,4$
ensembles \cite{Mehta} related to a contour $\gamma$. They may be
obtained as particular cases of $I^{(1)}$ as follows:

\noindent
 Integral $I^{(1)}_1(N)$ is a particular case of $I^{(1)}(N)$ where in the (A) case
 \be
 A(x_i,x_j)=\sgn(x_i-x_j), \quad a(x)=1
 \ee
  while in case (B)
 \be
A(x_k,x_j)= e^{-\tfrac {\pi i}2 } \sgn(\varphi_k-\varphi_j), \quad
a(x)=e^{-\tfrac {\pi i}4 },
 \ee
 with
$\varphi_i={\text{arg}}\, x_i$. To prove this we use:
 \bl
 \be
  \Pf\left[\sgn(x_k-x_j)\right]=\sgn\,\Delta(x),\quad x_k\in\mathbb{R},
 \ee
 \be
 \Pf\left[\sgn(\varphi_k-\varphi_j)\right]=
 \sgn\,\left(e^{-\tfrac {\pi i}4(N^2-N) }\Delta(x)\right),\quad x_k=e^{i\varphi_k}
  \ee
  where $k,j=1,\dots,N$.
 \el

Integral $I^{(1)}_{3}(N)$ is obtained from $I^{(1)}(N)$ by setting
 \be
A(x_i,x_j)=\frac{x_i-x_j}{x_i+x_j}, \quad  a(x)= 1.
 \ee
We use the fact that
  \be
\Delta^*(x) = \Pf\left[\frac{x_i-x_j}{x_i+x_j} \right]
  \ee

Integral $I^{(1)}_4(N)$ is obtained from $I^{(1)}(N)$ as follows. In case
(A) we set
  \be
  \label{delta_kernel}
  A(x_i,x_j)=\frac 12 \left(x_j\frac{\partial}{\partial
  x_j}\delta(x_i-x_j)-(x_i\leftrightarrow x_j)\right)
  \ee
   and in  case (B) we set
  \be
  A(x_i,x_j)=\frac{\partial}{\partial
  \varphi_j}\delta(\varphi_i-\varphi_j).
  \ee

To relate these integrals to the 2-BKP hierarchy we introduce
deformations $I^{(1)}(N)\to I^{(1)}(N;\bt,{\bar\bt})$ through the
following deformation of the measure
 \be d\mu({ x})\to d\mu({
x}|\bt,{\bar\bt})= b(\bt,\{ x\})b(-{\bar\bt},\{ x^{-1}\})d\mu({x})
\ee where
\begin{equation}
\label{ebb2} b(\bs,\bt)=\exp \sum_{n\ {\rm odd}} \frac{n}{2}
s_nt_n
\end{equation}
and
\begin{equation}
\label{bracketz} \{
z\}=(2z,\frac{2z^3}{3},\frac{2z^5}{5},\cdots)\, .
\end{equation}

Below, we show that the generating series obtained by
Poissonization (the grand partition function) \be Z_i(\mu \,
;\bt,{\bar\bt})\, =\,b(\bt,{\bar\bt})\sum_{N=0}^\infty
\,I_i(N;\bt,{\bar\bt}) \,\frac{\mu^N}{N!}\,,\quad i=1,2,3,4, \ee
are particular 2-BKP tau functions (\ref{2-nBKP}).

We also consider the following $2N$-fold integrals:
 \be\label{I5}
I_5(N;\bt^{(1)},\bt^{(2)},{\bar\bt}^{(1)},{\bar\bt}^{(2)}):=\int
\Delta^*_{N}(z)\Delta^*_{N}(y) \prod_{i=1}^{N}d\nu({ z}_i,
y_i|\bt^{(1)},\bt^{(2)},{\bar\bt}^{(1)},{\bar\bt}^{(2)}),
 \ee
where
 \be
  d\nu({
z},y|\bt^{(1)},\bt^{(2)},{\bar\bt}^{(1)},{\bar\bt}^{(2)})=
  \ee
 \[
b(\bt^{(1)},\{ z\})b(-{\bar\bt}^{(1)},\{ z^{-1}\}) b(\bt^{(2)},\{
y\})b(-{\bar\bt}^{(2)},\{ y^{-1}\})d\nu(z,y)
 \]
(here $d\nu(z,y)$ is an arbitrary bi-measure), and show that the
generating series
 \be\label{Z5}
Z_5(\mu\, ;\bt^{(1)},\bt^{(2)},{\bar\bt}^{(1)},{\bar\bt}^{(2)})\,
=\,b(\bt^{(1)},{\bar\bt}^{(1)})b(\bt^{(2)},{\bar\bt}^{(2)})\sum_{N=0}^\infty
\,I_5(N;\bt^{(1)},\bt^{(2)},{\bar\bt}^{(1)},{\bar\bt}^{(2)})
\,\frac{\mu^N}{N!}
 \ee
is a particular case of the two-component 2-BKP tau function
(\ref{2c-2-nBKP}).

 \br
 \label{Z2=Z5}
Note that \be Z_2(\mu \, ;\bt,{\bar\bt}) =Z_5(\mu \,
;\bt^{(1)},\bt^{(2)},{\bar\bt}^{(1)},{\bar\bt}^{(2)}) \ee
 if
 \be
 d\nu(z,y) = \delta(z-y)d\nu(z)d\nu(y), \quad
\bt=\bt^{(1)}+\bt^{(2)}, \quad
{\bar\bt}={\bar\bt}^{(1)}+{\bar\bt}^{(2)}. \ee
 \er
The integrals $Z_1(\mu \, ;\bt,{\bar\bt})$, $Z_2(\mu \,
;\bt,{\bar\bt})$, $Z_4(\mu \, ;\bt,{\bar\bt})$ and $Z_5(\mu \,
;\bt^{(1)},\bt^{(2)},{\bar\bt}^{(1)},{\bar\bt}^{(2)})$
 may be obtained as  continuous limits of
$S_1(\bt_\infty,\bt^*)$, $S_2(\bt_\infty,\bt_\infty,\bt^*)$,
$S_4(\bt_\infty,\bt^*)$ and $S_5(\bt_\infty,\bt_\infty,\bt^*)$,
respectively.

  \subsection{Partition functions  of the $\beta=1$ and $\beta=4$
  circular ensembles as tau functions \label{partition-section}}

As it well known \cite{Mehta} the integral over random orthogonal
($\beta=1$) and symplectic ($\beta=4$) $N$ by $N$ matrices may be
reduced to the $N$-ply integral over real lines
  \be\label{eigenvalue-int-orth}
Z_N^{\beta}:=\int_{\mathbb{R}^1} \cdots \int_{\mathbb{R}^1} \,
\prod_{n<m\le N}|x_n-x_m|^\beta \, \prod_{i=1}^N  \,
e^{\sum_{n=1}^\infty \, t_n x_i^n} \, d\mu(x_i)
  \ee
where $d\mu$ is an integration measure which yields the
probability weight, while the exponential factor in front of it is
a deformation of this measure with a given set of deformation
parameters $\bt=(t_1,t_2,\dots)$.

The wonderful result of \cite{AvM-Pfaff}-\cite{AMS} that
$Z_N(\bt)$ is a tau function of a certain hierarchy of integrable
equations which were introduced and studied in a series of papers
and named as Pfaff lattice  and which were found in \cite{AMS} to
be a tricky constraint of 2D Toda lattice hierarchy. Then it was
proven by J. van de Leur in \cite{L1} that $Z_N^{\beta},\,
\beta=1,4$ is a tau function of the charged BKP hierarchy
introduced in \cite{KvdLbispec} which may be viewed as a
modification of the DKP hierarchy described in \cite{JM}. Here we
remind and modify some of results of \cite{L1}.

 We generalize \eqref{eigenvalue-int-orth} as follows. First we
replace the integrals along the real ax $\mathbb{R}^1$ by the
integrals along any contour $\gamma$ on $\mathbb{C}^1$. Second we
add an additional set of deformation parameters: a set ${\bar
\bt}=({\bar t_1},{\bar t_2},\dots)$ and an integer $l$.

Thus we study
  \be\label{eigenvalue-int-gamma}
Z_N^{\beta}[\gamma]\,:=\,\int_{\gamma} \cdots \int_{\gamma}  \,\,
\left(\Upsilon_N(\bf{x})\right)^\beta \,\,\prod_{i=1}^N
\,d\mu(x_i\,;\,\bt,{\bar \bt},l)
  \ee
Here
 \be
\Upsilon_N(x)\,:=\,\prod_{n<m\le N}(x_n-x_m)
\sgn\left(\varsigma(x_n)-\varsigma(x_m)\right)
 \ee
where we introduce the parameter $\varsigma$ along the curve, and
where the deformed measure  $d\mu(x\,;\,\bt,{\bar \bt},l)$ is
 \be
d\mu(x\,;\,\bt,{\bar \bt},l)=x^l e^{\sum_{n=1}^\infty (t_n
x^n-{\bar t}_n x^{-n})} \, d\mu(x)
  \ee
  where $d\mu$ is any measure along $\gamma$ and $l,\bt,{\bar \bt}$ are deformation
  parameters.

Our main example will be circular beta-ensembles ($\beta=1,4$)
 \be\label{eigenvalue-int-cir}
Z_N^{C,\beta}(l,\bt,{\bar \bt}):=\oint \cdots \oint \,
\prod_{n<m\le N}|x_n-x_m|^\beta \, \prod_{i=1}^N \, x_i^l
\,e^{\sum_{n=1}^\infty (t_n x_i^n-{\bar t_n} x_i^{-n})} \,
d\mu(x_i)
  \ee
  which were not considered in \cite{L1}. We shall consider these
  ensembles in a way similar to \cite{L1}. Because of notational
  reasons we prefer to consider DKP hierarchy introduced in
  \cite{JM} rather than the charged BKP one introduced in
  \cite{KvdLbispec} in spite of the fact that BKP hierarchy seems to be more
  natural for such problems.

  \bp \label{matrix ens=tau BKP} \em Let   $d\mu(x)$ be any measure on the
  circle.
  \be\label{PropCirc1-BKP}
\tau_{N}(l,\bt,{\bar \bt}):=\,\l N+l|\,\g(\bt)\,e^{\frac 12
\oint\oint \psi(x)
\psi(y)\sgn\left(\arg(x)-\arg(y)\right)d\mu(x)d\mu(y)
}\,g^{-,0}\,{\bar \g}({\bar \bt})\,|l\r
  \ee
  \be\label{PropCirc2-BKP}
=(-)^{\frac{N^2-N}{2}}\frac{c}{N!}\oint \cdots \oint
\,\prod_{n<m\le N}|x_n-x_m| \, \prod_{i=1}^N \,  x_i^{l+\frac12
(N-1)}e^{\sum_{n=1}^\infty t_n x_i^n+{\bar t}_n x_i^{-n}}d\mu(x_i)
  \ee
 where $c=\exp \sum_{m=1}^\infty mt_m{\bar t}_m$ and
  \be
g^{-,0}:=e^{\oint \psi(x)d\mu(x)\phi_0\sqrt{2}}
  \ee

  \ep
  Here $\tau_{N}(l,\bt,{\bar \bt})$ is a tau function of the 2-BKP
  hierarchy. In particular it means that if we fix the variables $\bt=(t_1,t_2,\dots)$
  then $\tau_{N+l,l}(\bt,{\bar \bt})$ is a 2-BKP tau function with respect to the
  variables ${\bar \bt}=({\bar t}_1,{\bar t}_2,\dots)$. The complete set of Hirota
  equations for the 2-BKP is written down in the Appendix.

  The proof basically repeats the proof of \cite{L1} of the similar statement for
  ensembles \eqref{eigenvalue-int-orth} which we write down in a little bit more
  unsophisticated way as follows.

   \bl\em
    Let $\gamma$ be a contour on a complex plane whose points $x\in\gamma$ are
    parameterized by a parameter
    $\varsigma$: $x=x(\varsigma)$.
    Let $d\mu(x,y)=d\mu(y,x)$ be a symmetric bi-measure
   on a $\gamma \times \gamma$. Then, for any antisymmetric
   function $A(x,y)$ we have
   \be
e^{\frac{z}{2}\int_\gamma\int_\gamma
\psi(x)\psi(y)A(x,y)d\mu(x,y)}=\sum_{N=0}^\infty z^N I_N
   \ee
   where $I_N$ is the following $2N$-ply integral
   \be
I_N=\int_\gamma \cdots \int_\gamma \psi(x_1)\cdots \psi(x_N)\,\Pf
\left[A({\bf x})\right]
   \ee
   where $\varsigma_i:=\varsigma(x_i)$ and $\varsigma_1>\cdots>\varsigma_N$ and where
   \be
A_{nm}({\bf x})=A(x_n,x_m)d\mu(x_n,x_m),\quad n,m=1,\dots,N
   \ee
   \el
   The proof is straightforward. We have
   \[
\frac 12 \int_\gamma\int_\gamma
\psi(x)\psi(y)A(x,y)d\mu(x,y)=\int\int_{\varsigma(x)>\varsigma(y)}
\psi(x)\psi(y)A(x,y)d\mu(x,y)
   \]
   Now
   \[
I_N=\frac{1}{N!}\left( \int\int_{\varsigma(x)>\varsigma(y)}
\psi(x)\psi(y)A(x,y)d\mu(x,y)\right)^N=
 \]
 \[
\int\cdots\int \psi(x_1)\psi(x_2)\dots \psi(x_N)\sum_{\sigma\in
S^N} {\sgn\,\sigma}\,A(x_{\sigma(1)},x_{\sigma(2)} )A(
x_{\sigma(3)},x_{\sigma(4)})\cdots A(
x_{\sigma(N-1)},x_{\sigma(N)})
\]
where the integration domain is restricted by the cone
$\varsigma(x_1)>\varsigma(x_2)>\cdots > \varsigma(x_N)$ and where
the sum runs over all possible permutations conditioned by
$\sigma(2i-1)>\sigma(2i),\, i=1,\dots,\frac 12 N$ and by
$\sigma(1)>\sigma(3)>\cdots
>\sigma(N-1)$ (the last restrictions gives rise to the additional factor $N!$).

Now we see that the following statement is valid
 \bl{\em
 \be
\langle N+l|\,e^{\frac{z}{2}\int_\gamma\int_\gamma
\psi(x)\psi(y)A(x,y)d\mu(x,y)}\,e^{\int_\gamma
a(x)\psi(x)d\mu(x)\phi_{0}\sqrt{2}}\,|l\rangle
 \ee

 \be
=\frac{(N-1)!!}{N!}\int_\gamma \cdots\int_\gamma
\Delta_{N}(x)A(x_1,x_2)A(x_3,x_4)\cdots A(x_{N-1},x_N)
 \ee
} \el

Now we apply this Lemma. For circular ensembles we take $\arg\,x$
as the parameter $\varsigma$ used in the Lemma.

 To get standard circular ensembles we imply that
 \be\label{factor-mu}
 d\mu(x,y)=d\mu(x,\bt,{\bar \bt})d\mu(y,\bt,{\bar \bt})
  \ee
  and
   \be
A(x,y)=\sgn(\arg x -\arg y)
   \ee

  $N$ even. It is enough to consider the case $\bt={\bar \bt}=0$. Writing the exponential
  in \eqref{PropCirc1} as Taylor series and keeping the $N$-th term we obtain the
  following $2N$-ply integral
  \[
\frac{1}{2^{N}}\oint\cdots\oint  \l N+l|\, \psi(x_1)\cdots
\psi(x_{2N})\,|l\r\,
\Pf\left[\sgn\left(\arg(x_i)-\arg(x_j)\right)d\mu(x_i,x_j)\right]|_{i,j=1,\dots,N}
  \]
  \be
=\,\frac{1}{2^{N}}\oint\cdots\oint
\,\prod_{i<j}^N\,(x_i-x_j)\prod_{i=1}^N\, x_i^l
\,d\mu(x_i,\bt,{\bar \bt})
  \ee
  where the integration domain is restricted by $\arg x_1>\arg x_2>\dots
  >\arg x_N$.
  Here we took into account that
   \be\label{psi-Vand}
\l N+l|\, \psi(x_1)\cdots \psi(x_{2N})\,|l\r=\prod_{i=1}^N\,
x_i^l\,\prod_{i<j}^N\,(x_i-x_j),
    \ee
and the last Lemma. Now using
     \[
\prod_{n<m\le N}\frac{|x_n-x_m|}{x_n-x_m}=\prod_{i=1}^N
{(-x_i)}^{\frac{1-N}{2}}\prod_{n<m\le N}\sgn\left(\sin\frac{\arg
x_n-\arg x_m}{2}\right)
\]
we can get rid of the restriction $\arg x_1>\arg x_2>\dots
  >\arg x_N$ getting the factor $(N!)^{-1}$. Thus we re-write $I_N$ as
   \be
\label{IN} Z_N^{C\beta=1}(\bt,{\bar
\bt})=(-)^{\frac{N^2-N}{2}}\frac{1}{N!}\oint \cdots \oint \,
\prod_{n<m\le N}|x_n-x_m| \, \prod_{i=1}^N \, x_i^{l+\frac12
(N-1)} \, d\mu(x_i,\bt,{\bar \bt})
   \ee
in accordance with \eqref{PropCirc2}. In the similar way we prove
\eqref{PropCirc4}.

\paragraph*{$\beta=4$ circular ensemble.} Somehow circular ensembles
  were not considered
  in the paper \cite{L1}. The circular $\beta=4$ ensemble is also
  related to a DKP tau function which is rather similar to the the charged BKP (cBKP)
  tau function (found in
  \cite{L1})
  which gives rise to
  the integral over symplectic matrices:

   \bp \em Let   $d\mu(x)$ be any measure on the
  circle. We have
  \be\label{PropCirc1}
\tau_{2N+l,l}(\bt,{\bar \bt}):=\l 2N+l|\,\g(\bt)\,e^{ \oint
\frac{d\psi(x)}{dx} \psi(x)d\mu(x) }{\bar \g}({\bar \bt})\,|l\r
  \ee
  \be\label{PropCirc2}
={c}\oint \cdots \oint \,\prod_{n<m\le N}|x_n-x_m|^4 \,
\prod_{i=1}^N \,  x_i^{l+2 (N-1)}e^{2\sum_{n=1}^\infty t_n
x_i^n+{\bar t}_n x_i^{-n}}d\mu(x_i)
  \ee
   where $c=c(N)\exp \sum_{m=1}^\infty mt_m{\bar t}_m$
  \ep
For the proof we use \eqref{psi-Vand} where we consider the limit
 $x_{2i-1}\to x_{2i}$.

Expectation values similar to \eqref{double-beta=1A} and
\eqref{double-beta=1B} yields sums of integrals similar to
\eqref{2-matrixCauchy-un} and \eqref{2-matrixCauchy-Her} where the
interaction
$\frac{|\Delta_N(z)\Delta_N(z')|}{\prod_{i,k=1}^N(z_i-z_k')}$ is
replaced by
$\frac{|\Delta_N(z)\Delta_N(z')|^4}{\prod_{i,k=1}^N(z_i-z_k')^4}$.
Since I do not yet know links of such series to anything else I
shall not write down them explicitly.

 \subsection{ Perturbation series in coupling constants for $\beta=1,4$
 ensembles and discrete matrix models \label{perturbation-section}}.

The fermionic language provides the simplest method to convert
integrals into sums based on the equalities
  \[
\int_\gamma\int_\gamma\, A(x,y)\psi(x)\psi(y)=\sum_{n>m}\,
A_{nm}\psi_n\psi_m,
  \]
  \[
 \int_\gamma\int_\gamma\,
A(x,y)\phi(x)\phi(y)=\sum_{n>m}\, A_{nm}\phi_n\phi_m,
  \]
\[
\int_\gamma \, a(x)\psi(x)\phi_0=\sum_{n}\, a_{n}\psi_n\phi_0
  \]
  where
  \be\label{x-n}
A_{nm}=\int_\gamma\int_\gamma \,A(x,y)x^ny^m\, dxdy,\quad
a_{n}=\int_\gamma \,a(x)x^n\,dx
  \ee
In this way each multiple integral considered in Section
\ref{partition-section} may be converted into a multiple sum
considered either in Section \ref{DKPtau-section} or in Section
\ref{fermionic-BKP-section}.

\paragraph{}

 In this subsection we shall consider perturbation series for $\beta=1,4$ ensembles and
 in particular for $\beta=1,4$ circular  ensembles.
Perturbation series for each ensemble
(\eqref{eigenvalue-int-cir},\eqref{eigenvalue-int-orth},\dots) may
be written as
   \be\label{perturb}
Z_N^{\beta=1}(\bt,{\bar \bt})= \, c\sum_{\lambda\in\Pa\atop
\ell(\lambda)\le N} \, A_{\{h\}}(0,{\bar \bt}) s_\lambda(\bt)
   \ee
   where the sum ranges over all partition whose length (that is the number of
   non-vanishing parts $\lambda_i$) does not exceed $N$  and where $A_{\{h\}}$ are
   defined via a matrix ${\tilde A}$ and numbers $\{ a_n\}$ as
   follows

Let $N=2K$. Given ordered set $h:=(h_1,\dots,h_N)$ we define
$A_{\{h\}}$ as the Pfaffian of $2K\times 2K$ matrix $a=a(h)$
 \be\label{Aeven-cir}
A_{\{h\}}=\Pf\, [a],\quad
a_{nm}(h):=(A)_{h_nh_m}|_{n,m=1,\dots,2K}
 \ee
 For $N=2K+1$ we define $A_{\{h\}}$ as the Pfaffian of the $(2K+2)\times (2K+2)$
 matrix $\tilde{a}=\tilde{a}(h)$
 \be\label{Aodd-cir} A_{\{h\}}=\Pf\, [\tilde{a}],\quad
\tilde{a}_{nm}(h):=(A)_{h_nh_m}|_{n,m=1,\dots,2K+1},
 \ee
 \be
\tilde{a}_{n,2K+2}=-\tilde{a}_{2K+2,n}=\oint x^{n}d\mu(x,\bt,{\bar
\bt}),\quad n=1.\dots,2K+2
  \ee
  The matrix ${\tilde A}$ is defined in term of matrices of
  moments of each of ensembles separately:

\paragraph{For $\beta=1$ ensembles}

From
\[
e^{\frac 12 \int_\gamma\int_\gamma\, \psi(x)
\psi(y)\sgn\left(\varsigma(x)-\varsigma(y)\right)d\mu(x)d\mu(y)
}=g^{--}(A)
 \]
 we obtain
 \be\label{Aeven}
A_{nm}=A_{nm}(\bt,{\bar \bt})=\frac 12 \int_\gamma\int_\gamma \,
x^{n}y^{m}\sgn\left(\arg(x)-\arg(y)\right)d\mu(x,\bt,{\bar
\bt})d\mu(y,\bt,{\bar \bt})
   \ee
   Also
   \be
a_n(\bt,{\bar \bt})=\int_\gamma\, x^n d\mu(x,\bt,{\bar \bt})
 \ee

 In particular for circular ensemble we have
  \be\label{Aeven'}
A_{nm}=A_{nm}(\bt,{\bar \bt})=\frac 12 \oint\oint
x^{n}y^{m}\sgn\left(\arg(x)-\arg(y)\right)e^{\sum_{i=1}^\infty
(x^{i}+y^{i})t_i-(x^{-i}+y^{-i}){\bar t}_i}d\mu(x)d\mu(y)
   \ee
   Also
   \be
a_n(\bt,{\bar \bt})=\oint \, x^n d\mu(x,\bt,{\bar \bt})
 \ee

 There are two ways to get such series.
 \paragraph*{(I)} Given fermionic representation \eqref{PropCirc1} we
 have the following straightforward way.
 Here we put $l=0$ for the sake of simplicity.
 Because
 \[
e^{\frac 12 \oint\oint \psi(x)
\psi(y)\sgn\left(\arg(x)-\arg(y)\right)d\mu(x)d\mu(y) }=g^{--}(A)
 \]
 where $A$ is the matrix of moments
 \[
A_{nm}=A_{nm}(\bt,{\bar \bt})=
 \]
\be\label{Aeven''}
 \frac 12 \oint\oint
x^{n}y^{m}\sgn\left(\arg(x)-\arg(y)\right)e^{\sum_{i=1}^\infty
(x^{i}+y^{i})t_i-(x^{-i}+y^{-i}){\bar t}_i}d\mu(x)d\mu(y)
   \ee
   we obtain representation \eqref{BKP-tau-Schur}
   \be\label{perturb'}
Z_N^{C\beta=1}(\bt,{\bar \bt})= \, \sum_{\lambda\in\Pa\atop
\ell(\lambda)\le N} \, A_{\{h\}}(0,{\bar \bt}) s_\lambda(\bt)=\,
\sum_{\lambda\in\Pa\atop \ell(\lambda)\le N} \, A_{\{h\}}(\bt,0)
s_\lambda({\bar \bt})
   \ee
   where the sum ranges over all partition whose length (that is the number of
   non-vanishing parts $\lambda_i$) does not exceed $N$  and where $A_{\{h\}}$ is
   defined as  follows as in \eqref{BKP-tau-Schur}:

\paragraph*{(II)} Formula \eqref{perturb} may be also obtained in a different
way: via development of $Z_N$ in Taylor series in deformation
parameters which is written as series in the Schur functions,
usage of the Cauchy-Littlewood relation \cite{Mac} in form
 \be
e^{\sum_{n=1}^\infty t_n \sum_{i=1}^N x_i^n} = \sum_{\lambda\in\Pa}
\, s_\lambda(\bt)\frac{\det
\left(x_i^{h_k}\right)_{i,k=1,\dots,N}}{\prod_{n<m\le
N}(x_n-x_m)}\, ,\quad h_i=\lambda_i-i+N
 \ee
 and the following known (see \cite{Mehta}, section 14.3)
 \bl \em
 \be
\int \cdots \int
\,\prod_{i=1}^N\,d\mu(x_i)\,\det\left[\theta_i(x_j)
\right]\,\sgn\,\Delta(x)=N! \Pf\left[
a_{ij}\right]_{i,j=1,\dots,2m}
 \ee
 where $2m=N $ if $N$ is even and $2m=N+1$ if $N$ is odd, and
  \be
a_{ij}=\int\int_{x\le y}\,
d\mu(x)d\mu(y)\left[\theta_i(x)\theta_j(y)-\theta_j(x)\theta_i(y)\right],\quad
i,j=1,\dots,N
  \ee
  When $N$ is odd we have in addition $a_{N+1,N+1}=0$ and
  \be
a_{i,N+1}=-a_{N+1,i}=\int \theta_i(x)d\mu(x),\quad
i=1,\dots,N
  \ee
 \el

\,

The perturbation series for the {\bf orthogonal ensemble}
\eqref{eigenvalue-int-orth} is the same series \eqref{perturb}
where, now, the moment matrix is
   \be\label{Aeven-orth}
A_{nm}(\bt,{\bar \bt})=\int_R\int_R\,
x^{n}y^{m}\sgn(x-y)d\mu(x,\bt,{\bar \bt})d\mu(y,\bt,{\bar \bt})
   \ee
   and
 \be
a_n(\bt,{\bar \bt})=\int_R\, x^n d\mu(x,\bt,{\bar \bt})
 \ee
   where
   \be
d\mu(x,\bt,{\bar\bt}):=\,e^{\sum_{m=1}^\infty\,x^{m}{
t}_m}e^{-\sum_{m=1}^\infty\,x^{-m}{\bar t}_m}d\mu(x)
  \ee

Let $N=2K$. Given ordered set $h:=(h_1,\dots,h_N)$ we define
$A_{\{h\}}$ as the Pfaffian of $2K\times 2K$ matrix $a=a(h)$
 \be
A_{\{h\}}=\Pf\, [a],\quad
a_{nm}(h):=(A)_{h_nh_m}|_{n,m=1,\dots,2K}
 \ee
 For $N=2K+1$ we define $A_{\{h\}}$ as the Pfaffian of the $(2K+2)\times (2K+2)$
 matrix $\tilde{a}=\tilde{a}(h)$
 \be A_{\{h\}}=\Pf\, [\tilde{a}],\quad
\tilde{a}_{nm}(h):=(A)_{h_nh_m}|_{n,m=1,\dots,2K+1},
 \ee
 \be\label{Aodd-orth}
\tilde{a}_{n,2K+2}=-\tilde{a}_{2K+2,n}=\int x^{n}d\mu(x),\quad
n=1.\dots,2K+2
  \ee

{\bf Perturbation series for $\beta=4$ ensemble.}

First, we write
 \be
 \int_{\gamma} \,
 \frac{\psi(x)}{dx}\psi(x)\,d\mu(\bt,{\bar\bt},x)\,=
 \,\sum_{n,m}\,A_{nm}(\bt,{\bar\bt})\psi_{n}\psi_m
  \ee
  where
  \be
 A_{nm}(\bt,{\bar\bt})=\frac{n-m}{2}\,\int_\gamma \, x^{n+m-1}\,
 \,d\mu(\bt,{\bar\bt},x)
 \ee
and
 \be
a_n(\bt,{\bar \bt})=\int_R\, x^n d\mu(x,\bt,{\bar \bt})
 \ee
where for the symplectic ensemble we take $\gamma=R$ while for the
circular ensemble $\gamma=S^1$.
  \be
d\mu(\bt,{\bar\bt},x):=\,e^{\sum_{m=1}^\infty\,x^{m}{
t}_m}e^{-\sum_{m=1}^\infty\,x^{-m}{\bar t}_m}d\mu(x)
  \ee

 Then we obtain
 \[
\l
N+l|\,\Gamma(\bt)\,e^{\sum_{n,m}\,A_{nm}\psi_{n}\psi_m}\,e^{\sum_{n=0}^\infty\,a_n
\psi_n\phi_0}\,{\bar\g}({\bar\bt})\,|l\r\,=
  \]
 \[
\,\sum_{\lambda\in\Pa}\,A_{\{h\}}(0,{\bar\bt})s_\lambda(\bt)
 \]

\subsection{On character formulae}

\br There is a known relation (see \cite{Kr}) between the Schur
functions and the odd orthogonal character $so_\lambda$ of
rectangular shape as follows
 \be
\sum_{\lambda_1 \le p}\,s_\lambda(x_1,\dots,x_m)=(x_1\dots
x_m)^{\frac 12 p}\,so_{\left(\left(\frac
p2\right)^m\right)}(x_1^{\pm 1},\dots,x_m,x_m^{\pm 1},1)
 \ee
 The odd orthogonal characters $so_\lambda(x^{\pm 1}_1 , x^{\pm 1}_2 , \dots , x^{\pm 1}_m , 1)$,
  where $x^{\pm 1}_1$ is a shorthand
notation for $x_1, x^{-1}_1 , \dots$, and where $\lambda$ is an
$m$-tuple $(\lambda_1, \lambda_2, \dots , \lambda_m)$ of integers,
or of half-integers, is defined by
 \be\label{char-so-odd}
so_{\lambda}(x_1^{\pm 1},\dots,x_m^{\pm 1},1)\,:=\,
\frac{\det\left(x_j^{\lambda_i-i+m+\frac
12}-x_j^{-(\lambda_i-i+m+\frac
12)}\right)}{\det\left(x_j^{-i+m+\frac 12}-x_j^{-(-i+m+\frac
12)}\right)}
 \ee
  (see, say, (3.3) in \cite{Kr}).

  The even orthogonal character may defined as
   \be
so_{\lambda}(x_1^{\pm 1},\dots,x_m^{\pm 1})\,:=\,
\frac{\det\left(x_j^{\lambda_i-i+m}+x_j^{-(\lambda_i-i+m)}\right)+
\det\left(x_j^{\lambda_i-i+m}-x_j^{-(\lambda_i-i+m)}\right) }{\det
\left(x_j^{-i+m}+x_j^{-(-i+m)}\right)}
 \ee
 (see, say, (2.12) in \cite{Kr98})

The even symplectic character is defined as
 \be\label{char-sp-even}
sp_{\lambda}(x_1^{\pm 1},\dots,x_m^{\pm 1})\,:=\,
\frac{\det\left(x_j^{\lambda_i-i+m+1}-x_j^{-(\lambda_i-i+m+1)}\right)}{\det
\left(x_j^{-i+m+1}-x_j^{-(-i+m+1)}\right)}
 \ee
 (see, say, (4.5) in \cite{Kr})

 In  \cite{Proctor} there was also defined odd symplectic characters
 $sp_\lambda(x^{\pm 1}_1 , x^{\pm 1}_2 , \dots , x^{\pm 1}_n , 1)$,
 which are for example defined by
 \be
sp_\lambda(x^{\pm 1}_1 , x^{\pm 1}_2 , \dots , x^{\pm 1}_n , 1) =
\frac 12 \det_{1\le i,j\le n } \left( h_{\lambda_i-i+j}(x^{\pm
1}_1 , x^{\pm 1}_2 , \dots , x^{\pm 1}_n , 1) +
h_{\lambda_i-i-j+2}(x^{\pm 1}_1 , x^{\pm 1}_2 , \dots , x^{\pm
1}_n , 1) \right) ,
 \ee
  (see, say, (4.6) in \cite{Kr})
where $h_k(z_1, z_2, \dots , z_r)$ denotes the $k$-th complete
homogeneous symmetric function

\er

\paragraph{The fermionic approach. Odd orthogonal character.} Formula \eqref{char-so-odd} may be
written in form
 \be
 so_{\lambda}(x_1^{\pm 1},\dots,x_m^{\pm
1},1)\,:=\,\l 0|\,\Psi(x_1)\dots \Psi(x_m) \,|\{\lambda\}\r
 \ee
 \be
=\, \left( -\sqrt{2x_1}\right)\dots \left( -\sqrt{2x_m}\right)\l
0|\,\left(g^{--} \right)^{-1}\,\psi^\dag(x_1)\dots
\psi^\dag(x_m)\,g^{--} \,|\{\lambda,m\}\r
 \ee
 where
 \be
g^{--}=e^{\frac 12\oint \,\psi(x)\psi\left( -x^{-1}
\right)\,dx }\, \equiv\,e^{\frac 12 \sum_{n\in\mathbb{Z}}
\,(-)^n\psi_n \psi_{n+1}}
 \ee
 \be \Psi(x):= \sqrt{2x}\psi\left(-x^{-1} \right)-
\sqrt{2x}\psi^\dag(x)
 \ee and where for
$\lambda=(\lambda_1,\dots,\lambda_r )$
 \be
|\{\lambda,m\}\rangle \,:=\,{\phi}_{\lambda_1-1+m}\dots
{\phi}_{\lambda_r-r+m}\,|0 \r
 \ee
 \be
 {\phi}_n\, :=\,\frac{1}{\sqrt{2}}\left( \psi_n +(-)^n \psi^\dag _{-n} \right)
 \ee
 \paragraph{The fermionic approach. Even symplectic case.}
 \be
 sp_{\lambda}(x_1^{\pm 1},\dots,x_m^{\pm
1})\,:=\,(-1)^m\l
0|\,\left(\psi^\dag(x_1)+\psi\left(-x^{-1}_1\right)\right)\dots
\left(\psi^\dag(x_m)+\psi\left(-x^{-1}_m\right)\right)
\,|[\lambda,m ]\r
 \ee

 Now take

 \be
 |[\lambda,m
]\r \,:=\,{\tilde{\psi}}_{\lambda_1-1+m}\dots
{\tilde{\psi}}_{\lambda_r-r+m}\,|0 \r
 \ee
 \be
 {\tilde{\psi}}_n\, :=\,\frac{1}{\sqrt{2}}\left( \psi_n -(-)^n \psi^\dag _{-1-n} \right),
 \qquad {\tilde{\psi}}^\dag_n\, :=
 \,\frac{1}{\sqrt{2}}\left(\psi^\dag _{n} - (-)^n\psi_{-n-1}   \right)
 \ee

We have
 \be
[{\tilde{\psi}}^\dag_n,{\tilde{\psi}}^\dag_k  ]_+=0\,\quad
[{\tilde{\psi}}_n,{\tilde{\psi}}_k  ]_+=0\,\quad
[{\tilde{\psi}}^\dag_n,{\tilde{\psi}}_k ]_+=\delta_{nk}\,\quad
 \ee

\be
 {\psi}_n\, :=\,\frac{1}{\sqrt{2}}\left( \tilde{\psi}_n +(-)^n \tilde{\psi}^\dag _{-1-n} \right),
 \qquad {\psi}^\dag_n\, :=
 \,\frac{1}{\sqrt{2}}\left(\tilde{\psi}^\dag _{n} + (-)^n\tilde{\psi}_{-n-1}   \right)
 \ee

\paragraph{Bogolyubov transform.}
 Introduce fermionic operators $\tilde{\psi}_n$ and $\tilde{\psi}_n^\dag$ as
 the following Bogolyubov transform of the fermionic operators $\psi_n$ and $\psi_n^\dag$
 \be\label{Bogolyubov-psi}
 {\tilde{\psi}}_n\, :=e^{-\frac {\pi }{4}\tilde{J}_0^-}\psi_n e^{\frac {\pi }{4}\tilde{J}_0^-}
 =\,\frac{1}{\sqrt{2}}\left( \psi_n -(-)^n \psi^\dag _{-1-n} \right)
 \ee
 \be\label{Bogolyubov-psi-dag}
 {\tilde{\psi}}^\dag_n\, :=e^{-\frac {\pi }{4}\tilde{J}_0^-}
 \psi_n^\dag e^{\frac {\pi }{4}\tilde{J}_0^-}=
 \,\frac{1}{\sqrt{2}}\left(\psi^\dag _{n} - (-)^n\psi_{-n-1}   \right)
 \ee
 where
 \be\label{canonic-Bogolyubov}
\tilde{J}_0^-:=\,\sum_{n\in\mathbb{Z}}\,(-1)^n\left(\psi_{-n-1}\psi_n
+ \psi_{-n-1}^\dag\psi_{n}^\dag \right)
 \ee
The inverse transformation is
 \be
 {\psi}_n\, :=\,\frac{1}{\sqrt{2}}\left( \tilde{\psi}_n +(-)^n \tilde{\psi}^\dag _{-1-n} \right),
 \qquad {\psi}^\dag_n\, :=
 \,\frac{1}{\sqrt{2}}\left(\tilde{\psi}^\dag _{n} + (-)^n\tilde{\psi}_{-n-1}   \right)
 \ee

 Introduce fermionic fields
 \be\label{varphi-field}
{\tilde{\psi}}(x):=\,\sum_{} {\tilde{\psi}}_n x^n
=\frac{1}{\sqrt{2}}\left(\psi(x)-\psi^\dag(-x) \right)\,,\quad
{\tilde{\psi}}^\dag(x):=\,\sum_{} {\tilde{\psi}}_n^\dag x^{-n-1}
=\frac{1}{\sqrt{2}}\left(\psi^\dag(x)+\psi(-x) \right)
 \ee
 then
 \be
{\psi}(x):=\,\frac{1}{\sqrt{2}}\left(\tilde{\psi}(x)+\tilde{\psi}^\dag(-x)
\right)\,,\quad
{\psi}^\dag(x):=\,\frac{1}{\sqrt{2}}\left(\tilde{\psi}^\dag(x)-\tilde{\psi}(-x)
\right)
 \ee

By \eqref{Bogolyubov-psi}-\eqref{Bogolyubov-psi-dag} we obtain
 \be\label{varphi-vac}
\tilde{\psi}_n|m\r=\tilde{\psi}_{-n-1}^\dag|m\r=\psi_n|m\r=\psi_{-n-1}^\dag|m\r\,=0\,,\quad
n<m
 \ee

We obviously have
 \be
[{\tilde{\psi}}^\dag_n,{\tilde{\psi}}^\dag_k  ]_+=0\,,\quad
[{\tilde{\psi}}_n,{\tilde{\psi}}_k  ]_+=0\,,\quad
[{\tilde{\psi}}^\dag_n,{\tilde{\psi}}_k ]_+=\delta_{nk}
 \ee
and
 \be [{\tilde{\psi}}^(x),{\tilde{\psi}}^\dag(y)  ]_+=0\,,\quad
[{\tilde{\psi}}(x),{\tilde{\psi}}(y)  ]_+=0\,,\quad
[{\tilde{\psi}}^\dag(x),{\tilde{\psi}}(y)
]_+=[{\psi}^\dag(x),{\psi}(y) ]_+=\delta(x/y)\,\quad
 \ee

As one can verify, for any $N,n.m$
  \be
\l N|\, \tilde{\psi}_n\tilde{\psi}_m^\dag \,|N\r\,=\l N|\,
\psi_n\psi_m^\dag \,|N\r\,,\quad \l N|\,
\tilde{\psi}_n\tilde{\psi}_m \,|N\r\,=\l N|\,
\tilde{\psi}_n^\dag\tilde{\psi}_m^\dag \,|N\r\,=0
  \ee

  As one can see the normal ordering given by
  \be\label{ordering}
:\psi_n\psi_m^\dag: \, = \psi_n\psi_m^\dag \, -\,\l 0|
\psi_n\psi_m^\dag\,|0\r\,,\quad :\psi_n\psi_m: \, = \psi_n\psi_m
\,,\quad :\psi_n^\dag\psi^\dag_m: \, = \psi^\dag_n\psi^\dag_m
  \ee
  yields
   \be\label{tilde-ordering}
:\tilde{\psi}_n\tilde{\psi}_m^\dag: \, =
\tilde{\psi}_n\tilde{\psi}_m^\dag \, -\,\l 0|
\tilde{\psi}_n\tilde{\psi}_m^\dag\,|0\r\,,\quad
:\tilde{\psi}_n\tilde{\psi}_m: \, = \tilde{\psi}_n\tilde{\psi}_m
\,,\quad :\tilde{\psi}_n^\dag\tilde{\psi}^\dag_m: \, =
\tilde{\psi}^\dag_n\tilde{\psi}^\dag_m
  \ee

 Direct computation with the use of \eqref{varphi-field}, \eqref{varphi-vac} yields
 \be\label{Green}
\l N|\, \tilde{\psi}(x)\tilde{\psi}^\dag(y)
\,|N\r=\sum_{n=0}^\infty\,x^{-n-N-1}y^{n+N}= \l N|\,
\psi(x)\psi^\dag(y) \,|N\r
 \ee

\paragraph{Currents and flows.}
 \be\label{tilde-currents}
:\psi(x)\psi^\dag(x):=\sum_{n=-\infty}^{+\infty}\,J_n
x^{n-1}\,,\quad
:\tilde{\psi}(x)\tilde{\psi}^\dag(x):=\sum_{n=-\infty}^{+\infty}\,\tilde{J}_n
x^{n-1}
 \ee
By \eqref{tilde-ordering} and \eqref{ordering} both sets $\{ J_n
\}$ and $\{ \tilde{J}_n \}$ are the subjects of two Heisenberg
algebras
 \be\label{tilde-Heisenberg}
[\,J_n,J_m\,]=[\,\tilde{J}_n,\tilde{J}_m\,]=n\delta_{n+m,0}
 \ee

By \eqref{Bogolyubov-psi}-\eqref{Bogolyubov-psi-dag} and taking into account that $\sum
(-1)^i\psi_i\psi_{n-i}=\sum (-1)^i\psi_i^\dag\psi_{n-i}^\dag=0$
for even $n$ we obtain
 \be\label{odd-tilde-currents}
J_{2n-1}=\tilde{J}_{2n-1}
 \ee
  \be
{J}_{2n}=\frac 12 \sum_{i=-\infty}^{+\infty}\,(-1)^i
\left(\tilde{\psi}_i\tilde{\psi}_{-i-1-2n}+
\tilde{\psi}_{-i-1}^\dag \tilde{\psi}_{i+2n}^\dag \right)
 \ee
 One can compare the last expression with formulae \eqref{I^pm}.

Let us introduce
 \be
{J}_{2n}^{-}=\frac 12 \sum_{i=-\infty}^{+\infty}\,(-1)^i
\left(\tilde{\psi}_i\tilde{\psi}_{-i-1-2n}-
\tilde{\psi}_{-i-1}^\dag \tilde{\psi}_{i+2n}^\dag \right)
 \ee
 and use the notation ${J}_{2n}^{+}:={J}_{2n}$.
 Then
 \be\label{odd-tilde-commutation}
[\,\tilde{J}_{2n+1},{J}_{2m-1}\,]=(2n+1)\delta_{n+m,0}\,,\quad
[\,\tilde{J}_{2n+1},{J}_{2m}\,]=0
 \ee
  \be
[{J}_{2m}^\pm\,,\tilde{J}_{2n}]=\tilde{J}_{2n+2m}^\mp
  \ee

Similarly
  \be
\tilde{J}_{2n}^{+}\,:=\tilde{J}_{2n}\,= \,-\frac 12
\sum_{i=-\infty}^{+\infty}\,(-1)^i \left({\psi}_i{\psi}_{-i-1-2n}+
{\psi}_{-i-1}^\dag {\psi}_{i+2n}^\dag \right)
 \ee
  \be
\tilde{J}_{2n}^{-}\,:= \,-\frac 12
\sum_{i=-\infty}^{+\infty}\,(-1)^i \left({\psi}_i{\psi}_{-i-1-2n}-
{\psi}_{-i-1}^\dag {\psi}_{i+2n}^\dag \right)
 \ee
And we get
 \be
[\tilde{J}_{2m}^\pm\,,{J}_{2n}]={J}_{2n+2m}^\mp
  \ee

 \br

\bl

\el

One can consider the transformation
 \be\label{Bogolyubov-a-b}
 {\tilde{\psi}}_n\, :=\,c\left( \psi_n + a(-1)^n \psi^\dag _{-1-n} \right),
 \qquad {\tilde{\psi}}^\dag_n\, :=
 \,c\left(\psi^\dag _{n} +  a(-1)^n\psi_{-n-1}   \right)
 \ee
 \be\label{Bogolyubov-a-b'}
 {{\psi}}_n\, :=\,c\left( \tilde{\psi}_n - a(-1)^n \tilde{\psi}^\dag _{-1-n} \right),
 \qquad {{\psi}}^\dag_n\, :=
 \,c\left(\tilde{\psi}^\dag _{n} - a (-1)^n\tilde{\psi}_{-n-1}   \right)
 \ee
where $c=\frac{1}{\sqrt{1+a^2}}$. Instead of \eqref{varphi-field}
we have
 \[
{\tilde{\psi}}(x):=\,c\left(\psi(x)+a \psi^\dag(-x)
\right)\,,\quad {\tilde{\psi}}^\dag(x):=\,c\left(\psi^\dag(x)-a
\psi(-x) \right)
 \]
 \[
{\psi}(x):=\,c\left(\tilde{\psi}(x)-a\tilde{\psi}^\dag(-x)
\right)\,,\quad
{\psi}^\dag(x):=\,c\left(\tilde{\psi}^\dag(x)+a\tilde{\psi}(-x)
\right)
 \]
The previous case was related to the choice $a=-1$.

Then all properties \eqref{varphi-vac}-\eqref{Green} are still
true. Then again the currents \eqref{tilde-currents} satisfy the
Heisenberg algebra relations \eqref{tilde-Heisenberg}. However,
the explicit relations for currents are as follows. The relation
for odd components of the currents \eqref{odd-tilde-currents} are
still correct (therefore relations \eqref{odd-tilde-commutation}
are also correct), while for the even components we obtain
  \be
{J}_{2n}^+=\frac{1-a^2}{1+a^2}\,\tilde{J}_{2n}\,-\,\frac{a}{1+a^2}
\sum_{i=-\infty}^{+\infty}\,(-1)^i \left( \tilde{\psi}_{-i-1}^\dag
\tilde{\psi}_{i+2n}^\dag + \tilde{\psi}_i\tilde{\psi}_{-i-1-2n}
\right)\,,\quad n\in\mathbb{Z}
 \ee
  \be
\tilde{J}_{2n}^+(a)
=\frac{1-a^2}{1+a^2}\,{J}_{2n}\,+\,\frac{a}{1+a^2}
\sum_{i=-\infty}^{+\infty}\,(-1)^i \left({\psi}_{-i-1}^\dag
{\psi}_{i+2n}^\dag + {\psi}_i {\psi}_{-i-1-2n}\right)\,,\quad
n\in\mathbb{Z}
 \ee
   \be
{J}_{2n}^-\,=\frac 12
[{J}_{2n}^+,\tilde{J}_{0}^+]=-\frac{a}{1+a^2}
\sum_{i=-\infty}^{+\infty}\,(-1)^i \left( \tilde{\psi}_{-i-1}^\dag
\tilde{\psi}_{i+2n}^\dag - \tilde{\psi}_i\tilde{\psi}_{-i-1-2n}
\right)\,,\quad n\in\mathbb{Z}
 \ee
  \be
\tilde{J}_{2n}^- \,=\frac 12 [\tilde{J}_{2n}^+,J_0^+]
=\frac{a}{1+a^2} \sum_{i=-\infty}^{+\infty}\,(-1)^i
\left({\psi}_{-i-1}^\dag {\psi}_{i+2n}^\dag - {\psi}_i
{\psi}_{-i-1-2n}\right)\,,\quad n\in\mathbb{Z}
 \ee

 We notice that
  \be
\tilde{\psi}_n= e^{\alpha
\tilde{J}_0^-}\,{\psi}_n\,e^{-\alpha\tilde{J_0}^-}\,,\qquad
\tilde{\psi}^\dag_n=
e^{\alpha\tilde{J_0}^-}\,{\psi}^\dag_n\,e^{-\alpha\tilde{J_0}^-}\,,\quad
\tan\alpha = a
  \ee
and therefore for $n\,\neq \,0$
 \be
\tilde{J}_n^+=
e^{\alpha\tilde{J_0}^-}\,{J}_n^+\,e^{-\alpha\tilde{J_0}^-}
 \ee
 It is not true for $n=0$ thanks to the fact that the definition of $J_0$ includes the
 normal ordering.

 \er

 \paragraph{Symmetric polynomials.}
Let us notice that
 \be
 \l N_1|\,{\Gamma}(t)\,{g}\,|N_2 \r=\l N_1|\,\tilde{\Gamma}(t)\,\tilde{g}\,|N_2 \r
 \ee
 where
  \be
\tilde{\Gamma}(t)\,:=\, e^{\sum_{n=1}^\infty \tilde{J}_n t_n}
  \ee
  and $\tilde{g}$ is of form \eqref{DKPg}  where fermions $\psi$ and $\psi^\dag$ are
  replaced by the fermions $\tilde{\psi}$ and $\tilde{\psi}^\dag$.
As we see
 \[ {s}_\lambda(t):=\,\l 0|\,e^{\sum_{n=1}^\infty \tilde{J}_n(a)
t_n}\,| \{\lambda,a\} \r
 \]
Consider
 \be
\tilde{s}_\lambda(t,a):= \,\l 0|\,e^{\sum_{n=1}^\infty {J}_n
t_n}\,e^{\arctan(a) \tilde{J}_0^{-}}\,{|}\lambda \r=\,\l
0|\,e^{\sum_{n=1}^\infty \tilde{J}_n(-a) t_n}\,{|}\lambda \r =\,\l
0|\,e^{\sum_{n=1}^\infty \,{J}_n t_n}\,{|}\{\lambda,a\} \r
 \ee
 This polynomial is an example of BKP tau function. $\tilde{s}_\lambda(t,a)$ is a
 homogeneous polynomial of the same weight as ${s}_\lambda(t)$.
 We have $\tilde{s}_\lambda(t,0)=s_\lambda(t)$.

 The polynomial $\tilde{s}_\lambda(t)$ may be expressed as a
 linear combination of the Schur functions as follows. First, we
 need the Frobenius notation for $\lambda$. Let $\lambda
 =(\alpha|\beta)$. Given $\lambda$ one can construct another
 partition by a permutation of a certain number, say, $n$ of
 variables $\alpha_{n_i}, \, i=1,\dots,k$ and $\beta_{m_i}, \,
 i=1,\dots,k$. The set of all partitions obtained in this way will be denoted by
 $P_k(\alpha,\beta)$

  \be
\tilde{s}_\lambda(t,a)=\sum_{\mu\in
P_k(\alpha,\beta)}\,(-1)^{}a^{2k} s_\mu(t)
  \ee

  In case $t_2=t_4=t_6=\cdots =0$ all Schur functions in the sum
  are equal.

\subsection{Double series in the Schur functions}

\subsection{Matrix integrals}

Using
 \be\label{} \int_{O\in \mathbb{O}(N)} s_\lambda(O) d_*O =
\begin{cases}
1 \quad \lambda\,\mbox{is\,even}  \\
0 \quad\, \mbox{otherwise}
 \end{cases}\,,\qquad
\int_{S\in \mathbb{S}p(N)} s_\lambda(S) d_*S =
\begin{cases}
1 \quad \lambda^{tr}\,\mbox{is\,even}  \\
0 \quad\, \mbox{otherwise}
 \end{cases}
  \ee
we get
 \be
J_1(\bt,N):=\int_{O\in \mathbb{O}(N)} e^{\sum_{m=1}^\infty t_m\Tr O^m} d_*O = \sum_{\lambda\,\mbox{even}\atop \ell(\lambda)\le N} s_\lambda(\bt)
 \ee
 \be
J_2(\bt,N):=\int_{S\in \mathbb{S}p(2n)} e^{\sum_{m=1}^\infty t_m\Tr S^m} d_*S = \sum_{\lambda \in \FP \atop \ell(\lambda)\le 2n} s_\lambda(\bt)
 \ee

The right hand sides were obtained in \cite{OST-I} (see eq. (38) and for $N\to\infty$ see eqs.(83),(86) there) 
as examples of the BKP tau function.

There are three different fermionic representations of these integrals.

(1) There are representations for $S^{(1)}_4$ see eq.(162) in \cite{OST-I}

(2) The fermionic representation for $N=2n$ is
 \be
J_\alpha(\bt,2n)=b(\bt)\l 2n|\g(\bt) e^{\int_{\Lambda_2} A(z_1,z_2)
 w_\alpha(z_1) w_\alpha(z_2) dz_1dz_2}{\bar{\g}}({-\bt}) |0\r
 \ee
where for $w^{(\alpha)}$ see \eqref{w12} below. and where
 \be\label{A^{(alpha)}}
A^{(\alpha)}(z_1,z_2)=\frac{\left((z_1+z_1^{-1})^n-(z_2+z_2^{-1})^n \right)^2}{z_1+z_1^{-1} -z_2-z_2^{-1}}
 \ee
Here
 \be
 b(\bt):=e^{\sum_{m=1}^\infty m{ t}_m^2}
 \ee
To get the fermionic representation we use that for \eqref{A^{(alpha)}} 
$\Pf\left[ A(z_i,z_j) \right]=\prod_{i<j}^n\left(z_i+z_i^{-1}-z_j-z_j^{-1}\right)^2$, see Lemma 5.7 in\cite{Ishikawa}.

(3) The other way is to present the both matrix integrals as the following tau function of the two-component 2-KP

 \be
\l 2n,-2n|\g^{(1)}(\bt^{(1)})\g^{(2)}(\bt^{(1)}) 
e^{\oint  w_\alpha^{(1)}(z)w_\alpha^{\dag (1)}(z) }{\bar{\g}}({-\bt}){\bar{\g}}({-\bt}) |0\r
 \ee
for $w_\alpha$ see \eqref{w12} below, superscript shows the numerous of the component of two-component fermions.

\paragraph{Character expansion: KP tau function}
\be
\chi_\lambda(z,n):=\frac{\l n+N,\lambda|w(z_1)\cdots w(z_N)|n\r}{\l n+N|w(z_1)\cdots w(z_N)|n\r} \,=\,
\frac{\det\left[\l n|\psi_{n+h_i}^\dag w(z_j)|n\r \right]}{\det\left[\l n|\psi_{n+N-i}^\dag w(z_j)|n\r \right]}\,,
\quad h_i=\lambda_i-i+N
 \ee
where $w(z)$ is a linear combination of the Fermi fields.
Denote
 \be
\chi_\lambda^{(\alpha)}(z):=
\frac{\l N,\lambda|w_\alpha(z_1)\cdots w_\alpha(z_N)|0\r }{\l N|w_\alpha(z_1)\cdots w_\alpha(z_N)|0\r}
 \ee
where
 \be
w^{\alpha}(z)= z^{-\frac{\alpha}{2}}\left( Q_\alpha(z)\right)^{-1} \psi(z)  Q_\alpha(z)
\,,\quad Q_\alpha=\exp \sum_{n\in\mathbb{Z}} \psi_{-n}\psi^\dag_{n+\alpha}=
\exp\frac{1}{2\pi i}\oint z^\alpha \psi(z^{-1})\psi^\dag(z)dz
 \ee
namely
 \be\label{w12}
w_1(z)=\psi(z)z^{-\frac 12}-\psi(z^{-1})z^{\frac 12}\,,\quad 
w_2(z)=\psi(z)z^{-1}-\psi(z^{-1})z
 \ee
Then the odd character of the orthogonal group is
 \be\label{char-so-odd}
so_{\lambda}(z_1^{\pm 1},\dots,z_m^{\pm 1},1)\,=\,
\frac{\det\left(z_j^{\lambda_i-i+m+\frac
12}-z_j^{-(\lambda_i-i+m+\frac
12)}\right)}{\det\left(z_j^{-i+m+\frac 12}-z_j^{-(-i+m+\frac
12)}\right)}=\chi_\lambda^{(1)}(z)
 \ee
  (see, say, (3.3) in \cite{Kr}).

The even symplectic character is defined as
 \be\label{char-sp-even}
sp_{\lambda}(z_1^{\pm 1},\dots,z_m^{\pm 1})\,:=\,
\frac{\det\left(z_j^{\lambda_i-i+m+1}-z_j^{-(\lambda_i-i+m+1)}\right)}{\det
\left(z_j^{-i+m+1}-z_j^{-(-i+m+1)}\right)}=\chi_\lambda^{(2)}(z)
 \ee
 (see, say, (4.5) in \cite{Kr})

Now, introduce the following KP tau function
 \be
\tau^{(\alpha)}(\bt)=\l n+N|\g(\bt) w_\alpha(z_1)\cdots w_\alpha(z_N)|n\r \,=
\,\sum_\lambda \chi_\lambda^{(\alpha)}(z)s_\lambda(\bt)
 \ee
where the characters play the role of Plucker coordinates in the Sato formula for KP tau functions.

\paragraph{BKP tau functions related to $\mathbb{O}(N)$ and $\mathbb{S}p(N)$ characters}

 \be
\l 2n|\g(\bt) e^{\int_{\Lambda_2} A(z_1,z_2) w_\alpha(z_1)w_\alpha(z_2) 
dz_1dz_2}{\bar{\g}}({\bar\bt}) |0\r = 
 \ee
 \[
=b({\bar \bt})\sum_\lambda s_\lambda(\bt +{\bar\bt})
\int_{\Lambda_N} A({\bar z})\,\chi^{(\alpha)}_\lambda(z)\prod_{i=1}^N
e^{-\sum_{m=1}^\infty {\bar t}_m(z_i^m+z_i^{-m})} dz_i  
 \]
In case $\bt +{\bar\bt}=0$ and ${\bar A}$ chosen as in \eqref{A^{(alpha)}} we come to $J_\alpha(N,\bt)$.

\paragraph{Littlwood-Hall }

\cite{Vidya}

\section{Appendic. Mehta-Pandey integration trick.}

This is a short review of some facts from Chapters 2 and 14 of \cite{Mehta}.

 Consider a Hermitian matrix $H=R+iS$ where 
$R$ is real symmetric and $S$ is real anti-symmetric. Let the matrix $H$ and an auxilary matrix $X$ are $N$ by $N$ matrices.
 Below we shall consider four cases where $X$ is (1) symmetric, (2)
 anti-symmetric, (3) quaternionic self-dual and (4) quaternionic anti-self-dual.

The following simple relation will be of use below 
 \be\label{Gauss-trick}
\int e^{c Y_{jk}X_{kj} + t X_{kj}^2 } dX_{kj}= \frac{\pi}{\sqrt{t}} e^{-\frac{c^2}{4t}Y^2_{jk}}
 \ee
where $Y_{jk}$, $X_{kj}$ $c$ and $t$ are real.

\paragraph{Interpolation between Gauss unitary (GUE) and Gauss orthogonal ensembles GOE.} The integration measure for
the unitary ensemble of $N$ by $N$ matrices is given by
 \be\label{Haar-Hermitian}
  dH=\prod_{j\le k} d{ R}_{jk}\prod_{j < k} d{ S}_{jk}
 \ee
The so-called Mehta-Pandey $GUE-GOE$ interpolating ensemble is defined by the following Gauss probability measure
 \be\label{Mehta-Pandey-GUE-GOE}
  P(H)=const \exp \left[-\sum_{j,k}\left(\frac{\left({ R}_{jk}\right)^2}{4v^2} +
\frac{\left({ S}_{jk}\right)^2}{4v^2\alpha^2}\right)\right]
 \ee
In case $\alpha^2=1$ \eqref{Mehta-Pandey-GUE-GOE} yields GUE. In the limit $\alpha^2 \to 0$ Gauss fluctuations of 
the anti-symmetric part $S$ of the Hermtian matrix $H=R+iS$ is suppressed and statistics coincides with the statistic of
Gauss symmetric matrices. Therefore, the interval $0<\alpha^2<1$ is related to the GOE-GUE interpolation between Gauss
ensembles of Hermitian matrices and of symmetric ones, $alpha^2$ being 
the interpolation parameter. Similarly, the interval $1<\alpha^2<+\infty$ may be related to the GOE-GUE interpolation
between Gauss ensembles of Hermitian and of anti-symmetric matrices.  

(1) If $X$ is a symmetric matrix. For a given $i,j$ we have
 \[
\int e^{c R_{jk}X_{kj} + t X_{kj}^2 } dX_{kj}= \frac{\pi}{\sqrt{t}} e^{-\frac{c^2}{4t}R^2_{jk}}
 \]
and thanks to $\Tr HX =\Tr RX$ we obtain
 \be\label{MehtaPandey-symm}
\int e^{c\Tr HX +t\Tr X^2} dX = \left( \frac{\pi}{\sqrt{t}} \right)^{\frac{N^2+N}{2}} e^{-\frac{c^2}{4t}\Tr R^2}
 \ee

With the help of this relation the one can write the mean value of any function $f$ of entries of $H$ in the Mehta-Pandey
ensemble (MP1) as follows
 \be
\l f(H) \r_{MP1}:=\int f(H)
\exp \left[-\sum_{j,k}\left(\frac{\left({ R}_{jk}\right)^2}{4v^2} +
\frac{\left({ S}_{jk}\right)^2}{4v^2\alpha^2}\right)\right] \prod_{j\le k} d{ R}_{jk}\prod_{j < k} d{ S}_{jk}
 \ee
 \be
=\int f(\{H_{jk}\})\int e^{t_2\Tr H^2+ c\Tr HX +t_2'\Tr X^2} dH dX
 \ee
where $t_2=-\frac{1}{4v^2\alpha^2}$, $t_2'=-\frac{1}{4v^2(1-\alpha^2)}$, $c=-\frac{1}{2v^2\alpha^2}$.

(2) If $X$ is an anti-symmetric matrix then then $\Tr HX =\Tr SX$. Now we obtain in a similar way
 \be\label{MehtaPandey-anti-symm}
\int e^{-ic\Tr HX +t\Tr X^2} dX = \left( \frac{\pi}{\sqrt{t}} \right)^{\frac{N^2-N}{2}} e^{-\Tr S^2}
 \ee

\paragraph{Interpolation between Gauss unitary (GUE) and Gauss symplectic ensembles GSE.}

Here we use the standard notions of quaternions $e_i$, $i=0,1,2,3$ which may be viewed as two by two matrices, 
$e_0$ is a unity matrix and $e_i$, $i=1,2,3$ are anti-Hermitian Pauli matrices, $e_1$ is diagonal imaginary, 
$e_2$ is real anti-symmetric, $e_3$ is symmetric imaginary.

A $2n$ by $2n$ Hermitian matrix $H$ may be written as $n$ by $n$ matrix with quaternionic entries as follows
 \[
 {\bf H}_{jk}=\left[ {\bf R}_{jk} +i{\bf S}_{jk} \right] 
 \]
where $2n$ by $2n$ matrices
 \[
  {\bf R}={ R}^0 \cdot e_0 + \sum_{\mu=1}^3  { R}^\mu\cdot e_\mu \,,\qquad 
{ S}={ S}^0 \cdot e_0 + \sum_{\mu=1}^3  { S}^\mu\cdot e_\mu
 \]
are written via $n$ by $n$ matrices ${ R}^\mu$ and ${ S}^\mu$, $\mu=0,1,2,3$, where
${ R}^0$ and ${ S}^\mu$ are real symmetric while ${ S}^0$ and ${ R}^\mu$ are real anti-symmetric.

The operation $e_0\to e_0,\,e_i\to -e_i,\, i=1,2,3$ is called conjugation. 
A matrix, say ${\bf R}$, with quaternionic entries is called self-dual if ${\bf R}_{jk}$ is conjugated to  
${\bf R}_{kj}$. A matrix, say ${\bf S}$, with quaternionic entries is called anti-self-dual if ${\bf S}_{jk}$ is conjugated to  
$-{\bf S}_{kj}$.

In terms of matrices ${ R}^\mu$ and ${ S}^\mu$ the integration measure \eqref{Haar-Hermitian} of the Hermitian matrix $H$ 
may be written as
 \[
  dH=
\prod_{j\le k \le n} \left(d{ R}^0_{jk}\prod_{\mu=1}^3   d{ S}^\mu_{jk} \right)
\prod_{j < k \le n} \left(d{ S}^0_{jk}\prod_{\mu=1}^3d{ R}^\mu_{jk} \right)
 \]
The Mehta-Pandey enterpolating ensemble  is defined by the following probability measure
 \be
  P(H)=\exp \left[-\sum_{j,k} \sum_{\mu=0}^3\left(\frac{\left({ R}^\mu_{jk}\right)^2}{4v^2} +
\frac{\left({ S}^\mu_{jk}\right)^2}{4v^2\alpha^2}\right)\right]
 \ee

In case $\alpha^2=1$ \eqref{Mehta-Pandey-GUE-GOE} yields GSE: the Gauss ensemble of self-dual matrices. 
In the limit $\alpha^2 \to 0$ Gauss fluctuations of 
the anti-self-dual part $S$ of the Hermtian matrix $H=R+iS$ is suppressed and statistics coincides with the statistic of
Gauss self-dual matrices. Therefore, the interval $0<\alpha^2<1$ is related to the GSE-GUE interpolation between Gauss
ensembles of Hermitian matrices and of self-dual ones, $alpha^2$ being 
the interpolation parameter. Similarly, the interval $1<\alpha^2<+\infty$ may be related to the GSE-GUE interpolation
between Gauss ensembles of Hermitian and of anti-self-dual matrices.

(3) If $N=2n$ is even and $X$ is real quaternionic self-dual $n$ by $n$ matrix`

\section{Appendix. Some properties of vector $|\Omega\rangle$ \label{some-section}}

{\bf On ASEP and 1D Ising model}

chto eto?
$\,\prod_{i<j}(\cos \theta_i$

Introduce
 \be
n^{(\pm)}_\lambda:=\langle\lambda|\,J_{\pm 1}\,|\Omega\rangle
 \ee
 One can interprets  $n^{(+)}_\lambda$ and $n^{(-)}_\lambda$ as the numbers of respectively in- and
 out-incurvities of a Young diagram $\lambda$.
 Then
 \be\label{n+=n-+1}
n^{(+)}_\lambda =n^{(-)}_\lambda +1
 \ee
 In terms of Fock vector interpretation of the state $\lambda$ the
 number $n^{(+)}_\lambda$ ($n^{(-)}_\lambda$) counts the number of fermions which can hop
 up (resp. down) to a free neighboring site. The appearance of $1$ in the
 right-hand side of \eqref{n+=n-+1} is related to the fact that
 the bottom of Dirac see is fully packed.

Now introduce
 \be
V=\frac 12\,\sum_{i\in\mathbb{Z}}\, \,\left((2\hat{n}_i-
1)(2\hat{n}_{i+1}-1)-1\right), \qquad \hat{n}_i=\psi_i\psi^\dag_i
 \ee
Then obviously
  \be
\langle\lambda| \, V =n_\lambda \,\langle\lambda|
  \ee
where
  \be
 n_\lambda =
n^{(+)}_\lambda + n^{(-)}_\lambda = 2n^{(+)}_\lambda -1  =
2n^{(-)}_\lambda +1
  \ee
  is the number of free neighboring  sites for fermions in the state
  $\langle \lambda|$. \footnote{Notice that
   \be
\mathbb{Z}_{I}:=\langle\Omega_0|\,e^{V}\,|\Omega_0\rangle \equiv
\sum_\lambda \, e^{\beta {n_\lambda}}
   \ee
   may be interpreted as the partition function of the 1D Ising
   model, a parameter $\beta$ being the inverse temperature.}

   Introduce
   \be
|\Omega_a\rangle \, := \, \sum_{\lambda\in\Pa} \, |\lambda\rangle
\, e^{-a|\lambda|}
   \ee
Then
 \be
\langle\lambda|\,J_{\pm 1}\,|\Omega_a\rangle = e^{\pm
a}n^{(\pm)}_\lambda
 \ee
 or, the same,
\be J_{\pm 1}\,|\Omega_a\rangle  = \, \sum_{\lambda\in\Pa} \,
|\lambda\rangle \, e^{-a|\lambda|} e^{\pm a}n^{(\pm)}_\lambda
 \ee

 \be
\left(rJ_{- 1}+r^{-1}J_{ 1}-pV\right)\,|\Omega_a\rangle  = \,
\sum_{\lambda\in\Pa} \, |\lambda\rangle \, e^{-a|\lambda|}
\left(\frac 12 re^{- a}(n_\lambda -1) + \frac 12 r^{-1}e^{
a}(n_\lambda +1) - p n_\lambda\right)
 \ee
 when $p=\frac 12 re^{- a} + \frac 12 r^{-1}e^{
a}$ then
 \be
\hat{\mathbb{H}} \,|\Omega_a\rangle =\,0  ,\quad
\hat{\mathbb{H}}:= \left(rJ_{- 1}+r^{-1}J_{ 1}-\left(\frac 12
re^{- a} + \frac 12 r^{-1}e^{ a}\right)V + \left( \frac 12
re^{-a}-\frac 12 r^{-1}e^{ a}\right)\right)
 \ee
 At last we obtain
  \be\label{prob=1}
\langle\lambda |e^{\hat{\mathbb{H}}t}|\Omega_a\rangle \equiv 1
  \ee

   Via Jordan-Wigner transform $\hat{\mathbb{H}}$ may be
   identified with the Hamiltonian of a non-Hermitian spin chain
$\hat{\mathbb{H}} \to \sum_{} \, rS^-_i S^+_{i+1} +r^{-1}S^+_i
S^-_{i+1} + p S^z_i S^z_{i+1}$, which is the Hamiltonian of the
well-known XXX model if $r=1,\, p=\frac 14$. This representation
was used in \cite{GS1,GS2} for a description of the asymmetric
simple exclusion model (ASEP). Then relation \eqref{prob=1}
describes the fact that the sum of the probabilities to achieve
each of admissible states is equal to one. Then
 \[
p_{\lambda\to\mu}(t)=\langle
\lambda|e^{\hat{\mathbb{H}}t}|\mu\rangle
 \]
is the probability to achieve the state $\mu$ starting from the
state $\lambda$ after the lapse of the time $t$.

{\bf Determinants of infinite matrices}
 \be
\langle \Omega|e^{\sum_{n,m\ge 0}\,
\psi_n\psi^\dag_{-m-1}D_{nm}}|0\rangle=\det(\mathrm{1}+D)
 \ee

\section{Zonal functions \label{zonal-section}}

 \be
\oint\dots\oint \Delta^4(z) \,\prod_{i=1}^N
e^{2\sum_{m=1}^\infty\, z_i^mt_m}e^{2\sum_{m=1}^\infty\,
z_i^{-m}{\bar t}_m}dz_i
=\sum_{\lambda\in\Pa}\,\,e^{-F_\lambda^{({\frac
12})}}\,J^{({\frac 12})}_\lambda(\bt)J^{({\frac
12})}_\lambda({\bar\bt})
 \ee

 ==================

 is the particular case of

Thanks to
\[
\oint \cdots \oint \, \prod_{i\neq j
}^N\,(1-x_ix_j^{-1})^{\frac1\alpha} \,
J_\lambda^{(\alpha)}(x)\,J_\lambda^{(\alpha)}(x)\ \,
\prod_{i=1}^Nd x_i=\prod_{1 \le i<j\le n}\frac{\Gamma(\xi
_i-\xi_j+\frac1\alpha)\Gamma(\xi
_i-\xi_j-\frac1\alpha+1)}{\Gamma(\xi _i-\xi_j)\Gamma(\xi
_i-\xi_j+1)}
 \]
where $\xi_i:=\lambda_i+\frac1\alpha(N-1),\, 1\le i\le N$,
\cite{Mac} we have
 \be \oint\dots\oint \Delta_N^\beta(z)
\,\prod_{i=1}^N e^{2\sum_{m=1}^\infty\,
z_i^mt_m}e^{2\sum_{m=1}^\infty\, z_i^{-m}{\bar
t}_m}dz_i=\sum_{\lambda\in
\Pa}\,e^{-F_\lambda^{(\alpha)}}\,J^{(\alpha)}_\lambda(\bt)J^{(\alpha)}_\lambda({\bar\bt})
 \ee
 where
  \[
e^{-F_\lambda^{(\alpha)}}= .... \prod_{1 \le i<j\le
n}\frac{\Gamma(\xi _i-\xi_j+\frac1\alpha)\Gamma(\xi
_i-\xi_j-\frac1\alpha+1)}{\Gamma(\xi _i-\xi_j)\Gamma(\xi
_i-\xi_j+1)}
  \]

 \subsection{Appendix. On complex beta-ensembles \label{on complex-section}}

Below everything is wrong (kak polnyj kozel: v pyatyj raz
oshibayus' v odnom i tom zhe meste)

 In this section the bar means complex conjugation: ${\bar \bt}=({\bar t}_1,{\bar
  t}_2,\dots)$ denote the complex conjugated ${ \bt}=({ t}_1,{
  t}_2,\dots)$; ${\bar z}_i$ is the complex conjugated $z_i$.

Consider the following complex beta-ensemble:
 \be\label{CbetaE}
Z_N^{(\beta)}:=\int_C \cdots \int_C \, \prod_{n<m\le
N}|z_n-z_m|^{\beta} \,  e^{W(z_1,\dots,z_N)} e^{\frac 12
{\beta}\sum_{n=1}^\infty \left(t_n z_i^n +{\bar t}_n {\bar
z}_i^{n}\right)} \, d z_i d {\bar z}_i
 \ee
 where $W(z_1,\dots,z_N)$ is some potential, which is a symmetric
 function of variables $z_i$. Such integrals appear in the theory
 of quantum Hall droplets.

====

Introduce variables
 \be\label{t-z-N} \quad t_j=\frac
1j\,{\sqrt\frac{2}{\beta}}\,\,\sum_{n=1}^N \, z_n^j ,\quad
\beta=\frac{2}{\alpha}
  \ee
  As we see
  \[
d t_1\wedge \cdots \wedge d t_N= \left(\frac \beta 2\right)^{\frac
N2}\Delta_N(z) \, dz_1\wedge \cdots \wedge dz_N
  \]
In the large $N$ limit we want to replace integrals over
$z_i,{\bar z}_i,\, i=1,\dots,N$ by integrals over $t_i,{\bar
t}_i\, i=1,\dots,N$.

====

Introduce variables
 \be\label{t-z-N-gamma} \quad t_j=\frac
\gamma j\,\,\,\sum_{n=1}^N \, z_n^j
  \ee
  As we see
  \[
d t_1\wedge \cdots \wedge d t_N \, = \,\gamma ^{N}\,\Delta_N(z)\,
dz_1\wedge \cdots \wedge dz_N
  \]

Introduce
 \be\label{Cauchy-Littlewood-PQ}
\tau^{(\alpha)}(\bt,0,{\bar \bt}):= e^{\frac{2}{\alpha}
\sum_{n=1}^\infty nt_n{\bar t}_n}=\sum_\lambda \,
P_\lambda^{(\alpha)}(\bt)\,Q_\lambda^{(\alpha)}({\bar \bt})
 \ee
and
 \be \tau^{(\alpha)}(\bt,\bt^*,{\bar \bt}):=
 e^{-\sum_{n=1}^\infty \mathbb{H}^{(\alpha)}_n(\bt) t_n^*}\cdot e^{\frac{2}{\alpha}
\sum_{n=1}^\infty nt_n{\bar t}_n} =e^{-\sum_{n=1}^\infty
\mathbb{H}^{(\alpha)}_n(\bt) t_n^*}\cdot \sum_\lambda \,
P_\lambda^{(\alpha)}(\bt)\,Q_\lambda^{(\alpha)}({\bar \bt})
 \ee
 \be=\sum_\lambda \,
e^{-\sum_{n=1}^\infty\varepsilon^{(\alpha)}_n(\lambda)t_n^*}\,
P_\lambda^{(\alpha)}(\bt)\,Q_\lambda^{(\alpha)}({\bar \bt})\,
 \ee
 ????????????  where
 $\varepsilon^{(\alpha)}_\lambda$ is the "energy" of a
  configuration $\lambda=(\lambda_1,\dots,\lambda_\ell)$ defined as
   \be
\varepsilon^{(\alpha)}_\lambda:=\sum_{k=1}^{\ell(\lambda)}\,
\left(\varepsilon^{(\alpha)}_{\lambda_k-\frac12k\beta
}-\varepsilon^{(\alpha)}_{-\frac12k\beta }\right)
   \ee
 ??????????? where (???? n)
   \be
\varepsilon^{(\alpha)}_\lambda
P_\lambda=\mathbb{H}_n^{(\alpha)}P_\lambda
   \ee

As one can see for any $N$ we have the equality
 \be\label{Z-N-via-t}
\int\cdots \int \,\prod_{n=1}^N \,e^{\frac{2}{\alpha} \, nt_n'{
t}_n} \,e^{-\frac{2}{\alpha} \, nt_n{\bar t}_n}\,
e^{\frac{2}{\alpha} \, n{\bar t}_n t_n''}\,\,\frac{\alpha n \,\,
dt_nd{\bar t}_n}{2\pi \sqrt{-1} } =e^{\frac{2}{\alpha}
\sum_{n=1}^N nt_n'{ t}_n''}
 \ee
This follows from
 \be\label{t-lambda-ortogonality}
\int\cdots \int \,p_\lambda p_\mu\,\prod_{n=1}^N
e^{-\frac{2}{\alpha} \, nt_n{\bar t}_n} \,\,\frac{\alpha n \,
dt_nd{\bar t}_n}{2\pi \sqrt{-1} }
=\alpha^{\ell(\lambda)}z_\lambda\delta_{\lambda,\mu}
 \ee
 where for $\lambda=(\lambda_1,\dots,\lambda_k)$, $\mu=(\mu_1,\dots,\mu_s)$ it is supposed that
  $0\neq\lambda_k\le N $, $0\neq\mu_s\le N $, and where
 $p_\lambda:=p_{\lambda_1}\cdots p_{\lambda_k}$, $z_\lambda=\prod_i
 i^{m_i}m_i!$ where $m_i=m_i(\lambda)$ is the number of parts of $\lambda$
 equal to $i$ and $p_n:=nt_n,\,n=1,2,\dots$, $\ell(\lambda)=k$.

\begin{Remark} Changing variables via \eqref{t-z-N-gamma} we
obtain
 \be
\label{Z-N-via-z-gamma} N! \frac{ \alpha^N  }{(2\pi \sqrt{-1})^N }
\int\cdots \int \, \, \prod_{i,j=1}^N\,(1-z_i{\bar
z}_j)^{-\frac{2\gamma^2}{\alpha}}\,\,|\Delta_N(z)|^2\,\prod_{m=1}^N
\,e^{\frac{2\gamma}{\alpha} \,\sum_{n=1}^N\, (t_n'{ z}_m^n +
t_n''{\bar z}_m^n)}\, dz_md{\bar z}_m
 \ee
 \[
=e^{\frac{2}{\alpha} \sum_{n=1}^N nt_n'{ t}_n''}
 \]

\end{Remark}

 Then, relation \eqref{t-lambda-ortogonality} results in
 \be\label{P-Q-ortogonality}
\int\cdots \int \, P_\lambda Q_\mu\, \prod_{n=1}^N\,
e^{-\frac{2}{\alpha} \, nt_n{\bar t}_n} \,\,\frac{\alpha n \,
dt_nd{\bar t}_n}{2\pi \sqrt{-1} } =\delta_{\lambda,\mu}
 \ee
 which gives rise to \eqref{Z-N-via-t} via the second equality in
 \eqref{Cauchy-Littlewood-PQ}.

In the large $N$ limit we can write
 \be \int \,
\tau^{(\alpha)}(\bt',0,\bt) \tau^{(\alpha)}(\bt,0,{\bar \bt})
\tau^{(\alpha)}({\bar \bt},0,\bt'')\,\prod_{n=1}^\infty\,
\frac{\alpha n\, dt_nd{\bar t}_n}{2\pi \sqrt{-1}
}=\tau^{(\alpha)}(\bt',0,\bt'')
 \ee
 and thanks to the Hermitian property of Calogero Hamiltonians
 inside the integral we obtain
 \be\label{Z-N-as-integral-over-tau}
\int \, \tau^{(\alpha)}(\bt',{\bt^*}',\bt)
\tau^{(\alpha)}(\bt,\bt^*,{\bar \bt}) \tau^{(\alpha)}({\bar
\bt},{\bt^*}'',\bt'')\,\prod_{n=1}^\infty \frac{\alpha n \,
dt_nd{\bar t}_n}{2\pi \sqrt{-1} }
=\tau^{(\alpha)}(\bt',{\bt^*}'-\bt^*+{\bt^*}'',\bt'')
 \ee
  (For $\alpha=1$ it was marked in \cite{TauFuncMI}.)

wrong:

 In the large $N$ limit, $Z_N^{(\beta)}$ of \eqref{CbetaE}
  \[
  \prod_{n<m\le
\infty}|z_n-z_m|^{\beta} \,
e^{W(z_1,\dots,z_N)}\prod_{i=1}^\infty \, d z_i d {\bar
z}_i:=\tau^{(\alpha)}(\bt,\bt^*,{\bar
 \bt})\prod_{i=1}^\infty\,dt_id{\bar t}_i
  \]
and where variables $t_i,\,i=1,2,\dots, t_N$ are related to
variables $z_i,\,i=1,2,\dots,z_N$ via \eqref{t-z-N} where we send
$N$ to $\infty$.

Example. $\beta=2$. Choose $\bt^*$ in a way that
 \be
\tau^{(\alpha)}(\bt,\bt^*,{\bar
\bt})=\sum_{\lambda}\,\frac{1}{(N)_\lambda}
s_\lambda(\bt)s_\lambda({\bar \bt}) =\frac{\det\left[e^{z_i{\bar
z}_j}\right]}{|\Delta(z)|^{2}}
 \ee

***************************************

 \[
\oint \cdots \oint \, \prod_{i\neq j
}^N\,(1-x_ix_j^{-1})^{\frac1\alpha} \,
J_\lambda^{(\alpha)}(x)\,J_\lambda^{(\alpha)}(x)\ \,
\prod_{i=1}^Nd x_i=\prod_{1 \le i<j\le n}\frac{\Gamma(\xi
_i-\xi_j+\frac1\alpha)\Gamma(\xi
_i-\xi_j-\frac1\alpha+1)}{\Gamma(\xi _i-\xi_j)\Gamma(\xi
_i-\xi_j+1)}
 \]
where $\xi_i:=\lambda_i+\frac1\alpha(N-1),\, 1\le i\le N$,
\cite{Mac}.

\end{document}